\documentclass[lettersize,journal]{IEEEtran}

\IEEEoverridecommandlockouts
\usepackage{cite}
\usepackage{amsmath,amssymb,amsfonts}
\usepackage{algorithmic}
\usepackage{graphicx}
\usepackage{textcomp}
\usepackage[dvipsnames]{xcolor}
\usepackage{siunitx}

\usepackage{xspace}
\usepackage{tikz}
\usepackage[most]{tcolorbox}
\usepackage{graphbox}
\usepackage{mathtools}
\usepackage[most]{tcolorbox}
\usepackage{fancybox}
\usepackage{makecell}
\usepackage{multirow}
\usepackage{booktabs}
\usepackage[hyphens]{url}
\usepackage{hyperref}
\hypersetup{hidelinks}
\hypersetup{breaklinks=true}
\usepackage{cleveref}
\usepackage[normalem]{ulem} %
\usepackage{amssymb}
\usepackage{amsfonts}
\usepackage{tabularx}
\usepackage{makecell}

\definecolor{myBlue}{RGB}{0,133,255}
\newcommand\phase[1]{\tikz[baseline=(X.base)]\node [draw=black,fill=white,thick,rectangle,inner sep=2pt, rounded corners=2pt](X){\color{black}\textbf{#1}};}

\def\BibTeX{{\rm B\kern-.05em{\sc i\kern-.025em b}\kern-.08em
    T\kern-.1667em\lower.7ex\hbox{E}\kern-.125emX}}
    
\newcommand{\NAME}{\texttt{Diagnosis}\xspace}

\usepackage[framemethod=TikZ]{mdframed}
\mdfsetup{nobreak=false}
\newenvironment{Answer}[1][]{%
  \ifstrempty{#1}%
  {\mdfsetup{%
    frametitle={%
      \tikz[baseline=(current bounding box.east),outer sep=0pt]
      \node[line width=0pt,anchor=east,rectangle,draw=white,fill=white]
    ;}}
  }%
  {\mdfsetup{%
    frametitle={%
      \tikz[baseline=(current bounding box.east),outer sep=0pt]
      \node[anchor=east,rectangle,draw=white,fill=white]
    {\strut #1};}}%
  }%
  \mdfsetup{innertopmargin=-5pt,linecolor=black,%
            linewidth=0.5pt,topline=true,%
            frametitleaboveskip=\dimexpr-\ht\strutbox\relax,skipabove=\topskip,skipbelow=\topskip}
  \begin{mdframed}[nobreak=false]\relax
  }{\end{mdframed}}

\definecolor{col1}{RGB}{255,0,0}
\definecolor{col2}{RGB}{255,255,0}
\definecolor{col3}{RGB}{0,0,255}
\definecolor{col4}{RGB}{0,255,0}

\newcommand{\colorOne}{red\xspace}
\newcommand{\colorTwo}{yellow\xspace}
\newcommand{\colorThree}{blue\xspace}
\newcommand{\colorFour}{green\xspace}

\usetikzlibrary{calc}

\newcommand{\markboxB}[3]{%
  \tikz[baseline=(X.base)]{
    \node[
      draw,
      #1,                    %
      fill=col2!30,
    inner sep=1.2pt,
      minimum height=1.3em,
      line width=0.4pt
    ] (X) { #3};
  }%
}
\newcommand{\markboxC}[3]{%
  \tikz[baseline=(X.base)]{
    \node[
      draw,
      #1,                    %
      fill=col3!30,
    inner sep=1.2pt,
      minimum height=1.3em,
      line width=0.4pt
    ] (X) { #3};
  }%
}
\newcommand{\markboxD}[3]{%
  \tikz[baseline=(X.base)]{
    \node[
      draw,
      #1,                    %
      fill=col4!30,
    inner sep=1.2pt,
      minimum height=1.3em,
      line width=0.4pt
    ] (X) {#3};
  }%
}

\newcommand{\markB}[1]{\markboxB{dashed}{col2!30}{#1}}        %
\newcommand{\markC}[1]{\markboxC{dotted}{col3!30}{#1}}        %
\newcommand{\markD}[1]{\markboxD{}{col4!30}{#1}}      %

\usepackage{multirow}
\usepackage{graphicx}
\usepackage{textcomp}
\usepackage{listings}
\usepackage{caption}
\usepackage{subcaption}
\usepackage{longtable}
\usepackage{enumitem}
\newcommand\real{\mathbb{R}}

\definecolor{keywordcolor}{RGB}{127,0,85}
\newcommand{\lit}[1]{\textbf{\texttt{\textcolor{keywordcolor}{#1}}}}
\newcommand\synt[1]{\texttt{#1}}
\newcommand{\attt}{\ensuremath{\mathbin{\lit{@t}}}}
\newcommand{\atst}{\ensuremath{\mathbin{\lit{@i}}}}

\newcommand\trace{\ensuremath{ \pi}\xspace}
\newcommand\property{\ensuremath{ \phi}\xspace}

\newcommand\protrace{\ensuremath{\langle \trace,\property\rangle}\xspace}

\newcommand\budget{\ensuremath{b}}

\newcommand\True{\ensuremath{True}}

\newcommand\ourlogic{HLS\xspace}

\newcommand\tracerequirementcombinations{17\xspace}

\newcommand\nrequirementsARCH{16\xspace}

\newcommand\tracerequirementcombinationssimulink{16\xspace}

\newcommand\numexperiments{34\xspace}
\newcommand{\NAMESIMULINK}{\texttt{ATheNA-S}\xspace}

\newcommand{\simulink}{Simulink\xspace}

\usepackage[english]{babel}
\usepackage{amsthm}

\theoremstyle{definition}
\newtheorem{definition}{Definition}[]
\definecolor{keywordcolor}{RGB}{127,0,85}

\definecolor{background}{rgb}{0.94,0.95,0.96}

\usepackage{graphicx}
\usepackage[dvipsnames]{xcolor}
\usepackage{adjustbox}

\usepackage{calc}

\newlength{\DepthReference}
\newlength{\HeightReference}
\newlength{\Width}%

\newcommand{\MyColorBox}[2][red]%
{%
    \settowidth{\Width}{#2}%
    \colorbox{#1}%
    {%
        \raisebox{-\DepthReference}%
        {%
                \parbox[b][\HeightReference+\DepthReference][c]{\Width}{\centering#2}%
        }%
    }%
}
\usepackage{listings}

\begin{document}

\title{Search-based Trace Diagnostic for Cyber-Physical Systems}

\author{
Ricardo Caldas\thanks{This work has been submitted to the IEEE for possible publication. Copyright may be transferred without notice, after which this version may no longer be accessible.}\thanks{The first three authors contributed equally to this paper.}, 
Gabriel~Araujo\thanks{}, Federico~Formica\thanks{},  Genaína~Rodrigues, Patrizio~Pelliccione,~\IEEEmembership{Member,~IEEE}, Claudio~Menghi,~\IEEEmembership{Senior Member,~IEEE}
\thanks{
R.~Caldas, is with Gran Sasso Science Institute (GSSI), L'Aquila, Italy - e-mail: \{ricardo.caldas@gssi.it\}\newline
G.~Araujo, is with University of Brasília, Brasília,  Brazil - e-mail:
\{frutuoso.gabriel@aluno.unb.br\}\newline
F.~Formica is with McMaster University, Hamilton, Canada - e-mail: \{formicaf@mcmaster.ca\}\newline
G. Rodrigues is with University of Brasília, Brasília,  Brazil - e-mail:
\{genaina@unb.br\}\newline
P.~Pelliccione is with Gran Sasso Science Institute (GSSI), L'Aquila, Italy - e-mail: \{patrizio.pelliccione@gssi.it\}\newline
C.~Menghi is with the University of Bergamo, Bergamo, Italy and the McMaster University, Hamilton, Canada - e-mail: \{claudio.menghi@unibg.it\}\newline
}
}

\maketitle

\begin{abstract}
Cyber-physical systems (CPS) development requires verifying whether system behaviors violate their requirements.
This analysis often considers system behaviors expressed by execution traces and requirements expressed by signal-based temporal properties.
When an execution trace violates a requirement, engineers must solve the trace diagnostic problem---they need to understand the cause of the breach. 
Automated trace diagnostic techniques aim to support engineers in the trace diagnostic activity.

This paper proposes search-based trace diagnostic (SBTD), a novel trace diagnostic technique for CPS requirements.
Unlike existing techniques, SBTD relies on evolutionary search.
SBTD starts from a set of candidate diagnoses, applies an evolutionary algorithm to generate new candidate diagnoses (via mutation, recombination, and selection), and uses a fitness function to determine the qualities of these solutions. Then, a diagnostic generator step is performed to explain the cause of the trace violation.
We implemented \NAME, an SBTD tool for signal-based temporal logic requirements expressed using the Hybrid Logic of Signals (HLS).
We evaluated \NAME by performing \numexperiments experiments for \tracerequirementcombinations trace-requirement combinations for property violations.
We assessed the effectiveness of SBTD in producing informative diagnoses and its efficiency.
\NAME achieved expert-aligned diagnoses for 29/34 experiments and scaled to the full HLS benchmark, whereas state-of-the-art literature remained restricted to a subset due to performance and language limitations.
SBTD treats trace-checking as a black box, which makes the checker replaceable. Substituting our HLS checker for an STL monitor, e.g., RTAMT, reproduces on two requirements the diagnoses at two to three orders of magnitude lower per-check cost.

\end{abstract}


\section{Introduction}
Cyber-physical systems (CPS) must satisfy temporal requirements.
These requirements can be expressed as \emph{signal-based temporal properties}~\cite{boufaied2020trace,menghi2021trace}. 
Signal-based temporal properties are a convenient tool for specifying  CPS requirements because they enable engineers to define how the system should behave over time by using signals.
Signals are entities that capture the values assumed by the system variables over time, and they can record both the software and physical dynamics of the CPS under analysis~\cite{menghi2021trace}.

Developing complex CPS requires engineers to test their systems to search for requirements violations. 
Many testing techniques (e.g.,
\cite{ARIsTEO,formica2022search,Waga20,falsQBRobCAV2021,NNFal,peltomaki2023requirement,annpureddy2011s,psytalirotool}), compared by existing competitions (e.g.,~\cite{menghi2023arch}), rely on \emph{trace-checking} to assess whether a trace that records the system behavior for a specific input leads to a requirement violation.
For example, \texttt{ThEodorE}~\cite{menghi2021theodore} is a trace-checking tool that supports requirements expressed in the Hybrid Logic of Signals (HLS)~\cite{menghi2021trace}, an expressive logic to capture CPS requirements.
Trace-checking techniques typically take a trace and a property that represents a requirement, and return a Boolean verdict: \emph{True} if the trace satisfies the requirement, and \emph{False} otherwise. 
If the trace satisfies the requirement, testing tools automatically generate new test cases to search for a test case that violates it.
In the opposite case, engineers must inspect the trace to understand the cause of the violation. 

\emph{Trace diagnostic} approaches consider a trace and a requirement violated by the trace and aim to explain why the requirement is violated.
Existing approaches either isolate slices of traces explaining the requirement violation~\cite{ferrere2015trace,mukherjee2012computing,beer2012,nivckovic2020amt} or check whether traces show common behaviors that lead to the requirement violation~\cite{dawes2019,dou2018,boufaied2023,luo2014rv}. 
Recent work~\cite{boufaied2023} showed the applicability of the latter approach for producing diagnoses for signal-based temporal requirements.
The approach requires a language-specific library of violation causes and diagnoses. 
Then, it analyzes the causes of violations and diagnoses within the library to produce an explanation for the requirement violation.

Two challenges may hamper the applicability of trace diagnostic solutions in practical scenarios.
\begin{itemize}
\item Challenge \emph{C1}: a library of violation causes and diagnoses is typically required and often unavailable before execution;
\item Challenge \emph{C2}: a valid explanation for the requirement violation within the violation causes and diagnoses might be missing from the library, i.e., none of the violation causes and diagnoses are suitable to explain the requirement violation. 
\end{itemize}
This work mitigates these challenges by proposing search-based trace diagnostic (SBTD), a novel trace diagnostic framework for CPS. 
Unlike existing techniques, SBTD uses an evolutionary search to generate new candidate diagnoses. 
SBTD treats trace-checking as a black box and only exploits its verdict.
This makes the diagnosis layer largely independent of the specific requirement language and checker, enabling (in principle) substitution of the backend without changing the search-and-diagnosis logic.
We adopt a \emph{change-driven} view of diagnosis: a diagnosis is a small set of targeted changes to a selected portion of the requirement such that the resulting mutated requirement becomes satisfied by the observed trace.

This automated generation enables the dynamic creation of new diagnoses and provides two benefits.
First, it addresses challenge \emph{C1} since it does not require a library of predefined violation causes and diagnoses as input. 
This relieves engineers from defining such a library when it is unavailable, a time-consuming and error-prone task.
Second, it addresses the challenge \emph{C2} since the dynamic generation of new diagnoses mitigates the risk of terminating the trace diagnostic procedure without a valid explanation for the requirement violation.
We defined \NAME, an instance of SBTD that considers properties modeled using the Hybrid Logic of Signals (HLS)~\cite{menghi2021trace}.

Our objective is to support \emph{debugging} once a violation has been found: Given a violated requirement and trace, we aim to explain \emph{which sub-condition(s)} and \emph{which signals} are responsible for the violation. Although SBTD operationalizes this by searching over small, structured \emph{mutations} and learning an interpretable decision tree, the mutations are used as \emph{diagnostic probes} to localize the inconsistency, not as a recommendation to change the requirement. Therefore, even when requirements are immutable, SBTD complements trace-checking by providing actionable guidance on where to inspect the implementation and/or instrumentation.

We evaluated our solution by performing \numexperiments experiments involving \tracerequirementcombinations trace-requirement combinations that led to a property violation.
These trace-requirement combinations are from three different systems, i.e., two from the automotive domain and one from the robotic domain. 
We assess the effectiveness of our solution (\textbf{RQ1}), i.e., the capability of \NAME to produce informative diagnoses, and its efficiency (\textbf{RQ2}), i.e., the time required to produce them.
We also assess its cost-effectiveness (\textbf{RQ3}), i.e., how the choice of mutation space and budget affects the resulting diagnosis and its cost.

Overall, \NAME produced informative, expert-aligned diagnoses for {29} out of \numexperiments experiments within a practical time budget, and it applies to the full HLS benchmark.
In contrast, a state-of-the-art baseline (TD-SB-TemPsy~\cite{boufaied2023}) could return diagnoses only on a limited subset of our requirements due to performance and language limitations.
As expected with search-based procedures, a small number of cases may require substantially longer or may not converge within the time budget. 
The cost of these cases is dominated by trace-checking rather than by the search. Replacing our HLS checker with an STL monitor (i.e., RTAMT) reproduces the diagnoses at a substantially lower per-check cost, at the price of a less expressive requirement language.

We analyze these outliers in \Cref{sec:evalaution}.

To summarize, our contributions are as follows:
\begin{itemize}
    \item search-based trace diagnostic (SBTD), a novel trace diagnostic technique for CPS based on evolutionary search (\Cref{sec:approach});
    \item an SBTD framework that supports properties expressed using the HLS logical language (\Cref{sec:sbtforhls});
    \item the implementation of our SBTD framework, namely \NAME, which is publicly available;
    \item an extensive empirical evaluation of our solution (\Cref{sec:evalaution}).
\end{itemize}

Our paper is organized as follows.
\Cref{sec:background} presents our running and motivating example from the automotive domain.
\Cref{sec:approach} introduces SBTD.
{\Cref{sec:sbtforhls} presents \NAME, our instantiation of SBTD for HLS trace diagnosis.}
\Cref{sec:evalaution} evaluates our contribution.
\Cref{sec:discussion} reflects on our findings. 
\Cref{sec:related} discusses related work.
\Cref{sec:conclusion} presents our conclusions.
\section{Motivating Example}
\label{sec:background}
\begin{figure}[t]
    \centering
    \includegraphics[width=\columnwidth]{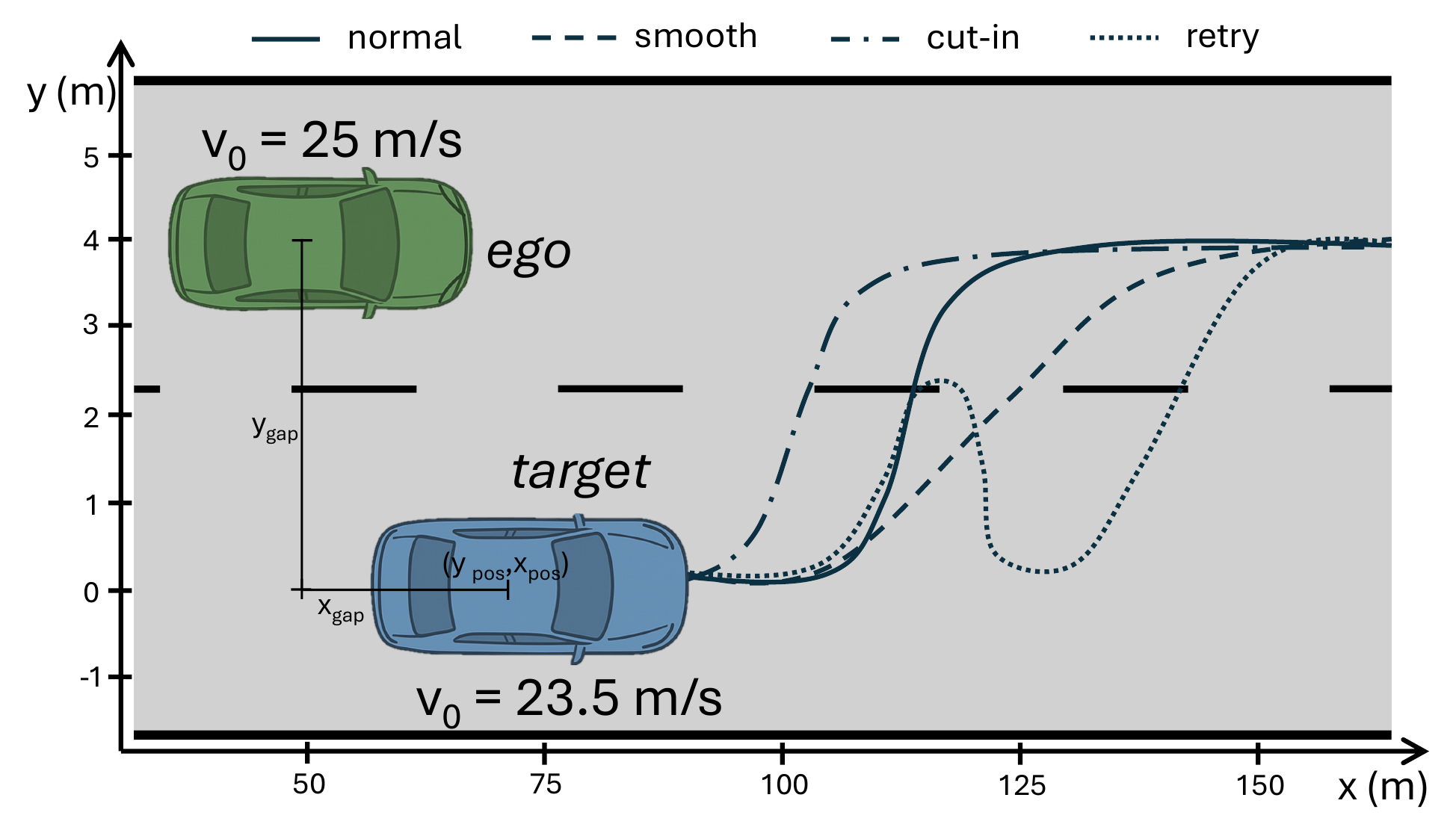}
    \caption{Example of failure-revealing scenario.  The solid, dashed, dash-dotted, and dotted lines represent the trajectories of the vehicle performing the cut-in in four different scenarios.}
    \label{fig:runningExample}
\end{figure}

Our running example concerns a lane change scenario. The {\it ego} vehicle should maintain the lane and avoid collisions with another vehicle in an adjacent lane, performing variations of a cut-in maneuver, such as the one from \Cref{fig:runningExample}.

\begin{figure}[t]
\footnotesize
\begin{tabular}{l@{\ }l@{\ }p{55mm}}
\toprule
\emph{Formula} &$ \synt{p}  \Coloneqq$ & $ \synt{tm}_1 \oplus \synt{tm}_2  \mid \lit{not}\ \synt{p} \mid \synt{p}_1\ \ominus \ \synt{p}_2  $\\
 & & $\mid $ $\triangleright\ \tau\ \lit{in}\ I_T\ \lit{such that}\  \synt{p}$ \\  & &$\mid $ $ \triangleright\ \sigma\ \lit{in}\ I_J \lit{such that}\ \synt{p}  $ \\
 & &  $\mid 
 \triangleright\  \rho\ \lit{such that}\ \synt{p} $ \\
 \midrule
\emph{Term} & $\synt{tm}   \Coloneqq $ & $ \synt{tt} \mid \synt{vt}  \mid \synt{it}$ \\
\midrule
\emph{Time Term} & $\synt{tt}  \Coloneqq$ & $\tau \mid t \mid \lit{i2t}(\synt{it})  \mid \synt{tt}_1 \odot \synt{tt}_2$   \\
\midrule
\emph{Index Term} & $\synt{it}  \Coloneqq$ & $\sigma \mid j  \mid 
\lit{t2i}(\synt{tt}) \mid \synt{it}_1 \odot \synt{it}_2$ \\
\midrule
\emph{Value Term} & $\synt{vt}   \Coloneqq$ & $  \rho \mid  x \mid (s \atst \synt{it}) \mid (s \attt \synt{tt})  \mid \synt{vt}_1 \odot \synt{vt}_2$\\
\bottomrule\\
\end{tabular}\\
$t,x \in \real,  
j \in \mathbb{N}^+, 
I_T \subseteq \real, 
I_J \subseteq \mathbb{N}^+,
\tau  \in \mathit{TV}, 
\sigma \in \mathit{IV}, 
\rho \in \mathit{RV}, s \in S$\\
$\odot \in \{+, -, * , /\}$; \\
$\oplus \in \{>, <, \leq, \geq, =,\not=\}$\\
$\ominus \in \{\lit{or}, \lit{and}, \lit{implies}\}$\\
$\triangleright \in \{\lit{exists},\lit{forall}\}$.

\caption{Syntax of HLS.}
\label{tab:grammar}
\end{figure}

\begin{figure*}[htb!]
    \centering

    \begin{minipage}[t]{0.49\linewidth}
        \centering

        \begin{subfigure}[t]{\linewidth}
            \centering
            \includegraphics[width=\linewidth]{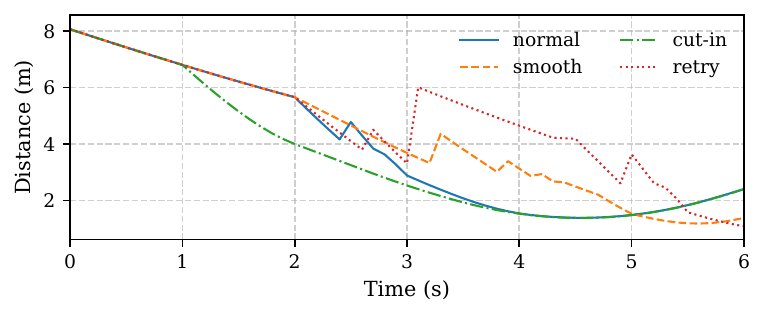}
            \caption{Distance between vehicles.}
            \label{fig:traceExample_dist}
        \end{subfigure}\vspace{0.4em}

        \begin{subfigure}[t]{\linewidth}
            \centering
            \includegraphics[width=\linewidth]{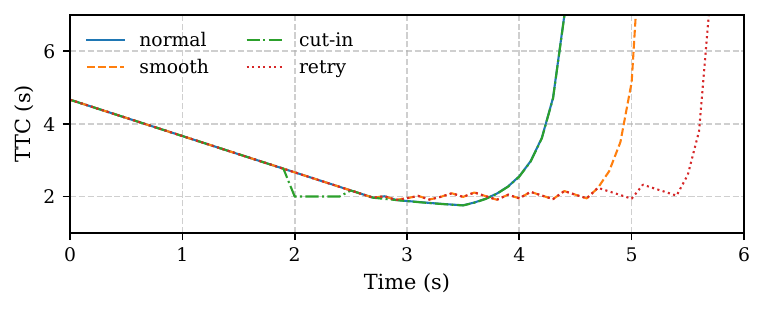}
            \caption{Time-to-collision (TTC).}
            \label{fig:traceExample_ttc}
        \end{subfigure}\vspace{0.4em}

        \begin{subfigure}[t]{\linewidth}
            \centering
            \includegraphics[width=\linewidth]{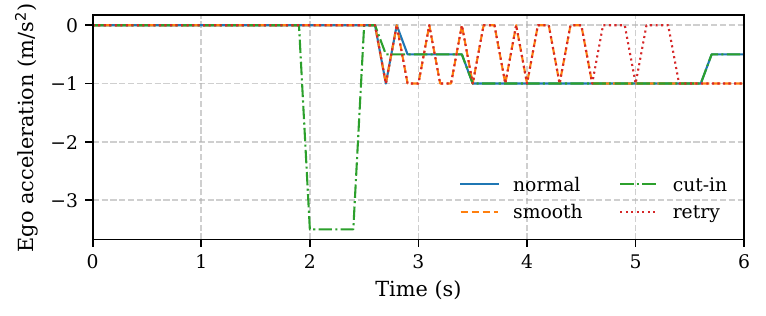}
            \caption{Ego acceleration.}
            \label{fig:traceExample_acc}
        \end{subfigure}
    \end{minipage}\hfill
    \begin{minipage}[t]{0.49\linewidth}
        \centering

        \begin{subfigure}[t]{\linewidth}
            \centering
            \includegraphics[width=\linewidth]{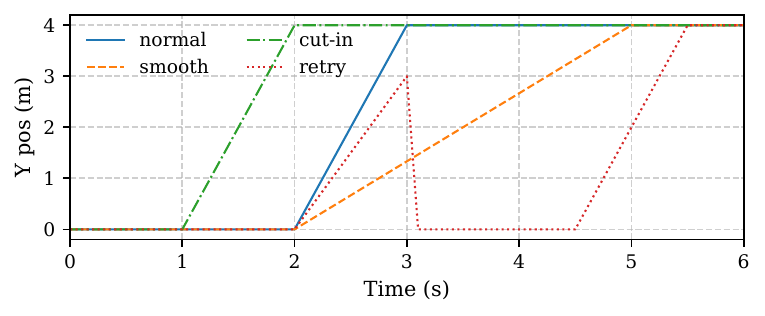}
            \caption{Target lateral position.}
            \label{fig:traceExample_targety}
        \end{subfigure}\vspace{0.4em}

        \begin{subfigure}[t]{\linewidth}
            \centering
            \includegraphics[width=\linewidth]{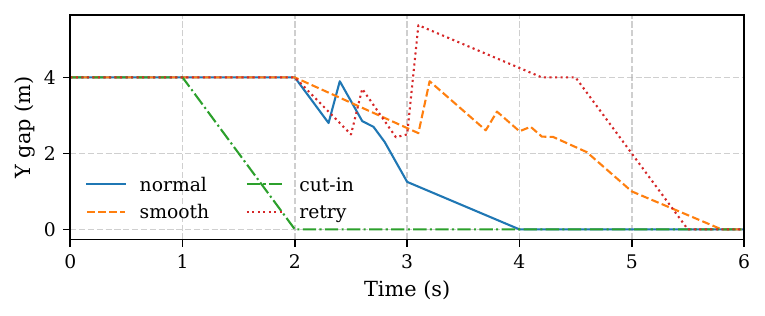}
            \caption{Lateral gap $|\Delta y|$.}
            \label{fig:traceExample_ygap}
        \end{subfigure}\vspace{0.4em}

        \begin{subfigure}[t]{\linewidth}
            \centering
            \includegraphics[width=\linewidth]{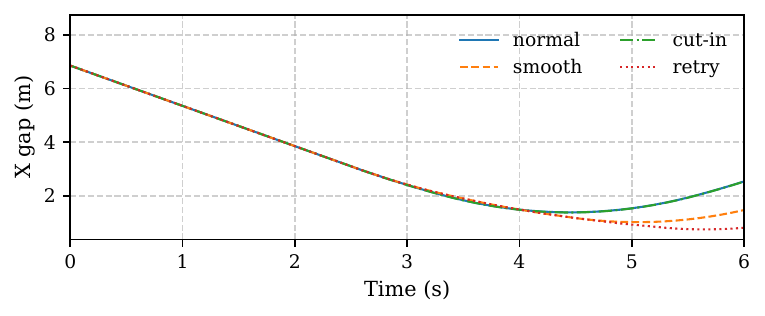}
            \caption{Longitudinal gap $|\Delta x|$.}
            \label{fig:traceExample_xgap}
        \end{subfigure}
    \end{minipage}

    \caption{Fragments of execution traces for the running example, shown for the four scenarios.}
    \label{fig:traceExample}
\end{figure*}

Engineers specify the system requirements using HLS.
\Cref{tab:grammar} presents the grammar of \ourlogic from~\cite{menghi2021trace}, where symbol ``$\mid$'' separates alternatives,
$\mathit{TV}$ is a set of timestamp variables, 
$\mathit{IV}$ is a set of index variables,  $\mathit{RV}$ is a set of real-valued variables, and 
$S$ is a set of signal variables.\footnote{We slightly revisited the presentation of the grammar to include derived operators (e.g., \lit{forall}, \lit{and}).} 
An HLS \emph{formula} (non-terminal \synt{p}) is a relational expression over terms,
a Boolean expression over formulae, 
or quantified formulae.
Quantified formulae support  quantification over  timestamp variables ($\triangleright$\ $\tau$ \lit{in} $I_T$ [\dots]), over index variables  ($\triangleright$\ $\sigma$ \lit{in} $I_J$ [\dots]), or over real-valued variables  ($\triangleright$\ $\rho$ [\dots]), where $\triangleright$\ represents the existential (\lit{exists}) or universal (\lit{forall}) quantifier.
A \emph{term} (non-terminal \synt{tm}) is a \emph{time term}, an \emph{index term}, or a \emph{value term}.
A \emph{time term} (non-terminal \synt{tt}) is a timestamp variable $\tau$, a literal denoting a value $t$,
the value returned by the operator \lit{i2t} (``index to timestamp''), or an arithmetic expression over these entities.
An  \emph{index term} (non-terminal \synt{it})  is an index variable $\sigma $, a literal denoting a value $j$, 
the value returned by the operator \lit{t2i} (``time to index''), or an arithmetic expression over these entities.
 A \emph{value term} (non-terminal \texttt{vt}) is a real-valued variable $\rho$, a literal denoting a value $x$,
the value of a signal returned by the operators \atst\ (``at index'')  and \attt\ (``at timestamp''), or an arithmetic expression over these entities.

Using HLS, engineers specify the requirements $\property_1$, $\property_2$, $\property_3$, $\property_4$ of the vehicle:
\begin{align*}
\phi_1::=\lit{f}&\lit{orall}\  \tau_0\ \lit{in}\  [0,\infty)\ \lit{such that}\ \texttt{ttc} \attt (\tau_0) < 2.5\text{s} \\ 
& \lit{implies}\ (\lit{exists}\ \tau_1\ \lit{in}\ [\tau_0, \tau_0+0.5]\ \\
& \lit{such that}\ \texttt{brake} \attt (\tau_1)  = 1)
\end{align*}
Requirement $\phi_1$ specifies that, for every time instant $\tau_0$ from the beginning (time $0$) to the end ($\infty$) of the simulation, whenever the time to collision ({\tt ttc}) is lower than a threshold value (2.5s), this implies that {\it ego} will activate braking (\texttt{brake} is set to the value $1$) within the next 0.5s. 
Moreover, $\phi_1$ ensures that the {\it ego} maintains a minimal time to collision.
\begin{align*}
\phi_2::=\lit{f}&\lit{orall}\  \tau_0\ \lit{in}\  [0,\infty)\ \lit{such that}\ \\ 
& \texttt{acc} \attt (\tau_0)  < -3m/s^2\ \lit{implies}\ \texttt{ttc} \attt (\tau_0)  < 1.5\text{s}
\end{align*}
Requirement $\phi_2$ specifies that, for every time instant $\tau_0$, an abrupt deceleration (acceleration below $-3~m/s^2$) is allowed when the time to collision is already tight (below 1.5s). Otherwise, {\it ego} would brake too aggressively without an immediate safe motivation.
\begin{align*}
\phi_3::=&\ \lit{forall}\ \tau_0\ \lit{in}\ 
[\colorbox{col1!30}{0},\colorbox{col1!30}{\(\infty\)})
\ \lit{such that}\\[2pt]
& \bigl(
    (\texttt{$y_{pos}$} \attt (\tau_0) > 0.5m)\ \lit{and}\\
&\qquad (\texttt{$y_{pos}$} \attt (\tau_0) < %
\markB{3.5m})\ \lit{and}\\
&\qquad (\texttt{$x_{gap}$} \attt (\tau_0) < 2.0m)
  \,\markC{\lit{and}}\\
&\qquad (\texttt{$y_{gap}$} \attt (\tau_0) < 1.5m) \bigr)\\[2pt]
&\lit{implies}\ 
  (\lit{exists}\ \tau_1\ \lit{in}\ [\tau_0, \tau_0+1]\ \lit{such that}\\
&\qquad \texttt{$y_{gap}$} \attt (\tau_1) > \markD{1.0m})
\end{align*}
 
Requirement $\phi_3$ specifies 
that if {\it ego} detects that the target vehicle is in the central portion of the lane (between 0.5 and 3.5 meters), and within a short longitudinal (\texttt{$x_{gap}$}) and lateral (\texttt{$y_{gap}$}) distance, then {\it ego} must take defensive action to increase lateral separation within 1 second. 
The semantics of the light-colored boxes will be described in \Cref{sec:approach}.
\begin{align*}
\phi_4::=\lit{f}&\lit{orall}\ \tau_0\ \lit{in}\ [0,\infty)\ \lit{such that}\  \\
& \texttt{dist} \attt (\tau_0) < 7m\ \lit{implies}\ \texttt{brake} \attt (\tau_0) = 1
\end{align*}
Requirement $\phi_4$ specifies
that if the distance to the leading vehicle falls below a critical threshold of 7 meters, {\it ego} must activate braking immediately (the value of the variable \texttt{brake} is set to the value 1). This hard constraint is intended to enforce a minimum safety buffer and avoid potential rear-end collisions.
The requirements constrain signals of two kinds: Those the controller sets directly, such as the \texttt{brake} command in $\phi_1$ and $\phi_4$, and those that describe the resulting environment, such as the gaps (\texttt{$x_{gap}$}, \texttt{$y_{gap}$}) in $\phi_3$. 
SBTD diagnoses a violation of the requirement; its goal is not to help engineers define the correct requirements.

Automotive engineers analyze whether their system behaves correctly by considering multiple driving scenarios, shown in Fig.~\ref{fig:traceExample}. The graphs report representative signals extracted from the execution traces of these scenarios (e.g., distance, time-to-collision (TTC), ego acceleration, target's lateral position (Y pos) and lateral/longitudinal gaps between vehicles). Across all graphs, scenarios are identified by line style: normal (solid), smooth (dashed), cut-in (dash-dotted), and retry (dotted).
Requirement violations may reveal failures in particular scenarios, as summarized in \Cref{tab:prop-trace_violations}. 
For example, $\phi_4$ is not satisfied in any of the scenarios: within the interval $[1,2]$s of simulation, the Euclidean distance between vehicles ({\tt dist}) is smaller than 7 meters, yet there are no brake attempts since {\it ego}'s acceleration remains constant at $0\ m/s^2$.

\begin{figure}[t]
\centering
\begin{tabular}{lcccc}
    \toprule
    \textit{trace/req.} & $\phi_1$ & $\phi_2$ & $\phi_3$ & $\phi_4$ \\
    \midrule
    \textbf{Normal}  & {\color{green!60!black} \checkmark} & {\color{green!60!black} \checkmark} & {\color{green!60!black} \checkmark} & {\color{red} $\times$} \\
    \textbf{Smooth}  & {\color{green!60!black} \checkmark} & {\color{green!60!black} \checkmark} & {\color{green!60!black} \checkmark} & {\color{red} $\times$} \\
    \textbf{Cut-in}  & {\color{green!60!black} \checkmark} & {\color{red} $\times$} & {\color{green!60!black} \checkmark} & {\color{red} $\times$} \\
    \textbf{Retry}   & {\color{green!60!black} \checkmark} & {\color{green!60!black} \checkmark} & {\color{red} $\times$} & {\color{red} $\times$} \\
    \bottomrule

\end{tabular}
\caption{Summary of trace–requirement violations.
}
\label{tab:prop-trace_violations}
\end{figure}

When engineers inspect the execution trace, they need to understand the causes of each failure. 
For some requirements, this is relatively straightforward. 
In the cut-in scenario of \Cref{fig:runningExample},
the violation of $\phi_2$ is easy to interpret from the cut-in traces in Fig.~\ref{fig:traceExample_ttc}--\ref{fig:traceExample_acc}: while {\it ego} brakes to defer from the other vehicle, its deceleration drops below $-3~m/s^2$ even though {\tt ttc} is still above {1.5s}.
Thus, the requirement reveals that the controller may brake too abruptly.
In contrast, other violations are much harder to explain by manual inspection alone. 
Consider the retry scenario, where $\phi_3$ is falsified. 
Here, the satisfaction of the left side of the implication depends on the lateral position of the target (\texttt{$y_{pos}$}, Fig.~\ref{fig:traceExample_targety}), the lateral gap (\texttt{$y_{gap}$}, Fig.~\ref{fig:traceExample_ygap}), the longitudinal gap (\texttt{$x_{gap}$}, Fig.~\ref{fig:traceExample_xgap}), the one-second reaction window, and the logical connective joining the \texttt{$x_{gap}$} and \texttt{$y_{gap}$} conditions. 
The interaction between these quantities makes the antecedent of $\phi_3$ hold only over a narrow time interval, and the violation is witnessed only in a short fragment of the retry trace.

This example illustrates a common challenge in trace diagnosis: While the trace checker confirms that $\phi_3$ is violated, it does not explain \emph{which sub-condition(s)} and \emph{which signal values} drive the violation, nor does it indicate \emph{where} in the trace the violation becomes inevitable.
Therefore, engineers need diagnostic guidance that (i)~localizes a small violating fragment of the trace, and (ii)~identifies tight boundary conditions (in time and signal space) that separate satisfaction from violation.

Violations caused by the combined effects of temporal bounds, spatial thresholds, and logical connectives, and manifested in a narrow time window, are difficult to identify from the plots alone.
Obtaining such explanations is usually challenging: Requirements (e.g.,  expressed in HLS~\cite{menghi2021trace} or SB-TemPsy-DSL~\cite{boufaied2020trace}) typically consist of many temporal operators and have a complex structure.
The goal of SBTD is to support engineers in automatically producing diagnostic information that helps understand the causes of failures.

\section{Search-based Trace Diagnostic}
\label{sec:approach}
\Cref{fig:approach} presents an overview of SBTD. 
In this section, we describe SBTD as a generic framework. 
In \Cref{sec:sbtforhls}, we instantiate our framework for HLS and present our implementation (i.e., \NAME).

The input of SBTD is a trace-requirement combination (\protrace) consisting of a requirement formalized as a property (\property) unsatisfied over the trace (\trace) and a time budget (\budget).
SBTD either successfully returns a \emph{change-driven diagnosis} $d$ (as defined in \Cref{def:diagnosis}) or informs the user that it could not find a diagnosis within the available budget.

SBTD assumes that some syntactic elements of the property can be mutated.
We use light-colored boxes to highlight precisely the syntactic elements of $\phi_3$ (from \Cref{sec:background}) that SBTD is allowed to mutate.\footnote{To improve accessibility in grayscale/black-and-white print, we complement the background colors of highlighted tokens with distinct border styles (none/solid/dashed/dotted).}
In particular, we use $\phi_3($\colorbox{col1!30}{0}$)$ and $\phi_3($\colorbox{col1!30}{\ensuremath{\infty}}$)$ for the lower and upper bounds of the universal time window, $\phi_3($%
\markB{3.5}$)$ for the upper bound of the central-lane condition on \texttt{$y_{pos}$}, $\phi_3($%
\markC{\lit{and}}$)$ for the logical connective between the \texttt{$x_{gap}$} and \texttt{$y_{gap}$} clauses, and $\phi_3(\markD{$1.0$})$
for the postcondition threshold on \texttt{$y_{gap}$}.  
These five colored tokens correspond exactly to the five mutation positions configured for this requirement.
SBTD considers these syntactic elements and works in three steps:

\tikzstyle{output} = [coordinate]

\begin{figure}[t]
\begin{tikzpicture}[auto,
 block/.style ={rectangle, draw=black, thick, fill=white!20, text width=5em,align=center, rounded corners},
 block1/.style ={rectangle, draw=blue, thick, fill=blue!20, text width=5em,align=center, rounded corners, minimum height=2em},
 line/.style ={draw, thick, -latex',shorten >=2pt},
 cloud/.style ={draw=white, thick, shape=rectangle, fill=white!20,
 minimum height=1em}
]
 \draw (0,0) node[output] (Input) {};
 
\node[block, right of=Input,node distance=1.5cm] (PropertyMutation){\phase{1} \footnotesize Generator \\ of Mutations};

\node[cloud] at ($(PropertyMutation.north)+(0,0.6)$) (TimeBudget) {\footnotesize Time Budget ($b$)}; %

\node [output, right of=PropertyMutation,node distance=2cm] (linktwob) {};

\node [output, below of=PropertyMutation,node distance=1.5cm] (linkone) {};

\node[block, right of=linkone,node distance=2cm] (TraceChecker){\phase{2} \footnotesize Trace-Checker};

\node [output, right of=TraceChecker,node distance=2cm] (linktwo) {};

\node[block, above of=linktwo,node distance=1.5cm] (DiagnosticGenerator){\phase{3} \footnotesize Diagnostic Generator};

\node [output, right of=DiagnosticGenerator,node distance=2.5cm] (outputnode) {};

\node [output, below of=TraceChecker,node distance=1.5cm] (inputtrace) {};

\draw[-stealth] (Input.east) -- (PropertyMutation)
    node[midway,below,text width=2cm]{\footnotesize \newline Violated \newline \footnotesize Requirements\newline ($\phi$)};

\draw[-stealth] (TimeBudget.south) -- ($(PropertyMutation.north)+(0,0.05)$)
    node[midway,right]{\footnotesize}; %

\draw[-stealth] (DiagnosticGenerator) -- (outputnode)
    node[midway,below,text width=1.55 cm]{\footnotesize \;Diagnosis($d$)\newline \phantom{x} or \newline \footnotesize `Not found'};    

\draw[-] (PropertyMutation) -- (linkone)
    node[midway,below,text width=1.5cm]{};   

\draw[-stealth] (linkone) -- (TraceChecker)
    node[midway,below,text width=2.5cm]{\footnotesize Mutated\newline \footnotesize Requirements\newline ($\Psi$)};  

\draw[-] (TraceChecker) -- (linktwo)
    node[midway,below,text width=5.45cm]{\footnotesize \phantom{xxxxxxxxxxxxxxxx} Trace-checking\newline \phantom{xxxxxxxxxxxxxxxx} Verdicts\newline 
    \phantom{xxxxxxxxxxxxxxxx} ($\Delta$)};   

\draw[-stealth] (linktwo) -- (DiagnosticGenerator)
    node[midway,below,text width=1.5cm]{}; 
    
\draw[-] (TraceChecker) -- (linktwob)
    node[midway,below,text width=1.8cm]{\footnotesize  \footnotesize };   

\draw[-stealth] (linktwob) -- (PropertyMutation)
    node[midway,below,text width=1.8cm]{\footnotesize  \footnotesize }; 
 
 \draw[-stealth] (inputtrace) -- (TraceChecker)
    node[midway,right,text width=1.5cm]{\footnotesize Trace\newline ($\pi$)};   

\draw[-stealth,dashed]
  (DiagnosticGenerator.north) -- ++(0,0.3)
  -- ++(-3.0,0)
  node[midway,above,yshift=-3pt, align=center,text width=2cm]
    {\footnotesize No Informative Diagnosis}
  -- (PropertyMutation.north);

 \end{tikzpicture}
\caption{Search-based trace diagnostic. }
\label{fig:approach}
\end{figure}

\begin{figure*}[ht!]
    \centering
    {\footnotesize
    \begin{tikzpicture}[level 1/.style={sibling distance=75mm},
                    level 2/.style={sibling distance=45mm},
                    level 3/.style={sibling distance=28mm},
                    level 4/.style={sibling distance=28mm},
                    level distance=40pt
]
\node{$\phi_3(\colorbox{col1!30}{\ensuremath{\infty}})$}
 child {
    node[blue]{\emph{True} (1396)}
    edge from parent node[fill=white,left] {$\leq 4.6998$s}
 }
 child {
    node[xshift=-5mm]{$\phi_3(%
    \markC{\lit{and}})$}
    child{
        node[xshift=-10mm]{$\phi_3(\colorbox{col1!30}{\ensuremath{\infty}})$}
        child{
            node[blue]{\emph{True} (160)}
            edge from parent node[fill=white,left] {$\leq 5.2003$s}
        }
        child{
            node[]{$\phi_3(%
            \markB{3.5})$}
            child{
                node[blue]{\emph{True} (64)}
                edge from parent node[fill=white,left] {$\leq 2.8006$m}
            }
            child{
                node[red]{\emph{False} (168)}
                edge from parent node[fill=white,right] {$> 2.8006$m}
            }
            edge from parent node[fill=white,right] {$> 5.2003$s}
        }
        edge from parent node[fill=white] {\lit{and}}
    }
    child{
        node[xshift=10mm]{$\phi_3(\colorbox{col1!30}{0})$}
        child{
            node[red]{\emph{False} (881/1)}
            edge from parent node[fill=white,left] {$\leq 4.9250$s}
        }
        child{
            node[]{$\phi_3(\colorbox{col1!30}{\ensuremath{\infty}})$}
            child{
                node[blue]{\emph{True} (7)}
                edge from parent node[fill=white,left] {$\leq 5.0639$s}
            }
            child{
                node[red]{\emph{False} (18)}
                edge from parent node[fill=white,right] {$> 5.0639$s}
            }
            edge from parent node[fill=white,right] {$> 4.9250$s}
        }
        edge from parent node[fill=white] {\lit{or}}
    }
    edge from parent node[fill=white,right] {$> 4.6998$s}
 };
\end{tikzpicture}
    \hspace{1.5cm}}
    {$^\ast$ \footnotesize \emph{Leaf counts report the number of training samples reaching the leaf (its empirical support).}}
    \caption[Diagnosis generated for our automotive example]{Diagnosis generated for our automotive example.
     \label{fig:runningexample_dt}
    }
\end{figure*}

\begin{itemize}

    \item[\phase{1}] The \emph{Generator of Mutations} step generates a set $\Psi$ of candidate mutated requirements from a (set of) requirement(s).
    Our SBTD framework generates mutated requirements with high similarity with the original requirement ${\phi}$, which can more likely be informative in explaining the cause of the violation.
    For example, given the requirement $\phi_3$ of our running example from \Cref{sec:background}, the generator synthesizes the following mutated requirement ${\phi_3^\prime}$ by changing the value $\phi_3($%
    \markB{3.5}$)$ reported within the \colorTwo colored background box (3.5m) into $2.8006$m. 
\end{itemize}
\begin{align*}
\phi_3^\prime::=&\ \lit{forall}\ \tau_0\ \lit{in}\ 
[\colorbox{col1!30}{0},\colorbox{col1!30}{\(\infty\)})\ \lit{such that}\\[2pt]
&\bigl(
    (\texttt{$y_{pos}$} \attt (\tau_0) > 0.5m)\ \lit{and}\\
&\qquad (\texttt{$y_{pos}$} \attt (\tau_0) < %
\markB{2.8006m})\ \lit{and}\\
&\qquad (\texttt{$x_{gap}$} \attt (\tau_0) < 2.0m)
  \bigr)\ %
  \markC{\lit{and}}\\
&\qquad (\texttt{$y_{gap}$} \attt (\tau_0) < 1.5m)\\[2pt]
&\lit{implies}\ 
  (\lit{exists}\ \tau_1\ \lit{in}\ [\tau_0, \tau_0+1]\ \lit{such that}\\
&\qquad \texttt{$y_{gap}$} \attt (\tau_1) > \markD{1.0m})
\end{align*}
\begin{itemize}
    \item[\phase{2}] The \emph{Trace-Checker} step takes as input a set of mutated requirements $\Psi$ and determines, for each requirement, whether it is satisfied or violated by the trace \trace. It produces a set of pairs, each linking a trace–requirement combination to a Boolean value.
    This value indicates whether the requirement holds or is violated on the given trace. 
    For our running example, when the trace-checker evaluates the mutated requirement $\phi_3^\prime$, it detects that the trace $\trace$ satisfies the requirement $\phi_3^\prime$ and produces the pair $\{\langle\trace,\phi_3^\prime\rangle,\True\}$.
    The \emph{Trace Checker} component produces a set $\Delta$ of pairs $\{\langle\trace,\property^\prime\rangle,\upsilon\}$ made by the trace $\trace$, the mutated requirement $\property^\prime$, and the corresponding trace checking verdict $\upsilon$. 
    However, to run the \emph{Diagnostic Generator} step, the set $\Delta$ should contain a sufficient number of satisfied and violated requirements to enable the \emph{Diagnostic Generator} to produce an informative diagnosis.
    Therefore, the \emph{Generator of Mutations} and the \emph{Trace-Checker} components are executed iteratively, and the set $\Delta$ is augmented with newly generated pairs until (at least) a given number of satisfied and violated requirements (defined by a configuration parameter) is present.
    
    \item[\phase{3}] The \emph{Diagnostic Generator} step analyzes the requirement \property and the pairs containing the trace-checking verdicts of the mutated requirements (e.g., $\{\langle\trace,\property^\prime\rangle,True\}$) to produce a diagnosis.
    If it can not produce an informative diagnosis, it starts another iteration by running step \phase{1} and by considering a new set of the mutated requirements. This is highlighted by a dashed arrow in Figure~\ref{fig:approach}. 
Otherwise, it returns the diagnosis to the user.
\end{itemize}
The algorithm stops as soon as the diagnosis stabilizes and outputs it or, if this does not happen within the time budget (\budget), it reports that SBTD could not produce a diagnosis within \budget.

To illustrate our methodology, \Cref{fig:runningexample_dt} presents a decision tree (DT) as the diagnosis of our SBTD for the running example.
Our search-based trace diagnosis (SBTD) builds such a diagnosis by exploring the space of values for the colored tokens in $\phi_3$, namely the bounds of the outer time window (\colorbox{col1!30}{0} and \colorbox{col1!30}{\ensuremath{\infty}}), the upper bound on \texttt{$y_{pos}$} (%
\markB{3.5}), the postcondition threshold on \texttt{$y_{gap}$} (\markD{{1.0}})
, and the connective %
\markC{\lit{and}} (which may be flipped to \lit{or}).
The diagnosis highlights which sets of changes for 
$\phi_3($\colorbox{col1!30}{0}$)$, $\phi_3($\colorbox{col1!30}{\ensuremath{\infty}}$)$, $\phi_3($
\markC{\lit{and}}$)$, and $\phi_3($%
\markB{3.5}$)$ can make the formula satisfied (and, conversely, which combinations consistently lead to a violation on the considered trace).

Note that a change-driven diagnosis does not propose a change to the requirement: It explains which sub-condition(s) and which signals are responsible for the violation that can be used as diagnostic probes to localize the inconsistency.
Each changed syntactic element identifies the aspect of the behavior (a temporal bound, a signal threshold, a logical connective) in which the implementation deviates, and may pinpoint a portion of the implementation responsible for it.

For example, within the branch where the connective remains %
\markC{\lit{and}}, SBTD learns a simple decision rule that can be expressed directly in terms of the original variables as
\[
\text{$\tau_0$}
  >
\text{\colorbox{col1!30}{$5.2003$}}\, \text{$\land$} \,
\text{%
\markC{\lit{and}}}\, \text{$\land$} \,
\texttt{$y_{pos}$}\attt(\tau_0)
  >
\text{%
\markB{$2.8006$}}
\Rightarrow \mathit{violation}
\]
The colors in this rule mirror those in $\phi_3$: \colorbox{col1!30}{\ensuremath{5.2003}}s refine the time-window bound, %
\markC{\text{$\lit{and}$}} corresponds to the mutated logical connective, and %
\markB{2.8006} refines the upper bound of the central-lane region.
Accordingly, when the time window extends beyond approximately $5.2$s, the DT shows that satisfaction can still be recovered by tightening the upper bound of the central-lane condition, i.e., by setting $\phi_3($%
\markB{3.5}$)$ to a value no larger than about $2.8006$m.
More generally, for the considered trace, the developer can maintain the \lit{and} logical operator for $\phi_3($%
\markC{\lit{and}}$)$ and restrict the (mutated) upper bound of the universal time window $\phi_3($\colorbox{col1!30}{\ensuremath{\infty}}$)$ so that it ends no later than approximately $4.7$s.
Finally, the DT contains a separate branch in which the $\phi_3($%
\markC{\lit{and}}$)$ Boolean operator is modified into an ``\lit{or}''.
By applying this change, the antecedent of the implication can be satisfied by either requiring the entire grouped proximity condition (involving \texttt{$y_{pos}$} and \texttt{$x_{gap}$}) to hold, or requiring that \texttt{$y_{gap}$} falls below its threshold.
Ultimately, the diagnosis is an actionable entry point to fix the implementation of the controller of the ego vehicle. 
The diagnostic information for $\phi_3($\colorbox{col1!30}{\ensuremath{\infty}}$)\!\leq\!4.6998$s shows the trace satisfies $\phi_3$ up to the target's \emph{second} lane-change attempt (which starts at $4.5$s), showing that the first intrusion was answered, while the second was not.
The $[\phi_3($\colorbox{col1!30}{\ensuremath{0}}$),\phi_3($\colorbox{col1!30}{\ensuremath{\infty}}$)]\approx\![4.93,5.06]$s window and $\phi_3($\markB{3.5}$)$ to $2.8006$m suggest that the failure occurs late in the target's lateral ramp. 
The \markC{\lit{and}}$ \to $\lit{or} node diagnostic information demonstrates that the violation depends on the conjunction of both gap conditions, while $\phi_1$ and $\phi_2$ holding on the same trace exonerate the braking logic.
This information helps the engineer localize the defect by guiding the debugging activity over the evasion rule's handling of the vehicle.
Specifically, this information provides a valuable guideline to the engineer for the identification of the \texttt{evasion\_completed} trigger as the source of the problem: The trigger was not reset when the intrusion condition is re-entered. 
Resetting the trigger fixes the source of the failure without editing any requirement. We applied the change to the model and re-ran the verification suite, and $\phi_1$--$\phi_4$ are satisfied in all four scenarios. This example shows that SBTD helps locate the cause of a failure even when the original formulation of the requirements is correct.

The diagnosis provided by SBTD can be customized, considering the information engineers require.
This is relevant in our approach since the definition of the diagnosis influences the behavior of the Generator of Mutations and the Diagnostic Generator components.
To that end, the Generator of Mutations generates requirements that most likely guide the search toward the generation of a suitable diagnosis.
Then, the Diagnostic Generator aggregates the pairs produced by the Trace Checker based on the type of the desired diagnosis.

\section{Search-based Trace Diagnostic for HLS}
\label{sec:sbtforhls}

In this section, we instantiate the generic SBTD framework from \Cref{sec:approach} for requirements expressed in HLS and present our SBTD implementation, namely \NAME.
Specifically, \NAME supports \emph{change-driven diagnosis} (\Cref{sec:diagnosisdefinition}), which is the type of diagnosis considered throughout this work.
Accordingly, we describe how the \emph{Generator of Mutations} (\Cref{sec:mutationgenerator}), \emph{Trace-Checker} (\Cref{sec:tracechecker}), and \emph{Diagnostic Generator} (\Cref{sec:DiagnosticGenerator}) components are specialized to support change-driven diagnosis for HLS.

\begin{table*}[!t]
    \caption{Mutation Operators: the table contains the original formula and the mutated formula}
    \label{tab:mutations}
    \begin{center}
    \begin{tabular}{l l l}
    \toprule
    \textbf{OP} & \textbf{Original Formula} & \textbf{Mutated Formula} \\
    \midrule
    \textbf{OP1} & \synt{p} & \lit{not}\ \synt{p} \\
    \textbf{OP2} & $ \synt{tm}_1 \oplus \synt{tm}_2$     &    $ \synt{tm}_1 \oplus^\prime \synt{tm}_2$ \\
    \textbf{OP3} & \lit{not}\ \synt{p} & \synt{p} \\
    \textbf{OP4} & $\synt{p}_1\ \ominus\ \synt{p}_2$ & $\synt{p}_1\ \ominus^\prime\ \synt{p}_2$\\
    \textbf{OP4a} & $\synt{p}_1\ \ominus\ \synt{p}_2$ & $\synt{p}_2\ \ominus\ \synt{p}_1$\\
    \textbf{OP5} & $\triangleright\ \tau\ \lit{in}\ I_T\ \lit{such that}\  \synt{p}$ & $\triangleright^\prime\ \tau\ \lit{in}\ I_T\ \lit{such that}\  \synt{p}$ \\
    \textbf{OP6} & $\triangleright\ \sigma\ \lit{in}\ I_J \lit{such that}\ \synt{p}  $ & $\triangleright^\prime\ \sigma\ \lit{in}\ I_J \lit{such that}\ \synt{p}  $ \\
     \textbf{OP7} &  $\triangleright\ \rho\ \lit{such that}\ \synt{p} $ & $\triangleright^\prime\ \rho\ \lit{such that}\ \synt{p} $ \\
    \bottomrule
    \end{tabular}
    \hspace{0.3cm}
        \begin{tabular}{l l l }
         \toprule
        \textbf{OP} & \textbf{Original Formula} & \textbf{Mutated Formula}\\
        \midrule
        \textbf{OP8} & $\synt{tt}_1 \odot \synt{tt}_2$ & $\synt{tt}_1 \odot^\prime \synt{tt}_2$ \\
         \textbf{OP9} & $\synt{it}_1 \odot \synt{it}_2$ & $\synt{it}_1 \odot^\prime \synt{it}_2$\\
           \textbf{OP10} & $\synt{vt}_1 \odot \synt{vt}_2$ & $\synt{vt}_1 \odot^\prime \synt{vt}_2$\\
                \textbf{OP11} & $t$ & $t^\prime$ \\
                 \textbf{OP12} & $j$ & $j^\prime$ \\
                   \textbf{OP13} & $x$ & $x^\prime$\\
                     \textbf{OP14} & $s \atst \synt{it}$ & $s^\prime \atst \synt{it}$\\
                      \textbf{OP15} & $s \attt \synt{tt}$ & $s^\prime \attt \synt{tt}$ \\
         \bottomrule
        \end{tabular}
    \end{center}
    $\oplus^\prime \in \{>, <, \leq, \geq, =,\not=\} \setminus \{\oplus\}$\\
$\ominus^\prime \in \{\lit{or}, \lit{and}, \lit{implies}\} \setminus \{\ominus\}$\\
$ \odot^\prime \in \{+, -, * , /\} \setminus \{\odot\}$; \\
$\triangleright^\prime \in \{\lit{forall}, \lit{exists}\} \setminus \{\triangleright\}$. \\
$t, t^\prime \in \mathbb{T}, \phantom{xxx}  j, j^\prime \in \mathbb{J}, \phantom{xxx} x, x^\prime \in \real,
\phantom{xxx} \tau  \in \mathit{TV}, 
\phantom{xxx} \sigma \in \mathit{IV}, 
\phantom{xxx} \rho \in \mathit{RV}, 
\phantom{xxx} s, s^\prime \in S$.
\end{table*}

\begin{figure*}[htb]
{\footnotesize    \centering
    \begin{subfigure}[b]{0.9\columnwidth}
 \centering
   \begin{tikzpicture}[level 1/.style={sibling distance=20mm},
                       level 2/.style={sibling distance=35mm},
                       level 3/.style={sibling distance=20mm},
                       level 4/.style={sibling distance=26mm},
                       level distance=35pt]
\node{$\lit{forall}\  \tau_0\ \lit{in}\ I_T \ \lit{such\ that}\ p$}
 child {node {$[\colorbox{col1!30}{0},\colorbox{col1!30}{$\infty$})$}}
 child {
    node[]%
    {\markC{\lit{and}}}
    child{
         node {\lit{and}}
         child{ node[xshift=7mm]{$\ldots$} }
         child{
                node[xshift=-3mm]{$<$}
                child{ node{$\texttt{$y_{pos}$} \attt (\tau_0)$}}
                child{ node[]%
                {\markB{$3.5$}}}
            }
         child{ node[xshift=-10mm]{$\ldots$} }
         }
         child{
            node{$<$}
            child{node{$\texttt{$y_{gap}$} \attt (\tau_0)$}}
            child{node{$1.5$}}
         }
    };
\end{tikzpicture}
    \caption{AST for the requirement \property of our motivating example.}
    \label{fig:astex1}
    \end{subfigure}
    \hspace{0.5cm}
    \begin{subfigure}[b]{0.9\columnwidth}
 \centering
\begin{tikzpicture}[level 1/.style={sibling distance=20mm},
                    level 2/.style={sibling distance=35mm},
                    level 3/.style={sibling distance=20mm},
                    level 4/.style={sibling distance=26mm},
                    level distance=35pt]
\node{$\lit{forall}\  \tau_0\ \lit{in}\ I_T \ \lit{such\ that}\ p$}
 child {node {$[\colorbox{col1!30}{0},\colorbox{col1!30}{$\infty$})$}}
 child {
    node[]%
    {\markC{\lit{and}}}
    child{
         node {\lit{and}}
         child{ node[xshift=7mm]{$\ldots$} }
         child{
                node[xshift=-3mm]{$<$}
                child{ node{$\texttt{$y_{pos}$} \attt (\tau_0)$}}
                child{ node[]%
                {\markB{$2.8$}}}
            }
         child{ node[xshift=-10mm]{$\ldots$} }
         }
         child{
            node{$<$}
            child{node{$\texttt{$y_{gap}$} \attt (\tau_0)$}}
            child{node{$1.5$}}
         }
    };
\end{tikzpicture}
    \caption{AST for the requirement $\property^\prime$.}
    \label{fig:astmutation}
    \end{subfigure}
    \label{fig:ast}
    \caption{AST of an HLS requirement and its mutation obtained by applying the mutation operator \textbf{OP13} to the node with the \colorTwo background.}
    \label{sec:examplemut}}
\end{figure*}

\subsection{Change-Driven Diagnosis}
\label{sec:diagnosisdefinition}
Change-driven diagnosis explains requirements' violations by describing which (set of) change(s) can lead to a requirement satisfied by the trace. 
For example, the decision tree (DT) from \Cref{fig:runningexample_dt} explains to engineers which changes to the requirement $\phi_3$ make it satisfied by the trace from \Cref{sec:background}. This information helps engineers understand that the violation is triggered by a short-lived configuration in which the target vehicle is in the central portion of the lane (i.e., \texttt{$y_{pos}$} is high), the gaps \texttt{$x_{gap}$} and \texttt{$y_{gap}$} are small, yet the postcondition requiring a rapid increase of lateral separation is not met within the one-second reaction window. 

In particular, the DT highlights that satisfaction can be recovered either by restricting the time window so that the antecedent is not evaluated after approximately $4.7$s, or, when the time window extends beyond approximately $5.2$s, by tightening the central-lane bound $\phi_3(%
\markB{3.5})$ to about $2.8$m while preserving the conjunction $\phi_3(%
\markC{\lit{and}})$ between the \texttt{$x_{gap}$} and \texttt{$y_{gap}$} clauses.

As our running example shows, engineers can select sub-portions of the requirements targeted by the changes.
For example, for the requirement $\phi_3$ from \Cref{sec:background}, the \colorOne, \colorTwo, \colorThree, and \colorFour labeled boxes identify the sub-portions of the formula the changes should refer to. The engineer is interested in how changes affect the satisfaction of the requirement, regarding: (i) the bounds of the universal time window ($\phi_3($\colorbox{col1!30}{0}$)$ and $\phi_3($\colorbox{col1!30}{\ensuremath{\infty}}$)$),
(ii) the upper bound of the central-lane condition on \texttt{$y_{pos}$} ($\phi_3($%
\markB{3.5}$)$), and (iii) the logical operator ``\lit{and}'' ($\phi_3($%
\markC{\lit{and}}$)$) that relates the grouped \texttt{$y_{pos}$} and \texttt{$x_{gap}$} conditions with the \texttt{$y_{gap}$} clause, and (iv) the postcondition threshold on \texttt{$y_{gap}$} ($\phi_3($%
\markD{1.0}$)$).

Intuitively, changes in the time and spatial thresholds enable engineers to understand when, and under which proximity conditions, the requirement antecedent becomes active. 
Changes of the ``\lit{and}'' Boolean operator enable engineers to understand whether both proximity conditions are required or if it is sufficient that only one of them holds.

\begin{definition}[Change-Driven Diagnosis]\label{def:diagnosis}
    \emph{Let $\protrace$ be a trace-requirement combination made by a requirement (\property) unsatisfied over the trace (\trace), and let $sub(\property)$ denote the portion of \property that changes are allowed to target.
    A \emph{change-driven diagnosis} $d$ is a (set of) change(s) to $sub(\property)$ such that applying $d$ to \property yields a mutated requirement $\property^{d}$ that is satisfied by $\trace$.}
\end{definition}

Notably, within the change-driven diagnosis paradigm, SBTD requires as input a \emph{mutation space} that specifies (i) which syntactic tokens of the requirement may be mutated, and (ii) the admissible ranges (or alternatives) for each token.
This 
{\em mutation-space specification}
captures domain knowledge about which parts of the requirement are plausible sources of mismatch with the observed traces (e.g., numeric thresholds, time-window bounds, or selected logical connectives), and it is \emph{not} a library of violation causes/diagnoses.
Unlike library-based trace diagnosis, SBTD does not assume a pre-defined catalog of explanations and matching rules. Instead, it uses the configured mutation space as a set of \emph{diagnostic probes} to automatically generate candidate explanations and learn an interpretable decision-tree diagnosis from their trace outcomes.

In the remainder of this section, we provide an in-depth perspective of \NAME, i.e., how SBTD is instantiated to generate change-driven diagnoses for HLS requirements.

\subsection{Generator of Mutations} 
\label{sec:mutationgenerator}
This component receives a (set of) requirement(s) as inputs and generates a set of mutated requirements by sequentially performing the mutation and crossover operations.

The \emph{mutation operations} component considers an HLS requirement and changes the portions of the abstract syntax tree (AST) that refer to the sub-portions of the requirements identified by the engineers.
For example, \Cref{fig:astex1} presents the AST for the requirement \property of our motivating example.
The portions of the abstract syntax tree (AST) referring to the sub-portions of the requirements identified by the engineers are identified by colored nodes.
Specifically, the nodes related to the logical operator ``\lit{and}'', the upper bound on \texttt{$y_{pos}$}, and the time-window bounds are with \colorThree, \colorTwo, and \colorOne background colors, respectively.

The operator has to select the number of nodes to mutate between zero and the total number of nodes of the AST. This selection is related to portions of the requirement that the engineers are interested in.
Then, it uses the mutation operators from \Cref{tab:mutations} to mutate the nodes of the AST. 
Depending on the specific application, engineers can specify a subset of operators to be used by the generator of mutations.
Operator \textbf{OP1} mutates the HLS requirement \synt{p} into its negation \lit{not}\ \synt{p}.
Operator \textbf{OP2} mutates the relational operator $\oplus$ by selecting another relational operator $\oplus^\prime$.
Operator \textbf{OP3} removes the negation operator from the HLS requirement $\lit{not}\ \synt{p}$.
Operator \textbf{OP4} mutates the Boolean operator $\ominus$ combining $\synt{p}_1$ and $\synt{p}_2$ into $\ominus^\prime$. When \textbf{OP4} mutates $\ominus$ into $\ominus'$, the operand order follows the AST produced by the parser.
Since \lit{implies} is not commutative, we add \textbf{OP4a} (operand swap) so that when \textbf{OP4} introduces \lit{implies}, the mutation space includes both operand orders, i.e., $\synt{p}_1 \Rightarrow \synt{p}_2$ and $\synt{p}_2 \Rightarrow \synt{p}_1$.
The operators \textbf{OP5}, \textbf{OP6}, and \textbf{OP7} mutate the existential quantifier \lit{exists} into the universal quantifier \lit{forall} and vice versa.
The operators \textbf{OP8}, \textbf{OP9}, and \textbf{OP10} mutate the arithmetic operator $\odot$ by selecting another arithmetic operator $\odot^\prime$.
The operators
\textbf{OP11}, \textbf{OP12}, and \textbf{OP13} mutate the time, index and value terms $t$, $j$, and $x$ by selecting new values $t^\prime$, $j^\prime$, and  $x^\prime$.
Finally, the operators \textbf{OP14} and \textbf{OP15} mutate the value terms $s \atst \synt{it}$ and $s \attt \synt{tt}$ into $s^\prime \atst \synt{it}$ and $s^\prime \attt \synt{tt}$ by selecting a new signal $s^\prime$.
With the exception of OP4a, the mutation operators do not change the structure of the AST, but only the content of its nodes.
OP4a swaps the two operands of a non-commutative Boolean operator (\lit{implies}), reordering the two branches under that node; preserving the structure for the other operators keeps the crossover alignment well-defined and the diagnoses comparable across requirements.

In our running example, engineers select the operator \textbf{OP4}, that can mutate the logical operator ``\lit{and}'', and the operators \textbf{OP11} and \textbf{OP13} that can mutate the time and value terms representing the time-window bounds and numeric thresholds (e.g., $\phi_3(%
\markB{3.5})$).
\Cref{fig:astmutation} presents the AST of the requirement $\phi_3^\prime$: An example of a mutation for the AST from \Cref{fig:astex1} of the requirement $\phi_3$ of our running example where the operator \textbf{OP13} replaces the value ``3.5'' with the value ``2.8''.

\begin{table*}[htb]
    \centering
    \caption{Configuration Parameters for our SBTD framework.}\label{tab:configurationParameters}
    \begin{tabular}{l l l}
    \toprule
    \textbf{ID} &  \textbf{Parameter}    & \textbf{Textual Description} \\
    \midrule
    CR &   Crossover rate   &  Probability of applying the crossover operator. \\
    MR &  Mutation rate   &  Probability of applying the mutation operator. \\
     PS & Population size   &  Number of requirements considered by the SBTD  framework at each iteration.\\
     SA & Selection Algorithm   & The algorithm to be chosen for the selection of the requirements (Elitism or Roulette Wheel).\\
     PTBC & Parents to Be Chosen & Number of requirements to be considered as a parent when using Elitism.\\
     MS & Min Samples   & Minimum number of samples in SBTD.\\
     TS & Tournament Size   & Number of requirements that compete to be selected as a parent.\\
     TCTO & Trace check time out & Maximum time allowed for trace check to check a requirement.\\
     PGTO & Program time out & Maximum time allowed for SBTD to find the requested solution. \\
    \bottomrule
    \end{tabular}
\end{table*}

\begin{figure*}[t]
{\footnotesize   \centering
    \begin{subfigure}[b]{0.3\textwidth}
 \centering
   \begin{tikzpicture}[level 1/.style={sibling distance=20mm},level 2/.style={sibling distance=20mm},level 3/.style={sibling distance=17mm},level 4/.style={sibling distance=30mm}, level distance=35pt
]
\node{$\lit{forall}\  \tau_0\ \lit{in}\ I_T \ \lit{such that}\ p$}
 child {node {[{0},{$\infty$})}}
 child {
    node {\lit{and}}
    child{ 
         node[fill=red]{$<$}
         child{
                node{$\texttt{$y_{pos}$} \attt (\tau_0)$}
            }
          child{node{$3.5$}}
         }
         child{ 
            node{...}
         }
    };
\end{tikzpicture}
    \caption{Requirement $\phi_3^{(1)}$.}
    \label{fig:prop1}
    \end{subfigure}
    \hspace{0.5cm}
    \begin{subfigure}[b]{0.3\textwidth}
 \centering
\begin{tikzpicture}[level 1/.style={sibling distance=20mm},level 2/.style={sibling distance=20mm},level 3/.style={sibling distance=17mm},level 4/.style={sibling distance=30mm}, level distance=35pt
]
\node{$\lit{forall}\  \tau_0\ \lit{in}\ I_T \ \lit{such that}\ p$}
 child {node {[{0},{$\infty$})}}
 child {
    node {\lit{or}}
    child{ 
         node[fill=red]{$\geq$}
         child{
                node{$\texttt{$y_{pos}$} \attt (\tau_0)$}
            }
          child{node{$2.8$}}
         }
         child{ 
            node{...}
         }
    };
\end{tikzpicture}
    \caption{Requirement $\phi_3^{(2)}$.}
    \label{fig:prop2}
    \end{subfigure}
    \hspace{0.5cm}
    \begin{subfigure}[b]{0.3\textwidth}
 \centering
\begin{tikzpicture}[level 1/.style={sibling distance=20mm},level 2/.style={sibling distance=20mm},level 3/.style={sibling distance=17mm},level 4/.style={sibling distance=30mm}, level distance=35pt
]
\node{$\lit{forall}\  \tau_0\ \lit{in}\ I_T \ \lit{such that}\ p$}
 child {node {[{0},{$\infty$})}}
 child {
    node {\lit{and}}
    child{ 
         node{$\geq$}
         child{
                node{$\texttt{$y_{pos}$} \attt (\tau_0)$}
            }
          child{node{$2.8$}}
         }
         child{ 
            node{...}
         }
    };
\end{tikzpicture}
    \caption{Generated requirement $\phi_3^{(3)}$.}
    \label{fig:genProp}
    \end{subfigure}
    \caption{Example of application of the crossover operator: the requirement $\phi_3^{(3)}$ is obtained from the requirement $\phi_3^{(1)}$ by swapping the subtree with the root node with a red background with the corresponding subtree from the requirement $\phi_3^{(2)}$.}
    \label{fig:crossover}}
\end{figure*}

The \emph{crossover operator} generates new candidate requirements by (a)~selecting a pair of requirements, and (b)~combining them. 
Next, we further explain how our algorithm selects the best pair of requirements by finding the best alignment between requirements. Then, we distill how we combine the best pair of requirements by swapping corresponding nodes from the AST trees representing each requirement. 

To select the best requirements pair, the crossover operator computes a fitness value for each mutated requirement. The fitness value of each mutated requirement is obtained by comparing the mutated requirement with the original requirement using the score function of a pairwise alignment algorithm, namely Smith–Waterman~\cite{SMITH1981195}. Therefore, the fitness value is the score of the best local alignment between requirements, such that the higher the fitness the more similar the mutated requirement to the original requirement is. Our choice for rewarding the similarity between requirements is grounded in the idea that fewer, but relevant, mutations in the requirements lead to fewer interactions between term changes, and consequently reduce the effect of the confounding bias~\cite{pearl2003statistics}. In other words, the higher the similarity between the originally violated and the mutated HLS requirements, the lower the chances of having spurious factors that could incorrectly imply causation between the term changes and the requirement satisfaction (or violation).

\Cref{fig:SWExample} demonstrates an example of calculating the fitness value of the mutated requirement $\phi_3^\prime$ using the score function of the Smith-Waterman algorithm. The algorithm compares the distance between the original requirement ($\phi_3$) and the mutated requirements ($\phi_3^\prime$), term by term. We use the algorithm as follows\footnote{Originally the SW algorithm computes the best local alignment to find the places where the term changes. However, for the purpose of informative diagnosis generation, we are concerned not only with the places of change but also with the domain and range of the values where the change takes place. Such step of our approach is further explained in~\Cref{sec:DiagnosticGenerator}.}: (i) mapping terms, (ii) calculating the initial scoring matrix, and (iii) collecting the score. 

    (i) \emph{Mapping terms}. The algorithm maps the terms that can be mutated from both requirements ($\phi_3$ and $\phi_3^\prime$) enriched with a ``null'' element to rows and columns of a scoring matrix. 
For example, \Cref{fig:SWExample} represents the scoring matrix $SM$ associated with the requirements $\phi_3$ and $\phi_3^\prime$, where the terms that can be mutated in $\phi_3$ (i.e., $\phi_3($\colorbox{col1!30}{\ensuremath{\infty}}$)$, $\phi_3($%
\markC{\lit{and}}$)$, $\phi_3($%
\markB{3.5}$)$) and $\phi_3^\prime$ (i.e., $\phi_3^\prime($\colorbox{col1!30}{\ensuremath{\infty}}$)$, $\phi_3^\prime($%
\markC{\lit{and}}$)$, $\phi_3^\prime($%
\markB{2.8}$)$) are respectively reported in the headers of its rows and columns.

(ii) \emph{Calculating the scoring matrix}. 
The value zero is associated with matrix cells from rows and columns labeled with the ``null'' elements. 
The values of the remaining cells are calculated according to \Cref{eq:sw}.

We adopt Smith--Waterman because crossover requires a structure-preserving notion of similarity between the sequences of targetable tokens in the original and mutated requirements.
\Cref{eq:sw} is the standard SW recurrence that computes the best local-alignment score. We use this score as the fitness value to bias crossover toward offspring that preserve long contiguous subsequences (small, localized edits) and to penalize fragmented alignments via the gap term $W$, which in turn reduces interactions among simultaneous changes and improves diagnosis interpretability.

\begin{equation}
\label{eq:sw}
SM[i,j] = max
     \begin{cases}
       SM[i-1,j-1] + s(\phi_i, \phi_j^\prime)\\
       SM[i-1,j] + W\\ %
       SM[i,j-1] + W\\ %
       0\\
     \end{cases}
\end{equation}
The equation specifies that the value of the scoring matrix $SM$ in position $i,j$, i.e., $SM[i,j]$, depends on the similarity score ($s(\phi_i,\phi_j^\prime)$) of requirement $\phi$ in position $i$ and requirement $\phi^\prime$ in position $j$, with gap score ($W$, a.k.a., penality gap).
The gap score penalizes formulae that require swapping many terms to be aligned. 
We set the similarity score $s(\phi_i,\phi_j^\prime)$ to the value $3$ when the terms from the column headers $i$ and $j$ coincide, to the value $-3$ otherwise. 
We considered the value $-2$ for the gap score ($W$). 

(iii) \emph{Collecting the score and measuring the fitness.} The score is the highest value in the scoring matrix. In the example from \Cref{fig:SWExample}, the score is 6 from cell $SM[3,3]$. Ultimately, we use the score as the fitness value in the following steps of the algorithm. 

\begin{figure}[t]
    \centering
    {\footnotesize
    \begin{subfigure}[b]{0.45\textwidth}
    \begin{tabular}{c | c c c c }
    \toprule
         & null & $\phi_3($\colorbox{col1!30}{\ensuremath{\infty}}$)$ & $\phi_3($%
         \markC{\lit{and}}$)$ & $\phi_3($%
         \markB{3.5}$)$   \\
         \midrule
      null                                          &  0  &  0  &  0 &  0 \\
      $\phi_3^\prime($\colorbox{col1!30}{\ensuremath{\infty}}$)$       &  0  &  3  &  1 &  0 \\
      $\phi_3^\prime($%
      \markC{\lit{and}}$)$&  0  &  1  &  \colorbox{gray!30}{6} &  4 \\
      $\phi_3^\prime($%
      \markB{2.8}$)$       &  0  &  0  &  4 &  3 \\
      \bottomrule
    \end{tabular}
    \centering 
\end{subfigure}
    \caption{Example of fitness calculation using the score function of the Smith-Waterman algorithm.}
    
    \label{fig:SWExample}}
\end{figure}

We implemented two selection methods that use these fitness values:
\begin{enumerate}
    \item Elitism~\cite{mitchell1999machine,goldberg1994genetic}: it
    selects two parents randomly between the best ten formulas following their \textit{fitness}.
    \item Roulette wheel~\cite{mitchell1999machine,goldberg1994genetic}: it
     selects two parents based on their fitness, where the higher the \textit{fitness}, the higher the probability of being selected.
\end{enumerate}

To combine the pair of HLS requirements, we randomly select a node from the AST of the first requirement and swap it with the corresponding node of the second requirement. 
For example, \Cref{fig:crossover} presents an example of an application of the mutation operator: The mutation operator selects the sub-tree from the requirement~$\phi_3^{(1)}$ with the red node as a root (\Cref{fig:prop1}) and swaps it with the corresponding sub-tree from the requirement~$\phi_3^{(2)}$ (\Cref{fig:prop2}), thus leading to requirement~$\phi_3^{(3)}$ (\Cref{fig:genProp}).
Notice that, since the mutation operators do not change the structure of the AST (aside from the OP4a operand swap), all the requirements have the same structure.

\Cref{tab:configurationParameters} lists the set of parameters to be configured by engineers to run the SBTD framework. 
For example, the crossover rate (CR) is the probability of applying the crossover operator.

The \emph{Generator of Mutations} component generates a set of candidate requirements $\Psi$, which are then considered by the \emph{Trace checker} component, as explained in the following.

For the generator of mutations (\phase{1}), we developed a Python script (i.e., ga.py) that implements the algorithm from \Cref{sec:mutationgenerator}.
We decided to implement this procedure (instead of using an external library) since this decision enables controlling the data structures used by the algorithms to represent HLS requirements. 
This decision simplified the implementation of the operators from \Cref{tab:mutations} and the fitness metric from \Cref{sec:mutationgenerator}.

\subsection{Trace-Checker}
\label{sec:tracechecker}
The trace checker component considers the trace \trace and the candidate requirements $\Psi$ and verifies which requirements hold on \trace.
This is done by considering each HLS requirement $\phi\in \Psi$, and by running an existing trace-checker that can verify whether the requirement holds or not on the trace \trace, i.e., whether $\trace \models \phi$.

The \emph{Trace Checker} component produces a set $\Delta$ of pairs $\{\langle\trace,\property^\prime\rangle,\upsilon\}$ made by the trace $\trace$, the mutated requirement $\property^\prime$, and the corresponding trace checking verdict $\upsilon$. 
These pairs are fed into the \emph{Diagnostic Generator}.

For the trace checker component (\phase{2}), we used the \texttt{ThEodorE} trace-checking tool~\cite{menghi2021theodore} since it supports requirements expressed in HLS.
\texttt{ThEodorE} can produce three possible verdicts: ``\emph{satisfied}'', if the trace satisfies the requirement, ``\emph{violated}'', if it does not, 
or ``\emph{unknown}'', if the SMT solver used by \texttt{ThEodorE} to solve the trace-checking problem can not deduce whether the requirement is satisfied or violated.
The ``\emph{unknown}'' verdict is returned when the underlying SMT technology used by the solver can not produce results for some specific instances of the problem~\cite{menghi2021trace}. Therefore, the diagnostic generator component will also create leaves and label {them with the {\tt undecided}}
verdict to explain cases where the trace-checker could not produce any verdict.
\subsection{Diagnostic Generator}
\label{sec:DiagnosticGenerator}

The \emph{Diagnostic Generator} summarizes the outcomes of the search phase into an interpretable change-driven diagnosis.
It (a) filters the evaluated mutated requirements and (b) learns a decision tree over the remaining labeled samples, whose internal nodes test mutated tokens and whose leaves predict whether the trace satisfies the mutated requirement.
These changes are diagnostic probes since they characterize the behavior the trace exhibits, and are not proposed edits to the requirement.
We implement this learner using J48~\cite{quinlan2014c4}, a widely used ML algorithm~\cite{witten2002data} that generates decision trees that classify training data.

The \emph{requirement filtering} step selects the requirement mutations that are more similar to the original requirement for the computation of the decision tree while ensuring that the number of satisfied and unsatisfied requirements is the same.
The requirement filtering ranks the mutated properties using the score function of the Smith–Waterman algorithm (as done by the cross-over operator explained in \Cref{sec:mutationgenerator}) for selecting the requirements to be combined.
Then, it selects a subset of requirement mutations with the highest fitness values. 
The number of selected requirement mutations is defined by the parameter Parents to Be Chosen (PTBC) specified by the user (see \Cref{tab:configurationParameters}).

The \emph{decision-tree} computation works in two steps: \emph{Data Preparation} and \emph{Learning}.

\begin{figure}
    \centering
    \input{figures/observations}
        \caption{Entries considered by the learning algorithm.}
    \label{fig:observations}
\end{figure}

\begin{itemize}
    \item \emph{Data Preparation}: Before running our learning technique, we have to prepare our data. Specifically, we have to represent the AST of each requirement in a format that is processable by a learning technique.
    Each line represents one observation: a requirement mutation and the corresponding trace-checking verdict (i.e., satisfied or violated). 
    For example, \Cref{fig:observations} presents a fragment of the input data representing a portion of the requirement mutations generated from $\phi_3$ (cf.\ \Cref{sec:background}), where each column corresponds to one mutation position selected by the engineer (e.g., the bounds of the outer time window, the logical connective that relates the grouped \texttt{$y_{pos}$}/\texttt{$x_{gap}$} conditions with the \texttt{$y_{gap}$} clause, and the numeric thresholds on \texttt{$y_{pos}$} and \texttt{$y_{gap}$}).
    We remark that the generator of mutations creates properties by not changing the structure of the AST.
\item \emph{Learning}: The learning algorithm processes the input file and classifies the requirements based on the trace-checking verdict (satisfied or violated).
We run J48. \Cref{fig:runningexample_dt} illustrates an example of a resulting decision tree where the mutated upper bound of the time window (e.g., $\phi_3($\colorbox{col1!30}{\ensuremath{\infty}}$)$) is one of the most informative split conditions, and subsequent splits refine the diagnosis by considering the logical operator $\phi_3($%
\markC{\lit{and}}$)$ and the upper bound on \texttt{$y_{pos}$}, i.e., $\phi_3($%
\markB{3.5}$)$ (cf.\ \Cref{sec:background}). Leaf nodes (\emph{True}, \emph{False}) are labeled with the frequency of whether the selected term results in the verdict.
 
\end{itemize}

A change-driven diagnosis ($d$), represented as a decision tree (DT), summarizes \emph{which token mutations} (within the configured mutation space) make the violated requirement become satisfied by the trace. It is interpreted top-down: each internal node tests one mutated token, edges correspond to test outcomes, and each root-to-leaf path defines a conjunction of constraints on the mutated tokens; the leaf label (\emph{True} or \emph{False}) is the predicted trace-checking verdict for mutations satisfying those constraints, while leaf counts report the number of training samples reaching the leaf (its empirical support). In \Cref{fig:runningexample_dt}, the root split is the mutated upper bound of the outer time window $\phi_3(\colorbox{col1!30}{\ensuremath{\infty}})$. Restricting this bound yields one direct sufficient condition for satisfaction, whereas the complementary branch indicates that additional constraints (e.g., on $\phi_3(\markC{\lit{and}})$ and $\phi_3(\markB{3.5})$) are required. The tree supports more than one reading: Information from tokens at higher level of the tree influences more trace-checking verdicts, while information offered by tokens at lower level provides engineers with more fine-grained information about the influence of changes in the tokens on trace-checking verdicts.

For the diagnostic generator component (\phase{3}), we used the Java implementation of the C4.5 algorithm~\cite{quinlan2014c4} available in Weka~\cite{hall2009weka}. 
We selected the C4.5 algorithm, since it is a widely used learning algorithm for decision trees~\cite{witten2002data}, and Weka, since it is a well-known library of machine learning algorithms~\cite{witten2005practical}.

\section{Evaluation}
\label{sec:evalaution}
Our evaluation assesses how SBTD can identify the correct cause for violated requirements. To this end, we consider three research questions: %
\begin{itemize}
    \item[] \textbf{RQ1}: How \emph{effective} is SBTD in producing informative diagnoses? (\Cref{sec:effectiveness})
\end{itemize}
To answer this question, we assess how useful the diagnoses produced by SBTD are in detecting the causes of the violations of the requirements, considering TD-SB-TemPsy~\cite{boufaied2023} as a baseline on the subset of requirements it supports.

\begin{table*}[t]
    \centering
    \caption{Requirements from our benchmark.}
    \label{tab:reqbenchmark}
    \footnotesize
    \begin{tabular}{p{0.7cm} p{16.6cm}}
    \toprule
       \textbf{ID}  &  \textbf{Textual description} \\
    \midrule
    AT1     & The vehicle's speed ($v$) shall be lower than  %
    \markC{$120$}$mph$ ($v\leq 120mph$) within [ \colorbox{col1!30}{ 0},  %
    \markB{ 20}]s.\\
    AT2     & The engine speed ($\omega$) shall be lower than  %
    \markC{$4750$}$rpm$ ($\omega\leq 4750rpm$) within [ \colorbox{col1!30}{0},  %
    \markB{10}]s.\\
    AT51    & If the transmission enters Gear $1$ within the time interval [ \colorbox{col1!30}{0},  %
    \markB{30}]s, it shall remain in that gear for the next $ %
    \markC{2.5}s$.\\
    AT52    & If the transmission enters Gear $2$ within the time interval [ \colorbox{col1!30}{0},  %
    \markB{30}]s, it shall remain in that gear for the next  %
    \markC{2.5}s.\\
    AT53    & If the transmission enters Gear $3$ within the time interval [ \colorbox{col1!30}{0},  %
    \markB{30}]s, it shall remain in that gear for the next $ %
    \markC{2.5}s$.\\
    AT54    & If the transmission enters Gear $4$ within the time interval [ \colorbox{col1!30}{0},  %
    \markB{30}]s, it shall remain in that gear for the next $ %
    \markC{2.5}s$.\\
    AT6a    & If the engine speed is lower than $ %
    \markB{3000}rpm$ within [$0, \colorbox{col1!30}{30}$]s, then the vehicle speed shall be lower than $ %
    \markD{35}mph$ within~[$0, %
    \markC{4}$]s.\\
    AT6b    &If the engine speed is lower than $ %
    \markB{3000}rpm$ within [$0, \colorbox{col1!30}{30}$]s, then the vehicle speed shall be lower than $ %
    \markD{50}mph$ within~[$0, %
    \markC{8}$]s.\\
    AT6c    &If the engine speed is lower than $ %
    \markB{3000}rpm$ within [$0, \colorbox{col1!30}{30}$]s, then the vehicle speed shall be lower than $ %
    \markD{65}mph$ within~[$0, %
    \markC{20}$]s.\\
    AT6abc  &The requirements AT6a, AT6b, and AT6c shall be simultaneously satisfied. (Same mutation parameters as AT6c)\\
    \midrule
    CC1     & Car 5 shall always be at most $ %
    \markC{40}$m ahead of car 4 within  [$ \colorbox{col1!30}{0}, %
    \markB{100}$]s\\
    CC2     & Within [$ \colorbox{col1!30}{0},70$]s, car 5 shall be at least $ %
    \markC{15}m$ ahead of car 4 at least once in the next [$ %
    \markB{0},30$]s.\\
    CC3     &  \colorbox{col1!30}{At all times} within [$0,80$]s, for the next $20s$, car 2 shall  %
    \markB{always} precede car 1 by at most $20m$,  %
    \markC{or} car 5 shall precede car 4 by $40m$ at least once.\\
    CC4     &  \colorbox{col1!30}{At all times} within [$0,65$]s, at least once in the next $30s$, car 5 shall  %
    \markB{always} be at least $ %
    \markC{8}m$ ahead of car 4 for the next $5s$.\\
    CC5     & Within [$0,72$]s, at least once in the next $8s$, if car 2 precedes car 1 by  \colorbox{col1!30}{more} than $ %
    \markB{9}m$ for $5s$, then car 5 shall precede car 4 by  %
    \markC{more} than $ %
    \markD{9}m$ in the next $15s$.\\
    CCx     & Within [$ \colorbox{col1!30}{0}, %
    \markB{50}$]s, all cars shall always be at least $ %
    \markC{7.5}m$ ahead of the car immediately behind it. (The mutation operator is applied only for the distance between cars 4 and 5).\\
    \midrule
   RR & From the beginning (time $0$) to the end ($\infty$) of the simulation, the following two conditions should hold:  the difference ($\texttt{d\_pos\_x} \attt (\tau_0) - \texttt{v\_pos\_x} \attt (\tau_0)$) between the desired position ($\texttt{d\_pos\_x}$) and the actual robot position ($\texttt{v\_pos\_x} $) in the x-axis at time $\tau_0$  is lower than a threshold value (\colorbox{col1!30}{$0.20$}m), %
   \markB{and} the Euclidean distance ($\texttt{d2obs}$) between the robot's border and the obstacle's border is greater than the threshold value (%
   \markC{$0.50$}m).\\
    \bottomrule
    \end{tabular}
    \begin{flushleft}
    \end{flushleft}
\end{table*}

\begin{table*}[ht]
    \centering
    \caption{HLS formalization for the requirements from \Cref{tab:reqbenchmark}.}
    \label{tab:reqbenchmarkHLS}
    \footnotesize
    \begin{tabular}{p{0.7cm} p{17cm}}
    \toprule
       \textbf{ID}  &  \textbf{HLS formalization} \\
    \midrule
    AT1     &   \lit{forall}\  $\tau_{0}$\ \lit{in}\ [\colorbox{col1!30}{0},%
    \markB{20}]\ \lit{such that}~v\attt ($\tau_0$) $\leq$ %
    \markC{120}.\\
    AT2     & \lit{forall}\  $\tau_{0}$\ \lit{in}\ [\colorbox{col1!30}{0},%
    \markB{10}]\ \lit{such that}~$\omega$\attt ($\tau_0$) $\leq$ %
    \markC{4750}.\\
    AT51    & \lit{forall}\  $\sigma_{0}$\ \lit{in}\ [\lit{t2i}(\colorbox{col1!30}{0})+1,\lit{t2i}(%
    \markB{30})]\ \lit{such that}~((gear\atst ($\sigma_0$-1) $\neq$ 1) \lit{and}\ (gear\atst ($\sigma_0$) = 1)) \lit{implies} ( \lit{forall}\ $\tau_{0}$\ \lit{in}\ [\lit{i2t}($\sigma_0$), \lit{i2t}($\sigma_0$)+%
    \markC{2.5}] \ \lit{such that}~ (gear \attt ($\tau_{0}$) = 1 )).\\
    AT52    & \lit{forall}\  $\sigma_{0}$\ \lit{in}\ [\lit{t2i}(\colorbox{col1!30}{0})+1,\lit{t2i}(%
    \markB{30})]\ \lit{such that}~((gear\atst ($\sigma_0$-1) $\neq$ 2) \lit{and}\ (gear\atst ($\sigma_0$) = 2)) \lit{implies} ( \lit{forall}\ $\tau_{0}$\ \lit{in}\ [\lit{i2t}($\sigma_0$), \lit{i2t}($\sigma_0$)+%
    \markC{2.5}] \ \lit{such that}~ (gear \attt ($\tau_{0}$) = 2 )).\\
    AT53    & \lit{forall}\  $\sigma_{0}$\ \lit{in}\ [\lit{t2i}(\colorbox{col1!30}{0})+1,\lit{t2i}(%
    \markB{30})]\ \lit{such that}~((gear\atst ($\sigma_0$-1) $\neq$ 3) \lit{and}\ (gear\atst ($\sigma_0$) = 3)) \lit{implies} ( \lit{forall}\ $\tau_{0}$\ \lit{in}\ [\lit{i2t}($\sigma_0$), \lit{i2t}($\sigma_0$)+%
    \markC{2.5}] \ \lit{such that}~ (gear \attt ($\tau_{0}$) = 3 )).\\
    AT54    & \lit{forall}\  $\sigma_{0}$\ \lit{in}\ [\lit{t2i}(\colorbox{col1!30}{0})+1,\lit{t2i}(%
    \markB{30})]\ \lit{such that}~((gear\atst ($\sigma_0$-1) $\neq$ 4) \lit{and}\ (gear\atst ($\sigma_0$) = 4)) \lit{implies} ( \lit{forall}\ $\tau_{0}$\ \lit{in}\ [\lit{i2t}($\sigma_0$), \lit{i2t}($\sigma_0$)+%
    \markC{2.5}] \ \lit{such that}~ (gear \attt ($\tau_{0}$) = 4 )).\\
    AT6a    & (\lit{forall}\  $\tau_{0}$\ \lit{in}\ [0,\colorbox{col1!30}{30}]\ \lit{such that}~$\omega$\attt ($\tau_0$) $<$ %
    \markB{3000}) \lit{implies}\ (\lit{forall}\  $\tau_{1}$\ \lit{in}\ [0,%
    \markC{4}]\ \lit{such that}~v\attt ($\tau_1$) $<$ %
    \markD{35}).\\
    AT6b    & (\lit{forall}\  $\tau_{0}$\ \lit{in}\ [0,\colorbox{col1!30}{30}]\ \lit{such that}~$\omega$\attt ($\tau_0$) $<$ %
    \markB{3000}) \lit{implies}\ (\lit{forall}\  $\tau_{1}$\ \lit{in}\ [0,%
    \markC{8}]\ \lit{such that}~v\attt ($\tau_1$) $<$ %
    \markD{50}).\\
    AT6c    & (\lit{forall}\  $\tau_{0}$\ \lit{in}\ [0,\colorbox{col1!30}{30}]\ \lit{such that}~$\omega$\attt ($\tau_0$) $<$ %
    \markB{3000}) \lit{implies}\ (\lit{forall}\  $\tau_{1}$\ \lit{in}\ [0,%
    \markC{20}]\ \lit{such that}~v\attt ($\tau_1$) $<$ %
    \markD{65}).\\
    AT6abc  & ((\lit{forall}\  $\tau_0$\ \lit{in}\ [0,30]\ \lit{such that}~$\omega$\attt ($\tau_0$) $<$ 3000) \lit{implies}\ (\lit{forall}\  $\tau_1$\ \lit{in}\ [0,4]\ \lit{such that}~v\attt ($\tau_1$) $<$ 35)) \lit{and}\ ((\lit{forall}\  $\tau_2$\ \lit{in}\ [0,30]\ \lit{such that}~$\omega$\attt ($\tau_2$) $<$ 3000) \lit{implies}\ (\lit{forall}\  $\tau_3$\ \lit{in}\ [0,8]\ \lit{such that}~v\attt ($\tau_3$) $<$ 50)) \lit{and}\ ((\lit{forall}\  $\tau_4$\ \lit{in}\ [0,\colorbox{col1!30}{30}]\ \lit{such that}~$\omega$\attt ($\tau_4$) $<$ %
    \markB{3000}) \lit{implies}\ (\lit{forall}\  $\tau_5$\ \lit{in}\ [0,%
    \markC{20}]\ \lit{such that}~v\attt ($\tau_5$) $<$ %
    \markD{65})).\\
    \midrule
    CC1     & \lit{forall}\  $\tau_0$\ \lit{in}\ [\colorbox{col1!30}{0},%
    \markB{100}]\ \lit{such that}~( (y5\attt ($\tau_0$) - y4\attt ($\tau_0$)) $\leq$ %
    \markC{40} ).\\
    CC2     & \lit{forall}\  $\tau_0$\ \lit{in}\ [\colorbox{col1!30}{0},70]\ \lit{such that}~(\lit{exists}\  $\tau_1$\ \lit{in}\ [$\tau_0$+%
    \markB{0},$\tau_0$+30]\ \lit{such that}~((y5\attt ($\tau_1$) - y4\attt ($\tau_1$)) $>$ %
    \markC{15}).\\
    CC3     & \colorbox{col1!30}{\lit{forall}}\  $\tau_0$\ \lit{in}\ [0,80]\ \lit{such that}~((%
    \markB{\lit{forall}}\  $\tau_1$\ \lit{in}\ [$\tau_0$,$\tau_0$+20]\ \lit{such that}~( (y2\attt ($\tau_1$) - y1\attt ($\tau_1$)) $<$ 20 )) %
    \markC{\lit{or}}\ (\lit{exists}\  $\tau_2$\ \lit{in}\ [$\tau_0$,$\tau_0$+20]\ \lit{such that}~( (y5\attt ($\tau_2$) - y4\attt ($\tau_2$)) $>$ 40 ))).\\
    CC4     & \colorbox{col1!30}{\lit{forall}}\  $\tau_0$\ \lit{in}\ [0,65]\ \lit{such that}~(\lit{exists}\  $\tau_1$\ \lit{in}\ [$\tau_0$,$\tau_0$+30]\ \lit{such that}~(%
    \markB{\lit{forall}}\  $\tau_2$\ \lit{in}\ [$\tau_1$,$\tau_1$+5]\ \lit{such that}~((y5\attt ($\tau_2$) - y4\attt ($\tau_2$)) $>$ %
    \markC{8}))).\\
    CC5     & \lit{forall}\  $\tau_0$\ \lit{in}\ [0,72]\ \lit{such that}~(\lit{exists}\  $\tau_1$\ \lit{in}\ [$\tau_0$,$\tau_0$+8]\ \lit{such that}~((\lit{forall}\  $\tau_2$\ \lit{in}\ [$\tau_1$,$\tau_1$+5]\ \lit{such that}~((y2\attt ($\tau_2$) - y1\attt ($\tau_2$)) \colorbox{col1!30}{$>$} %
    \markB{9})) \lit{implies}\ (\lit{forall}\  $\tau_3$\ \lit{in}\ [$\tau_1$+5,$\tau_1$+20]\ \lit{such that}~((y5\attt ($\tau_3$) - y4\attt ($\tau_3$)) %
    \markC{$>$} %
    \markD{9})))).\\
    CCx     & (\lit{forall}\  $\tau_0$\ \lit{in}\ [\colorbox{col1!30}{0},%
    \markB{50}]\ \lit{such that}~((y5\attt ($\tau_0$) - y4\attt ($\tau_0$)) $>$ %
    \markC{7.5})) \lit{and}\ (\lit{forall}\  $\tau_1$\ \lit{in}\ [0,50]\ \lit{such that}~((y4\attt ($\tau_1$) - y3\attt ($\tau_1$)) $>$ 7.5)) \lit{and}\ (\lit{forall}\  $\tau_2$\ \lit{in}\ [0,50]\ \lit{such that}~((y3\attt ($\tau_2$) - y2\attt ($\tau_2$)) $>$ 7.5)) \lit{and}\ (\lit{forall}\  $\tau_3$\ \lit{in}\ [0,50]\ \lit{such that}~((y2\attt ($\tau_3$) - y1\attt ($\tau_3$)) $>$ 7.5)).\\
    \midrule
    RR & \lit{forall}\  $\tau_0$\ \lit{in}\ [0,$\infty$]\ \lit{such that}~((\texttt{d\_pos\_x} \attt ($\tau_0$) - \texttt{v\_pos\_x} \attt ($\tau_0$) ) $<$ \colorbox{col1!30}{0.20} %
    \markB{\lit{and}}\ \texttt{d2obs} \attt ($\tau_0$) $>$ %
    \markC{0.50}).\\
    \bottomrule
    \end{tabular}
    \begin{flushleft}
    \end{flushleft}
\end{table*}

\begin{itemize}
    \item[] \textbf{RQ2}: How \emph{efficient} is SBTD in producing informative diagnoses? (\Cref{sec:efficiency})
\end{itemize}
To answer this question, we assess the time required by SBTD to produce the diagnoses and assess the execution time of the components from \Cref{sec:sbtforhls}. We also report the runtime of TD-SB-TemPsy on the requirements it supports, as a point of comparison.

\begin{itemize}
    \item[] \textbf{RQ3}: How do mutation choices and mutation budget affect the diagnosis stability, interpretability, and computational cost in SBTD? (\Cref{sec:cost-effective})
\end{itemize}
To answer this question, we performed a sensitivity study in which we varied the mutation space (i.e., which tokens in $sub(\property)$ are mutable and their admissible ranges) and the \NAME mutation budget (maximum number of token mutations per individual), and we measured the resulting runtime, the stability and  the complexity of the learned diagnostic decision trees.

\smallskip
To address our research questions, we used \NAME as a representative SBTD framework. Our findings derive from the benchmark selection, experimental setup, and tool configuration defined for \NAME. The implementation of \NAME is publicly available~\cite{Diagnosistool}.
An Appendix with a complete analysis of each experiment is also publicly available on Zenodo \cite{ZenodoAppendix}.
We also replicated TD-SB-TemPsy~\cite{boufaied2023} as a baseline on the subset of requirements it supports, and report a side-by-side comparison on the overlapping cases.

\subsection{Experiment Setting and Tool Configuration}
\label{sec:benchmark}
We considered \tracerequirementcombinations trace-requirement combinations, consisting of a trace and a violated requirement. 
Out of these combinations,  \tracerequirementcombinationssimulink trace-requirement combinations were generated by considering \nrequirementsARCH requirements from the ARCH 2023 Competition~\cite{menghi2023arch}, an international SBST competition for \simulink models.
The remaining trace-requirement combination is from a recent example concerning a robot that should follow a trajectory while avoiding collisions~\cite{Zhao2017}. 
 
The trace-requirement combinations from the ARCH 2023 Competition~\cite{menghi2023arch} were extracted from the replication package of two of the tools that participated in the competition (\texttt{ARIsTEO}\cite{ARIsTEO} and \NAMESIMULINK~\cite{formica2022search,fse2024}), and by considering a trace that violates the requirement that was returned by one of the tools. 
The traces have a large number of records (\emph{min}=1594, \emph{max}=10001, \emph{Avg}=6565.3, \emph{StdDev}=2656.9).
 \Cref{tab:reqbenchmark} contains a textual description of the requirements we considered in our evaluation.
 The column ID reports the identifier from the ARCH 2023 competition.
 Out of the seven models used in the competition, we considered only the Automatic Transmission (AT) and Chasing Cars (CC) since they have the highest number of requirements.
 The requirement identifiers from the AT and CC models start with ``AT'' and ``CC''.
For the robotic scenario, we considered one trace-requirement combination (RR).
The requirement specifies that the robot should follow a desired trajectory while avoiding collisions~\cite{Reynolds87}.

Since the requirements from the ARCH competition are formalized in Signal Temporal Logic (STL)~\cite{STL}, and \NAME supports HLS, we proposed an alternative specification of the requirements in HLS.
\Cref{tab:reqbenchmarkHLS} contains the HLS formalization for the requirements from \Cref{tab:reqbenchmark}.
Since HLS is more expressive than STL~\cite{menghi2021trace}, all the requirements could be expressed in HLS.

For each trace-requirement combination, we defined the terms from the requirements that should be considered to understand the causes of the violations. 
The parts of the requirements and their formalization considered to understand the cause of the violations are colored in \Cref{tab:reqbenchmark} and \Cref{tab:reqbenchmarkHLS}.
We performed two experiments for each trace-requirement combination, considering different subsets of terms to be mutated. 
The two columns of \Cref{tab:experiments} report the subset of terms considered for each trace-requirement combination.
\Cref{tab:experiments} shows that our benchmark primarily exercises numeric-token mutations: OP13 appears in 23/34 experiments (67.6\%) and OP11 in 18/34 (52.9\%), while non-numeric domains are used in 6/34 experiments (17.6\%).
Five operator families are exercised by the configured mutation spaces in this study (OP11, OP13, OP5, OP4, and OP2), whereas the remaining operators in \Cref{tab:mutations} (e.g., OP1, OP3, OP6--OP10, OP12, OP14, and OP15) are implemented but not covered by the current benchmark.
This distribution reflects the available subject requirements and our expert-guided choice to probe domain-plausible thresholds and a limited set of discrete operator flips.

Our benchmark's mutation spaces are dominated by numeric ranges, so they exercise only a subset of the operators of \Cref{tab:mutations} (OP2, OP4, OP5, OP11, and OP13). The mutations are dictated by the structure of the considered requirements, it is not a limitation of the mutation operators provided by our solution.

Considering two subsets of terms to be mutated for each trace-requirement combination led to \numexperiments experiments ($\tracerequirementcombinations \times 2$) marked in \Cref{tab:experiments} with the identifiers \emph{exp1}, \emph{exp2}, \ldots, \emph{exp34}.
The mutation operators to be used for each experiment and the value ranges to be considered to mutate the values of the real-valued variables are reported in \Cref{tab:experiments}.
For example, for the requirement AT1 and experiment \emph{exp1} the tool operator considered the mutation operator \textbf{OP13} for changing AT1(%
\markC{120}) with threshold values of [100, 140]mph; for experiment \emph{exp2} the operator considered the mutation operators \textbf{OP11}, with value ranges of [0, 10]s for AT1(\colorbox{col1!30}{0}) and [10, 30]s for 
 AT1(%
 \markB{20}), and \textbf{OP13}, with value range of [100, 140]mph for AT1(%
 \markC{120}).

For each violated trace--requirement pair, an expert (without access to \NAME's output) is given the requirement, the violating trace, and the mutation-space specification, and produces a reference diagnosis as a decision tree (or equivalent rules) over the mutable tokens predicting the satisfaction (or violation) for mutated requirements.

To answer the research questions of the evaluation, we configured \NAME as detailed in \Cref{tab:ga_bench_params}. 

\emph{Search dynamics.} We set the crossover rate (CR) to $0.95$ as in~\cite{nunez2007fitting} and the mutation rate (MR) to $0.90$ to favor the generation of new mutations.
We set the population size (PS) to $50$ and used roulette wheel selection (SA) as in~\cite{o2009riccardo}; we set the parents-to-be-chosen (PTBC) parameter to $10$ and the tournament size (TS) to $50$.

\emph{Stopping.} We stop the search on a \emph{precision--recall} criterion: after each generation we train a decision tree on the samples collected so far and stop once its cross-validated precision and recall both stay above $0.995$ for two consecutive generations, so the search ends as soon as the diagnosis stabilizes rather than after a fixed number of samples. As a safeguard we cap the search at $1000$ satisfied and $1000$ violated mutated requirements (MS) and at $100$ generations. The criterion requires two consecutive generations, so the search never stops within a single generation and the tree is never accepted on a trivially small sample.

\emph{Budget limits.} {Each trace-checking call is bounded by a two-tier timeout (TCTO), a short budget of $60$s that, if the verdict is still inconclusive, is escalated once to a high tier $600$s. \NAME stops if it cannot produce a diagnosis within five days (PGTO).} When the per-call trace-checking timeout (TCTO) is reached, the trace-checker returns \texttt{undecided}. We treat such outcomes as inconclusive, track them separately, and exclude them from the \texttt{sat}/\texttt{unsat} pools and from all precision and recall computations. 
{T}he five-day budget (PGTO) {is} an experimental cap to bound computational cost, chosen for the computational resources available for our study. Runs exceeding PGTO are \emph{timeouts}. %

As quantified in \Cref{sec:cost-effective} (\textbf{RQ3}), the dominant drivers of runtime are {the trace-checking budgets (TCTO/PGTO)}, together with the breadth of the configured mutation space.

We ran the experiments on the \emph{nibi} cluster, where each experiment used a single-node allocation of $8$ CPU cores and $16$\,GB of RAM (Intel Xeon~6972P)---with the support from Compute Ontario (\href{https://www.computeontario.ca/}{computeontario.ca}) and the Digital Research Alliance of Canada (\href{https://alliancecan.ca/}{alliancecan.ca}).

\begin{table}[t]
    \centering
    \begin{tabular}{l l l l}
    \toprule
    \textbf{Parameter} & \textbf{Value} & \textbf{Parameter} & \textbf{Value}\\
    \midrule
    CR & 0.95 & PTBC & 10\\
    MR & 0.90 & MS & $1000$ sat + $1000$ violated\\
    PS & 50 & TS & 50\\
    TCTO & {$60{\to}600$s} & SA & Roulette Wheel\\
    PGTO & 5 days\\
    \bottomrule
    \end{tabular}
\caption{Values for the configuration parameters of \NAME from \Cref{tab:configurationParameters}.}
    \label{tab:ga_bench_params}
    
\end{table}

\begin{table*}
    \caption{Mutation operators (Operators) and ranges for the value terms (Ranges) of each experiment (Exp).}
    \label{tab:experiments}
    \footnotesize
    \scalebox{0.93}{
    \begin{tabular}{p{1cm} | p{0.7cm} p{2cm} p{4cm} |  p{0.7cm} p{3cm} p{5.0cm}}
    \toprule
       \textbf{Req. ID}  & \textbf{Exp.} & \textbf{Operators} &  \textbf{Valid Range} & \textbf{Exp.} & \textbf{Operators} &  \textbf{Valid Range}  \\
    \midrule
    AT1     & \emph{exp1} & %
    \markC{OP13} & %
    \markC{[100,140]mph} 
    &\emph{exp2} & \colorbox{col1!30}{OP11},  %
    \markB{OP11},  %
    \markC{OP13} &  \colorbox{col1!30}{[0,10]s},  %
    \markB{[10,30]s},  %
    \markC{[100,140]mph} \\
    \midrule
    AT2     & \emph{exp3} & %
    \markC{OP13} & %
    \markC{[4700,4800]rpm} &\emph{exp4} & \colorbox{col1!30}{OP11},  %
    \markB{OP11},  %
    \markC{OP13} &$ \colorbox{col1!30}{[0,5]s}, %
    \markB{[5,15]s}, %
    \markC{[4700,4800]rpm}$ \\
    \midrule
    AT51    & \emph{exp5} & %
    \markC{OP11} & %
    \markC{[0,5]s}  
    &\emph{exp6} & \colorbox{col1!30}{OP11},  %
    \markB{OP11},  %
    \markC{OP11} & \colorbox{col1!30}{[0,15]s}, %
    \markB{[15,45]s}, %
    \markC{[0,5]s}\\
    \midrule
    AT52    & \emph{exp7} & %
    \markC{OP11} & %
    \markC{[0,5]s}  
    &\emph{exp8} & \colorbox{col1!30}{OP11},  %
    \markB{OP11},  %
    \markC{OP11} & \colorbox{col1!30}{[0,15]s}, %
    \markB{[15,45]s}, %
    \markC{[0,5]s}  \\
    \midrule
    AT53    & \emph{exp9} & %
    \markC{OP11} & %
    \markC{[0,5]s} &\emph{exp10} & \colorbox{col1!30}{OP11},  %
    \markB{OP11},  %
    \markC{OP11} & \colorbox{col1!30}{[0,15]s}, %
    \markB{[15,45]s}, %
    \markC{[0,5]s}\\
    \midrule
    AT54   & \emph{exp11} & %
    \markC{OP11} & %
    \markC{[0,5]s} 
    &\emph{exp12} & \colorbox{col1!30}{OP11},  %
    \markB{OP11},  %
    \markC{OP11} & \colorbox{col1!30}{[0,15]s}, %
    \markB{[15,45]s}, %
    \markC{[0,5]s} \\
    \midrule
    AT6a    & \emph{exp13} & %
    \markB{OP13},  %
    \markD{OP13} & %
    \markB{[2800,3200]rpm}, %
    \markD{[30,40]mph} 
    &\emph{exp14} & \colorbox{col1!30}{OP11},  %
    \markB{OP13},  %
    \markC{OP11},  %
    \markD{OP13} &  \colorbox{col1!30}{[20,40]s}, %
    \markB{[2800,3200]rpm}, %
    \markC{[2,6]s}, %
    \markD{[30,40]mph} \\
    \midrule
    AT6b    & \emph{exp15} & %
    \markB{OP13},  %
    \markD{OP13} & %
    \markB{[2800,3200]rpm}, %
    \markD{[40,60]mph} 
    &\emph{exp16} & \colorbox{col1!30}{OP11},  %
    \markB{OP13},  %
    \markC{OP11},  %
    \markD{OP13} & \colorbox{col1!30}{[20,40]s}, %
    \markB{[2800,3200]rpm}, %
    \markC{[4,12]s}, %
    \markD{[40,60]mph}  \\
    \midrule
    AT6c    & \emph{exp17} & %
    \markB{OP13},  %
    \markD{OP13} & %
    \markB{[2800,3200]rpm}, %
    \markD{[50,80]mph}
    &\emph{exp18} & \colorbox{col1!30}{OP11},  %
    \markB{OP13},  %
    \markC{OP11},  %
    \markD{OP13} & \colorbox{col1!30}{[20,40]s}, %
    \markB{[2800,3200]rpm}, %
    \markC{[15,25]s}, %
    \markD{[50,80]mph}\\
    \midrule
    AT6abc  & \emph{exp19} & %
    \markB{OP13},  %
    \markD{OP13} & %
    \markB{[2800,3200]rpm}, %
    \markD{[50,80]mph} 
    &\emph{exp20} & \colorbox{col1!30}{OP11},  %
    \markB{OP13},  %
    \markC{OP11},  %
    \markD{OP13} & \colorbox{col1!30}{[20,40]s}, %
    \markB{[2800,3200]rpm}, %
    \markC{[15,25]s}, %
    \markD{[50,80]mph} \\
    \midrule
    CC1     & \emph{exp21} & %
    \markC{OP13} & %
    \markC{[30,50]m}  
    &\emph{exp22} & \colorbox{col1!30}{OP11},  %
    \markB{OP11},  %
    \markC{OP13} &\colorbox{col1!30}{[0,50]s}, %
    \markB{[50,100]s}, %
    \markC{[30,50]m} \\
    \midrule
    CC2     & \emph{exp23} & \colorbox{col1!30}{OP11} & \colorbox{col1!30}{[0,20]s} 
    &\emph{exp24} & \colorbox{col1!30}{OP11},  %
    \markB{OP11},  %
    \markC{OP13} & \colorbox{col1!30}{[0,20]s}, %
    \markB{[0,10]s}, %
    \markC{[12,18]m} \\
    \midrule
    CC3     & \colorbox{brown!30}{\emph{exp25}$^{*}$} & \colorbox{col1!30}{OP5} & \colorbox{col1!30}{\{\lit{forall},\lit{exists}\}}
    & \emph{exp26}$^{*}$ & \colorbox{col1!30}{OP5}, %
    \markB{OP5}, %
    \markC{OP4} &$\colorbox{col1!30}{\{\lit{forall},\lit{exists}\}}$, $%
    \markB{\{\lit{forall},\lit{exists}\}}$, $%
    \markC{\{\lit{and}, \lit{or}\}}$ \\
    \midrule
    CC4     & \emph{exp27} & %
    \markC{OP13} & %
    \markC{[6,10]m} 
    &\emph{exp28} & \colorbox{col1!30}{OP5},%
    \markB{OP5},%
    \markC{OP13} & \colorbox{col1!30}{\{\lit{forall},\lit{exists}\}}, %
    \markB{\{\lit{forall},\lit{exists}\}}, %
    \markC{[6,10]m} \\
    \midrule
    CC5     & \emph{exp29} & %
    \markB{OP13},  %
    \markD{OP13} & %
    \markB{[7,11]m}, %
    \markD{[7,11]m} 
    &\emph{exp30} & \colorbox{col1!30}{OP2},  %
    \markB{OP13},  %
    \markC{OP2},  %
    \markD{OP13} &$ \colorbox{col1!30}{\{$>,<$\}}$, $%
    \markB{[7,11]m}$, $ %
    $\markC{\{$>,<$\}}$
    $, 
    $%
    \markD{[7,11]m}$\\
    \midrule
    CCx     & \emph{exp31} & %
    \markC{OP13} & %
    \markC{[5,10]m}     
    &\emph{exp32} & \colorbox{col1!30}{OP11},  %
    \markB{OP11},  %
    \markC{OP13} &$ \colorbox{col1!30}{[0,25]s}, %
    \markB{[25,75]s}, %
    \markC{[5,10]m}$ \\
    \midrule
    RR & \emph{exp33} &\colorbox{col1!30}{OP13}, %
    \markB{OP4} &$\colorbox{col1!30}{[5,7]m}$, $%
    \markB{\{\lit{and}, \lit{or}, \lit{implies}\}}$  &\emph{exp34} &\colorbox{col1!30}{OP13}, %
    \markB{OP4}, %
    \markC{OP13} &$\colorbox{col1!30}{[5,7]m}$, $%
    \markB{\{\lit{and}, \lit{or}\}}$, $%
    \markC{[0,2.5]m}$\\
    \bottomrule
    \end{tabular}}
\end{table*}

\begin{figure*}[t]
{\footnotesize
\begin{subfigure}[b]{0.5\textwidth}
    \centering
    \begin{tikzpicture}[level 1/.style={sibling distance=60mm},level 2/.style={sibling distance=40mm}, level distance=40pt]
\node[]{AT1(%
\markD{120})}
 child {
    node[red]{\emph{False (1013)}} edge from parent node[fill=white,left] {$\leq 120.006093$ mph}
 }
 child {
    node[blue]{\emph{True (1013)}} edge from parent node[fill=white] {$> 120.006093$ mph}
 };
\end{tikzpicture}
    \caption{Tool Diagnostics.}
\label{fig:at1_example_decisiontree}
\end{subfigure}
\begin{subfigure}[b]{0.5\textwidth}
    \centering \begin{tikzpicture}[level 1/.style={sibling distance=60mm},level 2/.style={sibling distance=40mm}, level distance=40pt]
\node[]{AT1(%
\markD{120})}
 child {
    node[red]{\emph{False}} edge from parent node[fill=white,left] {$\leq 120.022620$ mph} 
 }
 child {
    node[blue]{\emph{True}} edge from parent node[fill=white] {$> 120.022620$ mph} 
 };
\end{tikzpicture}
    \caption{Expert Prediction.}  \label{fig:at1_example_predictiontree}
\end{subfigure}
\caption{Diagnostic and prediction for the experiment \emph{exp1}.}
}
\label{sec:diagn}
\end{figure*}

\subsection{Baseline: TD-SB-TemPsy}
\label{sec:baseline}
A meaningful baseline for our experiments should take as input a violated signal-based temporal requirement and a trace, and return an automated \emph{diagnosis} that explains the violation at the level of the requirement (e.g., which conditions and signals are responsible).
We therefore compare \NAME against TD-SB-TemPsy~\cite{boufaied2023}, which explicitly targets trace diagnosis for signal-based temporal requirements and outputs root-cause diagnoses for violated requirements.
TD-SB-TemPsy's specification language corresponds to a subset of the temporal constructs supported by \NAME, since all TD-SB-TemPsy requirements can be expressed in  HLS~\cite{menghi2021trace}.

Re-running TD-SB-TemPsy from the public repository required additional configuration beyond the default instructions, in particular: Java~7, Eclipse Mars~4.5.2, EMF~2.11.x, and OCL Classic~3.5.0 (we also tested OCL~6.0.1).
It did not run out-of-the-box with newer Java/Eclipse/EMF/OCL versions, and we fixed one issue in the provided OCL file.
With support from the original authors, we confirmed this toolchain and replicated the baseline results; we documented the configuration and fix in a public pull request.\footnote{\href{https://github.com/SNTSVV/TD-SB-TemPsy/pull/2}{https://github.com/SNTSVV/TD-SB-TemPsy/pull/2}}

We report baseline outcomes and a side-by-side comparison with \NAME under \Cref{sec:effectiveness} (RQ1), and discuss baseline runtime behavior under \Cref{sec:efficiency} (RQ2).

\begin{table*}[t]
    \centering
    \caption{Comparison on the 5 requirements where TD-SB-TemPsy returns a diagnosis. TD-SB-TemPsy reports a signal and timestamp witness, while \NAME returns a threshold rule over mutated tokens. {\bf Note}: in {\bf RR}, $\delta_x = d\_pos\_x - v\_pos\_x$.}
    \label{tab:baseline_comp_table}
    \renewcommand{\arraystretch}{1.15}
    \setlength{\tabcolsep}{3pt}
    \begin{tabularx}{\linewidth}{@{} l X X X @{}}
    \toprule
    \textbf{Req.} & \textbf{Expert Prediction} & \textbf{TD-SB-TemPsy} & \textbf{\NAME} \\
    \midrule
    {\bf AT1} & If $Speed \le 120.022620$mph: False
        & Speed @ 1.999E7$\mu s$  = 120.00171mph
        & If $Speed \le 120.006093$mph: False \\
    \midrule
    {\bf AT2} & If $Engine \le 4754.401200$rpm: False
        & Engine @ 7.31E6$\mu s$  = 4751.7885rpm
        & If $Engine \le 4754.377727$rpm : False \\
    \midrule
    {\bf CC1} & If $y5-y4 \le 41.180040$m: False
        & $y5-y4 @ 2.119E7\mu s$ = 40.03113m
        & If $y5-y4 \le 41.167538$m: False \\
    \midrule
    {\bf CCx} & If $y5-y4 \le 7.389400$m: False
        & $y5-y4$ @ 1.818E7$\mu s$  = 7.4824m
        & If $y5-y4 \le 7.366460$m: False \\
    \midrule
    {\bf RR} &
    \makecell[cl]{If AND: \\
    ~~$\delta_x \le 5.48$m : False\\ 
    ~~$\delta_x>5.48$m : \\
    ~~~~$dist2obj>0.687$m : False\\
    Else (OR):  \\
    ~~$\delta_x \le 5.48$m : \\ 
    ~~~~$dist2obj>0.687$m : False}
    &
    \makecell[cl]{$\delta_x$  @ 3.06E6$\mu s$ 
    = 2.5833682m}
    &
    \makecell[cl]{If AND: \\ 
    ~~$dist2obj>0.6864$m : False \\
    ~~$dist2obj\le0.6864$m : \\ 
    ~~~~$\delta_x\le5.480303$m : False\\
    Else (OR):  \\ 
    ~~$\delta_x\le5.475945$m : \\
    ~~~~$dist2obj>0.7038$m : False}
    \\
    \bottomrule
    \end{tabularx}
\end{table*}

\subsection{Effectiveness (RQ1)}
\label{sec:effectiveness}
Our research hypothesis is that SBTD is effective in producing informative diagnoses.
We assessed how effective SBTD is in producing informative diagnoses to validate our hypothesis. We compare diagnostics produced using \NAME to the causes that led to requirement violation, according to an expert, and to the ones produced by TD-SB-TemPsy.

\emph{Comparison with TD-SB-TemPsy}. 
We ran TD-SB-TemPsy on the same violated trace--requirement pairs from our benchmark. On the 17 requirements, TD-SB-TemPsy (i) returns a diagnosis for 5/17 (AT1, AT2, CC1, CCx, RR), (ii) loads the specification but prints no diagnosis within our time limit for 4/17 (AT51--AT54), and (iii) cannot be used for 8/17 due to toolchain limitations in our setup (AT6a/AT6b/AT6c/AT6abc: \emph{Before}/\emph{Between} lead to a diagnosis-undefined crash; CC2--CC5: nested temporal operators). We therefore compare both tools side-by-side on the five overlapping cases (Tab.~\ref{tab:baseline_comp_table}).

{Each of the TD-SB-TemPsy diagnoses} has the form
\texttt{Tuple\{vSignal, Records, Interval, violationType\}} and consistently reports (i) a single signal (\texttt{vSignal}) and (ii) a concrete witness sample (\texttt{Records}) containing a timestamp and the corresponding value; 
{in our five cases, the witnesses are single timestamps, therefore the \texttt{Interval} field is \texttt{null}. The output does not provide additional causal structure (e.g., contributing sub-conditions or a responsibility interval).}
Concretely, TD-SB-TemPsy pointed to the signals \texttt{Speed} (AT1), \texttt{Engine} (AT2), \texttt{y5-y4} (CC1/CCx), and \texttt{d\_pos\_x - v\_pos\_x ($\delta_x$)} (RR), each with a single timestamp witness.
This makes the baseline explanations useful as \emph{localization cues} (what signal and 
{at which timestamp}
).
{As shown in Tab.~\ref{tab:baseline_comp_table}, for AT1/AT2/CC1/CCx both tools agree on the relevant signal and yield consistent threshold-level explanations (TD-SB-TemPsy as a point-wise witness and \NAME as an explicit split threshold). For RR, TD-SB-TemPsy localizes a witness on the derived signal $\delta_x$, whereas \NAME exposes additional structure by splitting on both $dist2obj$ and $\delta_x$.}
When the reported signal is derived (e.g., \texttt{y5-y4} in CC1/CCx), the output does not attribute the violation to underlying signals (e.g., \texttt{y5} or \texttt{y4}) or sub-conditions, which can limit its usefulness for debugging.

Four additional requirements (AT51--AT54) were loaded successfully by TD-SB-TemPsy, but produced no output of any kind. Unlike \NAME, which writes a file and a progress message for each generation of mutated requirements, TD-SB-TemPsy emits nothing until it terminates, so we could not distinguish a long computation from a stalled one. We interrupted these runs after one hour without output, and we do not count them among the diagnoses returned by the baseline.
Moreover, eight requirements could not be handled by the public TD-SB-TemPsy pipeline in our setup: AT6a/AT6b/AT6c/AT6abc require the \emph{Before}/\emph{Between} scopes, for which the tool reports diagnosis as undefined and crashes (OCL evaluation returns \texttt{null}, leading to a \texttt{NullPointerException}); CC2--CC5 require nesting of temporal operators, which is not supported by the SB-TemPsy toolchain.
Since HLS is more expressive than the SB-TemPsy fragment supported by TD-SB-TemPsy, all properties expressible for TD-SB-TemPsy can also be encoded in \NAME, while \NAME additionally supports the constructs needed by the other requirements in our benchmark.

\emph{Comparison with Expert Expectations}. We compared the diagnoses produced by \NAME with expert expectations to assess whether SBTD is effective (\textbf{RQ1}).
Specifically, 
\NAME generated a diagnostic decision tree for each experiment, while an expert independently produced a trace-grounded reference diagnosis for the same requirement and mutation space (tokens and admissible ranges). 
The reference is derived from concrete trace witnesses (e.g., extrema/crossings in the relevant time windows) and validated by re-checking representative mutated requirements on the same trace.
We then compared the tool-generated diagnoses against these expert predictions. 

The participants were two authors, one acted as the \NAME operator, while the other served as the domain expert (with extensive prior familiarity with the benchmark models and requirements). 
The two authors did not exchange information about the experiments during the experimental set, and the expert did not have access to \NAME's outputs when producing the reference diagnosis.
The experimental set followed two steps: (i) cause derivation and (ii) diagnostics and prediction comparison. The experimental set is summarized in \Cref{tab:experiments}, which maps requirement IDs to independent variables (namely Operators), and valid ranges exercised in each experiment. The colored background in \Cref{tab:experiments} maps terms from the \Cref{tab:reqbenchmark} to mutated operators. We include two (nested) mutation-space configurations per pair (\Cref{tab:experiments}), one focused and one broader, to assess sensitivity to the engineer-configured mutation space.

(i) {\em  Cause Derivation}. The {\em tool operator} and the {\em expert} worked separately to derive the causes of the violated requirements. 
The {\em tool operator} configured \NAME using the configuration parameters in \Cref{tab:ga_bench_params}. As a result, the {\em tool operator} collected one decision tree for each experiment. For example, \Cref{fig:at1_example_decisiontree} reports the diagnosis for the experiment \emph{exp1} that considers the impact of the value AT1(%
\markC{120}) on the satisfaction of the requirement AT1. 
The decision tree shows that setting the value of AT1(%
\markC{120}) higher and lower than $120.006093$ respectively, makes the property satisfied or violated since the signal reaches the value $120.006093$.
Specifically, the diagnosis reports the identity of the variable and the corresponding boundary value leading a change in the trace-checking verdict. The diagnosis information informs the engineer that the speed exceeds the $120$mph bound by only about $0.006$mph, so the violation is a near-miss rather than a gross exceedance---an information that guides the engineer in the debugging activity.
Note that the DT leaves contain the same number (1013) of satisfied and unsatisfied requirements since the requirement filtering step ensures that the number of satisfied and unsatisfied requirements corresponds. 

The {\em expert} analyzed the violated requirement according to their experience and manually synthesized a prediction. 
To synthesize the prediction the expert also plotted the trace and tried to reverse-engineer the cause of the violation and express it as a DT.
For example, \Cref{fig:at1_example_predictiontree} reports the prediction for the experiment \emph{exp1}.
Note that, for this example, since the expert can inspect the trace, they can identify the exact condition ($>$120.022620) that turns the property from violated to satisfied.

{\em (ii) Diagnostics and Prediction Comparison.} We compared the diagnostics ({\it tool operator}'s decision trees) and the predictions ({\it expert}'s decision trees). 

For our experiments, the DT produced by the tool operator and the expert can be significantly different: values can be considered in multiple orders and be split various times by each decision tree.
{To compare DTs consistently, we evaluate them on the same finite set of candidate mutated requirements, obtained by discretizing each continuous admissible mutation range (e.g., a numeric threshold interval) into a uniform grid.}
We therefore approximate each DT’s classification behavior by evaluating both on the same finite set of candidate properties.
Following~\cite{gaaloul2021combining}, we use a uniform grid of 101 values per mutated numeric token over its configured admissible range (endpoints included), as a pragmatic resolution-cost trade-off under combinatorial blow-up.
For discrete mutation tokens (e.g., logical operators), we enumerate all admissible values from the mutation configuration (rather than sampling), since these domains are finite and small.
For example, for \emph{exp2} a set of properties is generated by considering $101$ assignments for each variables:
for AT1(\colorbox{col1!30}{0}) values from $0$ to $10$ with increments of $0.1$, 
for AT1(%
\markB{20}) values from $10$ to $30$ with increments of $0.2$, and 
for AT1(%
\markC{120}) values from $100$ to $140$ with increments of $0.4$.
Considering the combinations of these values leads to a total of \num{1030301} properties.
When the mutations also involved logical operators (e.g., \emph{exp26}) this procedure was replicated for all the possible assignments of the logical operators.
For example, for \emph{exp26} the procedure assigned CC3(\colorbox{col1!30}{\lit{forall}}), CC3(%
\markB{\lit{forall}}), CC3(%
\markC{\lit{and}}) to both \{\lit{forall}, \lit{forall}, \lit{and}\},
\{\lit{forall}, \lit{forall}, \lit{or}\}, and \{\lit{forall}, \lit{exists}, \lit{and}\}, and all the remaining combinations of logical operators.

For each property, we assessed whether the property was expected to be satisfied or violated according to the DTs produced by the tool operator and the expert. 
This was done by assessing whether the leaf of the DT associated with that formula was labeled with a \emph{True} or a \emph{False} value. 
A true positive (TP)  is when the property is satisfied according to both the DTs (the one from the tool operator and the one from the expert).
A true negative (TN)  is when the property is violated according to both DTs.
A false positive (FP) is when the property is satisfied by the DT returned by the tool operator and violated by the one produced by the expert.
Finally, a false negative (FN) is when the property is violated by the DT returned by the tool operator and satisfied by the one produced by the expert.
We analyzed the precision and recall of the method. {In our setting, precision/recall quantify expected debugging effort under a bounded budget (not a soundness guarantee): lower precision implies more false leads to inspect, while lower recall indicates missing contributing conditions and require a follow-up run with a larger budget and/or expanded mutation space.}

Note that the \texttt{ThEodorE} trace-checker returns a three-valued verdict: it reports \texttt{sat} when Z3 finds a counterexample (i.e., the requirement is violated by the trace), \texttt{unsat} when no counterexample exists (i.e., the requirement is satisfied), and \texttt{undecided}
when satisfiability cannot be determined within the resource limits.
We treat \texttt{undecided} as inconclusive, track it separately, and exclude it from the satisfied or violated sets used by \NAME and from all effectiveness computations (TP/TN/FP/FN, precision, and recall).
Since \NAME's diagnosis predicts only a definitive verdict (\texttt{sat} or \texttt{unsat}), \texttt{undecided} is not a diagnostic class but a missing label: a candidate whose verdict the trace-checker could not establish within the per-call timeout. Precision and recall compare the tool and expert decision trees over the same discretized candidate grid, on which each tree predicts a definitive verdict, therefore, \texttt{undecided} verdicts can not be considered  for the precision/recall computation. 

The precision--recall stopping criterion ends each run when the diagnosis stabilizes, so the decision tree is learned automatically from the samples collected at that point
A run whose learned tree has no split is a \emph{degenerate diagnosis}: the collected samples are essentially single-class, so the tree predicts the same verdict everywhere and identifies no responsible token. We report such runs separately and exclude them from the precision and recall computation, since they contain no split to compare against the expert.

\emph{Results}. Running our experiments would have required approximately 24 days. The time was reduced to five days by exploiting the parallelization facilities of our computing platform. Each seed run used one node, and independent experiment--seed runs were scheduled concurrently; the platform did not parallelize a single diagnosis. The per-run wall-clock evidence is reported in \Cref{tab:time} and in the seed-level replication record.

We set five days as the global time budget (PGTO) for a run.  Runs that exceed this budget are reported as timeouts.

{\NAME produced a diagnosis, scored against the expert, for $29$ of the \numexperiments configurations: 27 did so in every seed, \emph{exp30} in nine of ten seeds, and \emph{exp32} in four of ten. Of the remaining five, \emph{exp25} and \emph{exp26} completed with degenerate trees, and \emph{exp27}, \emph{exp28}, and \emph{exp31} timed out in all ten seeds.}

\begin{figure}
    \centering
   \includegraphics[width = \columnwidth]{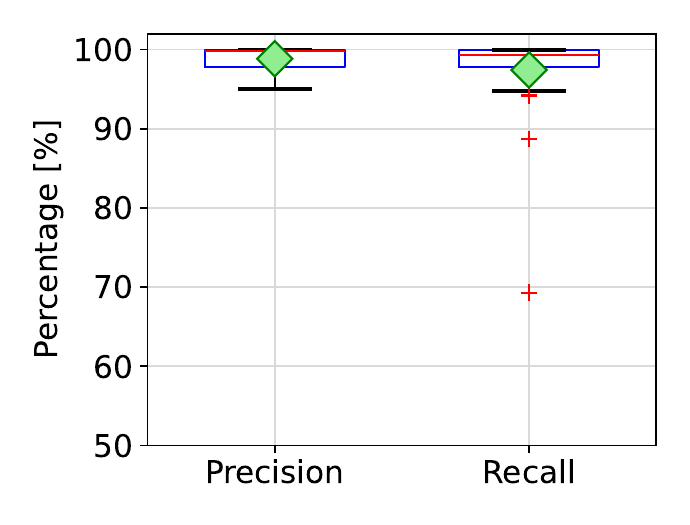}
    \caption{Precision and recall of SBTD across the different experiments. Diamonds depict the average, red lines are the median, and pluses depict the outliers.}
    \label{fig:boxplot}
\end{figure}

The boxplot from \Cref{fig:boxplot} presents the precision ($\frac{TP}{TP+FP}$) and recall ($\frac{TP}{TP+FN}$) of SBTD across the different experiments.
SBTD shows a considerable precision (\emph{min}={95.0}\%, \emph{max}=100.0\%, \emph{Avg}={98.8}\%, \emph{StdDev}={1.5}\%) across the medians of the scored experiments, showing that the value ranges for which the requirements are satisfied are confirmed by the expert.
SBTD shows a considerable recall (\emph{min}={69.3}\%, \emph{max}=100.0\%, \emph{Avg}={97.4}\%, \emph{StdDev}={5.9}\%) across the medians of the scored experiments, showing that SBTD can identify most of the values for which the requirements are satisfied. The \emph{exp30} and \emph{exp32} values summarize only their nine and four scorable seeds, respectively.

The diagnosis produced by SBTD provides a prioritization aid for debugging by highlighting signals and sub-formula fragments associated with violations.
Imperfect precision may increase inspection effort (false leads), while imperfect recall may miss contributing factors and require a follow-up run (e.g., expanded mutation space or budget). We therefore report precision/recall to quantify expected effort and residual uncertainty.

{For example, \emph{exp25} (}identified with a \colorbox{brown!30}{brown} background in Table~\ref{tab:experiments}), SBTD {did not produce an informative diagnosis: the learned tree is degenerate (no split)}.
{For this case, the CC3 mutation space is discrete and small, so \NAME collects too few distinct trace-checking verdicts to learn an informative split, and the tree degenerates to a single leaf.}
As %
reported by the authors~\cite{menghi2021theodore,menghi2021trace}, while supporting an expressive logic (HLS),  \texttt{ThEodorE} inherits the limitations of the SMT technology used to solve the trace-checking problem, which can require considerable time to solve the satisfiability problem and terminate with an 
{\tt undecided}
result. 
Note that we assumed that an 
{\tt undecided} result confirmed the violation of a property only when for some of the generated trace-requirement combinations the trace-checker could confirm that the property was satisfied by the trace. 

\newpage
\begin{Answer}[RQ1 - Effectiveness]

The baseline TD-SB-TemPsy returned a diagnosis on only 5 out of the 17 requirements in our benchmark, whereas \NAME can be applied to the full HLS benchmark. On the 5 requirements, both approaches identify consistent threshold-based explanations that align with the expert prediction (Tab.~\ref{tab:baseline_comp_table}).

We found that \NAME produced an expert-aligned diagnosis in {29} out of \numexperiments configurations against the expert-labeled causes
 for the same trace--requirement pairs.

{Two of the remaining five collected essentially single-class samples, so no discriminating tree could be learned, and three reached the budget without producing a tree.}
\end{Answer}

\subsection{Efficiency (RQ2)}
\label{sec:efficiency}
We assessed how efficient SBTD is in producing informative diagnoses as follows.

\emph{Methodology}. 
We consider the experiments executed to answer RQ1. 
We recorded the time \NAME, and its components (see \Cref{sec:approach}), required to produce the diagnoses and analyzed it.

\begin{table*}[t]
    \centering
       \caption{Median per-experiment wall-clock time across ten seeds (including elapsed timeout runs), broken down into the mutation engine (Tool), trace-checking (Tr.-check), and decision-tree generation (DT), together with the median number of trace-checker calls (TC). ``$^\ddagger$'' marks a configuration with no discriminating diagnosis in any seed; ``$^\dagger$'' marks a configuration in which at least one seed reached the PGTO/wall guard.}
    \label{tab:time}
    \setlength{\tabcolsep}{3pt}
    \footnotesize
    \begin{tabular}{l l r r r r r | l r r r r r}
        \toprule
        ID    & Exp.  &Tool (total)  &Tool \phase{1} &Tool \phase{2} &Tool \phase{3} & Tr.-check Calls & Exp.     &Tool (total)  &Tool \phase{1} &Tool \phase{2} &Tool \phase{3} & Tr.-check Calls \\ 
        \midrule
        AT1 & \emph{exp1} & 23min & 2s & 23min & $<$1s & 232 & \emph{exp2} & 58min & 42s & 56min & $<$1s & 1028\\
        AT2 & \emph{exp3} & 13min & 2s & 12min & $<$1s & 236 & \emph{exp4} & 13min & 4s & 13min & $<$1s & 291\\
        AT51 & \emph{exp5} & 15min & 3s & 15min & $<$1s & 184 & \emph{exp6} & 18min & 4s & 18min & $<$1s & 266\\
        AT52 & \emph{exp7} & 13min & 3s & 13min & $<$1s & 184 & \emph{exp8} & 27min & 7s & 26min & $<$1s & 276\\
        AT53 & \emph{exp9} & 28min & 3s & 28min & $<$1s & 186 & \emph{exp10} & 35min & 5s & 35min & $<$1s & 299\\
        AT54 & \emph{exp11} & 24min & 3s & 23min & $<$1s & 186 & \emph{exp12} & 43min & 5s & 42min & $<$1s & 298\\
        AT6a & \emph{exp13} & 1.6h & 5s & 1.6h & $<$1s & 346 & \emph{exp14} & 3.7h & 79s & 3.6h & $<$1s & 1014\\
        AT6b & \emph{exp15} & 2.4h & 7s & 2.4h & $<$1s & 411 & \emph{exp16} & 4.0h & 77s & 4.0h & $<$1s & 1014\\
        AT6c & \emph{exp17} & 1.9h & 7s & 1.9h & $<$1s & 368 & \emph{exp18} & 4.1h & 81s & 4.0h & $<$1s & 1022\\
        AT6abc & \emph{exp19} & 2.3h & 12s & 2.2h & 3s & 323 & \emph{exp20} & 6.1h & 4min & 6.0h & 37s & 1020\\
        \midrule
        CC1 & \emph{exp21} & 31.3h & 4s & 31.3h & $<$1s & 280 & \emph{exp22} & 33.6h & 11s & 33.6h & $<$1s & 537\\
        CC2 & \emph{exp23} & 6.6h & 3s & 6.6h & $<$1s & 254 & \emph{exp24} & 14.7h & 7s & 14.6h & $<$1s & 345\\
        CC3 & \emph{exp25} & 14min$^\ddagger$ & 11s & 14min & $<$1s & 2 & \emph{exp26} & 1.1h$^\ddagger$ & 10s & 1.1h & $<$1s & 8\\
        CC4 & \emph{exp27} & 119.8h$^{\dagger\ddagger}$ & 56s & 117.0h & $<$1s & 904 & \emph{exp28} & 119.8h$^{\dagger\ddagger}$ & 3min & 118.4h & $<$1s & 1176\\
        CC5 & \emph{exp29} & 51.3h & 15s & 51.3h & $<$1s & 370 & \emph{exp30} & 55.6h$^{\dagger}$ & 12s & 55.6h & 1s & 364\\
        CCx & \emph{exp31} & 119.8h$^{\dagger\ddagger}$ & 55s & 116.5h & $<$1s & 774 & \emph{exp32} & 119.8h$^{\dagger}$ & 49s & 115.3h & $<$1s & 754\\
        \midrule
        \midrule
        RR     & \emph{exp33} & 14min & 2s & 14min & $<$1s & 192 & \emph{exp34} & 14min & 3s & 14min & $<$1s & 237\\
        \bottomrule
    \end{tabular}
\end{table*}

\emph{Results}. \Cref{tab:time} reports, for each experiment, the median wall-clock time and median number of trace-checker calls across ten seeds. {Across the 34 per-experiment medians, the median runtime is $1.8$h and the interquartile range is $23$\,min--$12.6$h. The mutation engine and decision-tree learning take less than a few seconds, while trace-checking requires most of the time. Twenty-nine configurations produce an informative diagnosis, twenty-seven of them in all ten seeds. \emph{exp25} and \emph{exp26} complete in all seeds but produce degenerate trees, since the discrete CC3 mutation space yields only $2$--$8$ distinct verdicts, too few to support a split. \emph{exp27}, \emph{exp28}, and \emph{exp31} reach the PGTO/wall guard in all seeds without a final tree. 
These runs are timeouts, not converged diagnoses.} 

{To contextualize the Phase~2 bottleneck, we also measured TD-SB-TemPsy wall-clock runtime on a commodity laptop\footnote{MacBook Air, Apple~M3, 24\,GB RAM, macOS~26.0.1} with a one-hour cutoff per requirement. When applicable, TD-SB-TemPsy is very fast: it produced diagnoses in 327--1243\,ms (RR: 327\,ms; AT2: 469\,ms; AT1: 556\,ms; CC1: 1071\,ms; CCx: 1243\,ms). In contrast, four additional requirements (AT51--AT54) loaded successfully but did not emit a diagnosis within one hour (baseline timeouts). Finally, eight requirements could not be handled by the public pipeline in our setup: AT6a/AT6b/AT6c/AT6abc require \emph{Before}/\emph{Between} scopes (the tool reports diagnosis as undefined and crashes due to a \texttt{null}/\texttt{NullPointerException}), and CC2--CC5 require nesting of temporal operators, which is unsupported by the TD-SB-TemPsy toolchain. This highlights a trade-off: TD-SB-TemPsy is computationally lightweight when it applies (pattern-based diagnosis), whereas \NAME is more general but its end-to-end runtime is dominated by SMT-based trace-checking in Phase~2.}

\begin{Answer}[RQ2 - Efficiency]
Trace-checking dominates \NAME's runtime ($97$--$100$\%); the mutation engine and tree learning cost less than a few seconds. Most experiments converge within the $120$h budget (IQR $23$\,min--$12.6$h, median $\approx\!1.8$h); the \emph{CC4} and \emph{CCx} configurations reach it. 
Two configurations produce degenerate trees (single-class samples), and three reach the budget without producing a tree.
\end{Answer}

\subsection{Cost--Effectiveness (RQ3)}
\label{sec:cost-effective}

We assessed the cost-effectiveness of selecting tokens and their ranges for deriving diagnoses using SBTD as follows.

\emph{Methodology}. We performed a targeted sensitivity study on {five} violated requirements (the three motivating examples $\phi_2$--$\phi_4$ from \Cref{sec:background} and two benchmark requirements, AT53
and RR). For each requirement, we ran \NAME five times under multiple mutation-space configurations:
(i) a \emph{broad} configuration (B) {that uses the widest admissible ranges and, where an operator is mutable, the most alternatives},
(ii) a \emph{guided} configuration (G) {that narrows those ranges and alternatives to a domain-plausible subset},
(iii) \emph{interaction} configurations (G2 and G3) that allow up to 2 or 3 simultaneous token mutations (to probe multi-factor explanations), and
{(iv) \emph{omit} variants (O*) that intentionally exclude one or more plausibly causal terms from the selected mutable sub-formula, to test how sensitive the learned diagnosis is to missing mutation targets.}
A configuration assigns \NAME's mutation-space parameters: the mutable token positions, the admissible range for each position, the operators enabled (numeric perturbation and relational, logical, or quantifier flips), and the maximum number of simultaneous mutations. B, G, G2, and G3 keep the same token positions and differ only in the remaining parameters, whereas each O$^*$ drops one or more positions. The exact per-requirement assignments---positions, ranges, operators, and seeds---are provided as machine-readable configuration files (one per requirement and configuration) in the replication package.
We recorded (a)~\NAME wall-clock time (\texttt{computation time}) and (b)~decision-tree complexity proxies (i.e., size, depth, number of leaves) and root-feature stability across runs. 

\smallskip
\emph{Results}. The study highlights a clear cost--effectiveness trade-off, described in Table~\ref{tab:sensitivity}.
{Each row reports one requirement and one mutation-space configuration: broad (B), guided (G), interaction-enabled (G2/G3, up to 2/3 simultaneous token mutations), or omit (O*), averaged over 5 independent runs.
``Split?'' indicates whether a run produced a non-degenerate DT (i.e., at least one internal split), ``Comp.~time'' is wall-clock time, ``Tree size'' is the number of nodes in the DT, and ``Root split (mode)'' is the most frequent root decision across the 5 runs (token and split `operator'@`value'). For numeric entries reported as $avg \pm sd$, $avg$ is the average over the 5 runs and $sd$ is the standard deviation.}

First, the effect of a \emph{guided} mutation space depends on the requirement. It reduces end-to-end time for $\phi_2$, $\phi_3$, and RR, has little effect for $\phi_4$, and does not reduce time for AT53.

{For example, on $\phi_3$, moving from the \emph{broad} to the \emph{guided} configuration reduces the average computation time (over 5 runs) from $749.4$s to $604.7$s, while keeping a stable root split and yielding smaller trees.}
{For $\phi_2$, the corresponding averages are $487.7$s and $437.5$s. AT53 does not show a reduction: its guided average is $4427.0$s, compared with $4360.2$s for the broad configuration.}
This supports the idea that engineers can often start from a narrow, domain-plausible set of probes (tokens and ranges) and obtain comparable diagnoses at lower cost.

\begin{table}[t]
\centering
\caption{Sensitivity of SBTD to mutation-space configuration and mutation budget (5 runs per config). End-to-end wall-clock time is reported as mean$\pm$sd. ``Split?'' indicates whether the learned DT is non-degenerate (i.e., one or more internal splits).}
\label{tab:sensitivity}
\scriptsize
\setlength{\tabcolsep}{2.2pt}
\renewcommand{\arraystretch}{1.10}
\begin{tabular*}{\columnwidth}{@{\extracolsep{\fill}} l l c r r l @{}}
\toprule
\textbf{Req.} & \textbf{Cfg.} & \textbf{Split?} & \textbf{Comp. time [s]} & \textbf{Tree size} & \textbf{Root split (mode)} \\
\midrule
$\phi_2$ & B & Yes & $487.7\pm 90.8$ & $5.0\pm0.0$ & NUM3 @ $1.998\pm 0.001$ \\
$\phi_2$ & G & Yes & $437.5\pm 210.2$ & $5.0\pm0.0$ & NUM3 @ $1.998\pm 0.001$ \\
$\phi_2$ & G2 & Yes & $486.5\pm 8.7$ & $5.0\pm0.0$ & NUM3 @ $1.999\pm 0.001$ \\
$\phi_2$ & G3 & Yes & $494.0\pm 21.9$ & $5.0\pm0.0$ & NUM3 @ $1.999\pm 0.001$ \\
$\phi_2$ & O & Yes & $447.1\pm 203.2$ & $3.0\pm0.0$ & NUM2 @ $-3.501\pm 0.001$ \\
\midrule
$\phi_3$ & B & Yes & $903.3\pm 387.1$ & $33.4\pm7.1$ & NUM1 @ $5.193\pm 0.072$ \\
$\phi_3$ & G & Yes & $825.1\pm 383.1$ & $15.8\pm2.2$ & NUM1 @ $4.971\pm 0.241$ \\
$\phi_3$ & G2 & Yes & $1202.9\pm 114.6$ & $10.0\pm0.0$ & NUM1 @ $4.699\pm 0.003$ \\
$\phi_3$ & G3 & Yes & $1151.1\pm 62.4$ & $10.0\pm0.0$ & NUM1 @ $4.698\pm 0.006$ \\
$\phi_3$ & O1 & Yes & $850.9\pm 42.3$ & $10.4\pm2.2$ & NUM1 @ $5.196\pm 0.003$ \\
$\phi_3$ & O2 & Yes & $638.6\pm 56.0$ & $3.0\pm0.0$ & NUM3 @ $2.799\pm 0.000$ \\
$\phi_3$ & O3 & No & $160.0\pm 1.5$ & -- & -- \\
\midrule
$\phi_4$ & B & Yes & $491.5\pm 34.0$ & $3.0\pm0.0$ & NUM2 @ $4.958\pm 0.674$ \\
$\phi_4$ & G & Yes & $478.8\pm 49.3$ & $3.0\pm0.0$ & NUM1 @ $16.000\pm 0.000$ \\
$\phi_4$ & G2 & Yes & $510.2\pm 45.5$ & $5.0\pm0.0$ & NUM2 @ $6.780\pm 0.017$ \\
$\phi_4$ & G3 & Yes & $546.5\pm 59.9$ & $12.2\pm1.8$ & NUM1 @ $15.250\pm 0.957$ \\
$\phi_4$ & O & No & $507.1\pm 17.9$ & -- & -- \\
\midrule
AT53 & B & Yes & $4360.2\pm 107.7$ & $3.0\pm0.0$ & QUANTIFIERS1 @ ForAll \\
AT53 & G & Yes & $4427.0\pm 250.6$ & $3.0\pm0.0$ & QUANTIFIERS1 @ ForAll \\
AT53 & G2 & Yes & $5613.3\pm 176.9$ & $3.0\pm0.0$ & QUANTIFIERS1 @ ForAll \\
AT53 & G3 & Yes & $6225.0\pm 151.9$ & $6.0\pm0.0$ & QUANTIFIERS1 @ ForAll \\
AT53 & O1 & No & $4638.1\pm 219.3$ & -- & -- \\
AT53 & O2 & No & $2130.2\pm 15.3$ & -- & -- \\
\midrule
RR & B & Yes & $4644.9\pm 182.1$ & $5.2\pm1.1$ & LOGICALS2 @ And \\
RR & G & Yes & $2875.4\pm 106.5$ & $4.0\pm0.0$ & LOGICALS2 @ And \\
RR & G2 & Yes & $2085.4\pm 31.3$ & $6.0\pm0.0$ & LOGICALS2 @ And \\
RR & G3 & No & $1696.7\pm 4.3$ & -- & -- \\
RR & O & Yes & $7273.5\pm 107.4$ & $4.0\pm0.0$ & LOGICALS2 @ And \\
\bottomrule
\end{tabular*}
\end{table}

Second, \emph{over-specification} (broader $sub(\property)$, wider ranges, or enabling multi-token interactions) tends to increase search cost and may either increase diagnosis complexity or surface competing roots.
For $\phi_3$, moving from a guided space (G) to interaction-enabled spaces (G2 and G3) increases computation time (from $825.1$s to $1202.9$s and $1151.1$s) while producing \emph{smaller} but different trees (from $15.8$ nodes in G to $10.0$ in G2 and G3) with a shifted dominant split threshold (NUM1 at $\approx 4.97$ in G vs.\ NUM1 at $\approx 4.70$ in G2 and G3), indicating that additional mutability can yield alternative explanations at higher cost.

Third, \emph{under-specification} can either (i) force the diagnosis to pivot to a proxy explanation or (ii) prevent learning an informative diagnosis altogether.
For $\phi_2$, the omit variant (O) still produces an informative tree, but the dominant split changes (NUM3 $\rightarrow$ NUM2), i.e., the diagnosis shifts qualitatively.
In contrast, for $\phi_3$ omitting key probes (O3) yields no successful diagnosis, and for $\phi_4$ and AT53 the omit variants similarly collapse, showing that some tokens are necessary for SBTD to generate a discriminative {\tt sat} or {\tt unsat}
dataset and learn a meaningful DT within budget.
Finally, RR exhibits the same trade-off: guided configurations (G and G2) substantially reduce time (from $4644.9$s in B to $2875.4$s in guided and $2085.4$s in G2) while producing informative trees, whereas an under-specified variant (G3) yields a degenerate tree and a broader variant (O) produces an informative tree at markedly higher cost ($7273.5$s).

We use \textbf{Split?} to indicate that the learned DT contains at least one informative split, otherwise the run yields a degenerate tree and we omit the size and root.
The \textbf{Tree size} reflects diagnosis complexity. Smaller trees are typically easier to inspect, while larger trees indicate a more fragmented decision boundary under the chosen mutation space. Conversely, consistently small trees across runs can also indicate that the explanation is robust to small variations in the sampled mutated requirements (e.g., for $\phi_2$, the stable size of 5 suggests a simple, consistently learned boundary).

The sensitivity results should be interpreted in light of the operator coverage, discussed in \Cref{sec:discussionsub}, since most configurations mutate numeric thresholds and only a minority includes discrete operator choices.

\begin{Answer}[RQ3 - Cost--Effectiveness]
We found that \NAME’s mutation-space configuration directly affects runtime and diagnosis stability (Table~\ref{tab:sensitivity}). Guided configurations reduce end-to-end time for three of the five requirements, but not for AT53. For $\phi_2$, B, G, G2, and G3 all produce a non-degenerate 5-node tree in $\sim$438--494s with a stable NUM3 root. Broader or interaction-enabled configurations can change tree complexity and the dominant explanation (e.g., for $\phi_4$, G3 grows to $12.2$ nodes and shifts the root). Omitting key mutable tokens can yield only degenerate trees within the fixed campaign (e.g., $\phi_3$ O3, $\phi_4$ O, AT53 O1/O2, and RR G3).
\end{Answer}

    \section{Discussion and Threats to Validity}
    \label{sec:discussion}
    
    {This section discusses the main practical implications and limitations of our SBTD framework, and summarizes the key threats to validity affecting our evaluation.}
    
    \subsection{Discussion}
    \label{sec:discussionsub}
    {We discuss (i) limitations of our DT-based diagnosis representation, (ii) how SBTD complements expert analysis, (iii) sensitivity to mutation-space choices and baseline applicability, and (iv) black-box versus grey-box trade-offs.}
    
    \smallskip
    
    \noindent{\em DT-based diagnosis limits.}
    {Our SBTD pipeline relies on SMT-based trace checking to label mutated requirements as \texttt{sat}/\texttt{violated}/\texttt{undecided}. As a result, it inherits practical limitations of current SMT technology in this domain: for some requirements and mutation spaces, solver calls may be expensive and can return \texttt{unknown} within the timeout, which reduces the number of usable labeled samples and can prevent reaching the minimum-sample threshold within the allotted budget (cf.\ RQ2). Independently of SMT cost, our approach represents the final diagnosis using \emph{decision trees} (DTs), which constrains expressiveness: a DT is a conjunction of axis-aligned tests where each node refers to a single mutable term. Therefore, DTs cannot directly capture more complex multi-term relations between inputs, such as the quadratic dependency between the upper temporal limit AT1(\markB{20}) and the upper speed limit AT1(\markC{120}) observed in $exp2$. We plan to address these limitations by (i) improving trace-checking efficiency/robustness and (ii) exploring richer, still-interpretable diagnosis models (e.g., genetic programming) in future work.}

    \smallskip
    \noindent{\em Complementing experts and parallelization.} {SBTD is intended to support and complement expert designers by automatically synthesizing an initial, interpretable diagnosis and helping validate trace-grounded witnesses. Moreover, runs can be parallelized across trace--requirement pairs and configurations, which helps scale analysis when multiple violations must be investigated. In practice, SBTD complements by-hand analysis by providing an initial, interpretable prioritization of candidate requirement fragments and time windows for inspection, which engineers can validate and iteratively refine by adjusting the mutation space or budget.}

    \smallskip
    \noindent{\em From diagnosis to code.} The syntactic element changed by the diagnosis indicates the class of defect to investigate: shifted temporal bounds implicate the handling of events and state over time; shifted signal thresholds implicate the constants the implementation enforces; changed connectives implicate the structure of the guards.
    In our running example, the time boundaries of \Cref{fig:runningexample_dt} localize a stale-state defect (an evasion flag that is never reset). We implemented the reset fix and executed the four-scenario verification in the replication package; all four running-example requirements remain satisfied in every scenario.

\smallskip
    \noindent{\em Sensitivity to mutation-space choices.} The choice of mutation positions and their admissible ranges influences both the computational cost of the search and the structure of the learned decision tree (i.e., which features appear as dominant splits and how complex the resulting diagnosis is).
    To quantify this effect, we performed a focused sensitivity study across multiple falsified requirements by varying mutation ranges (broad vs.\ guided) and mutation subsets (omit variants), and by increasing the allowed number of simultaneous mutations. The results indicate that (i) guided ranges can substantially improve interpretability and reduce cost, (ii) removing key mutation terms can change the dominant diagnostic split, and (iii) overly restrictive subsets may yield degenerate datasets where no diagnosis can be learned.

\smallskip
    \noindent\emph{Operator coverage.} Since our experiments do not exercise every operator, we ran a dedicated coverage check: we applied each operator OP1--OP15 as an abstract-syntax-tree transformation to minimal AT-derived requirements and checked every mutant with the same trace-checker. All fifteen produce well-formed mutants. This is a completeness check of the catalog rather than an effectiveness comparison, and it does not imply that the search samples every operator autonomously. The uncovered operators therefore reflect the structure of the subject requirements, not a limitation of the catalog.

\smallskip
    \noindent\emph{Engineer input vs.\ diagnosis libraries.}
    {While \NAME\ requires a mutation space specification provided by the engineers, this input is \emph{requirement-local} (selecting tokens already present in the requirement plus admissible domains) and is \emph{not} a hand-crafted library of violation causes/diagnoses; unlike library-based approaches, we do not assume a pre-defined catalog of explanations and matching rules, but use the mutation space as diagnostic probes to synthesize candidate explanations and learn an interpretable DT.}
    The mutation space is thus a lighter, requirement-local form of expert input, not an equivalent replacement of one library by another.
    The engineer therefore specifies where the diagnosis may look, not which explanations it may return. More importantly, \NAME\ synthesizes candidates within that space and can report combinations of tokens and thresholds that were not anticipated when the space was defined.

\smallskip
    \noindent{\em Baseline applicability.} We additionally compared \NAME with TD-SB-TemPsy on the subset where the baseline is applicable. TD-SB-TemPsy returned diagnoses for 5/17 requirements, where it provides useful localization cues, while \NAME applies to the full HLS benchmark and, on the overlapping subset, returns expert-aligned rule diagnoses (Tab.~\ref{tab:baseline_comp_table}). Our baseline comparison is partial because TD-SB-TemPsy does not apply to 12 requirements from our benchmark (unsupported operators), so we report baseline results only where it produces an output and interpret them as 
    {an applicability comparison, i.e., which requirements each tool can handle,} rather than a full head-to-head evaluation.
    On the requirements both tools handle, their threshold explanations agree with the expert (\Cref{tab:baseline_comp_table}). The distinction is the \emph{kind} of diagnosis---change-driven versus pointwise---not the quality of the explanation. \NAME's broader coverage instead follows from expressiveness, HLS subsumes the SB-TemPsy fragment the baseline supports, so the additional requirements are ones the baseline cannot express.

\smallskip
    \noindent{\em Black-box vs.\ grey-box trade-off.} The black-box design of SBTD provides generality and modularity, but its dominant cost {comes from repeatedly trace-checking} mutated properties (Component~\phase{2}).
    A more grey-box variant could exploit checker- or trace-specific feedback (e.g., violation localization or counterexamples, unsat cores, incremental solving, or caching) to steer mutation generation and reduce evaluations.
     We leave these optimizations to future work because they are backend-dependent and would reduce the checker-agnostic scope of the current framework.

\smallskip
    \noindent{\em Checker-agnostic generality.} To turn the modularity claim from design intent into evidence, we ported \NAME's trace checker from \texttt{ThEodorE}/Z3 to RTAMT~\cite{nivckovic2020rtamt}, an STL robustness monitor, leaving the search, mutation, and DT learning unchanged, and re-ran two requirements (AT1, CCX). RTAMT reproduces Z3's \texttt{sat}/\texttt{violated} verdicts at $\sim$153$\times$ (AT1) and $\sim$1250$\times$ (CCX) lower per-check cost while preserving DT accuracy (\Cref{tab:rtamt_generality}), confirming that the checker is a replaceable oracle.
    In practice, the whole diagnosis for AT1 completes in $124$s instead of $8.8$h, and for CCX in $353$s instead of $4.4$ days, while the accuracy against the expert prediction is unchanged ($97.2\%$ against $98.0\%$ for AT1, $99.9\%$ against $99.5\%$ for CCX). We ran both engines under the same search configuration and on the same machine, so these results isolate the effect of the checker and are not comparable with the campaign runtimes of \Cref{tab:time}.
    To check that the swap preserves the diagnosis and not merely the runtime, we re-checked every candidate with both engines and compared the verdicts pairwise, over \num{4350} distinct AT1 candidates the two engines never disagreed without an explanation.
    This generality is bounded by expressiveness: RTAMT supports STL, a fragment of the \ourlogic\ that \NAME\ supports, so only STL-expressible requirements can be delegated to it; i.e., faster checkers improve performance but come with a cost of lower language expressiveness.

    \begin{table}[t]
    \caption{Swapping the checker (Z3\,$\to$\,RTAMT) on AT1 and CCx: the search and diagnosis logic are unchanged; only the per-check cost differs, while DT accuracy against the analytical ground truth is preserved.}
    \label{tab:rtamt_generality}
    \centering
    \setlength{\tabcolsep}{4pt}
    \renewcommand{\arraystretch}{1.1}
    \begin{tabular}{@{}llrrrrr@{}}
    \toprule
    \textbf{Req.} & \textbf{Engine} & \textbf{Calls} & \textbf{TC time} & \textbf{s/call} & \textbf{Diag.\ time} & \textbf{DT acc.} \\
    \midrule
    AT1 & Z3 & 2300 & 8.6\,h & 13.5 & 8.8\,h & 97.97\% \\
    AT1 & RTAMT & 1024 & 90\,s & 0.0877 & 124\,s & 97.24\% \\
    \midrule
    CCx & Z3 & 200 & 4.4\,d & 1916.7 & 4.4\,d & 99.54\% \\
    CCx & RTAMT & 229 & 349\,s & 1.52 & 352\,s & 99.88\% \\
    \bottomrule
    \end{tabular}
    \end{table}

    \subsection{Threats to Validity}
    {We discuss threats related to (i) the expert reference and DT-comparison metric (RQ1), (ii) tool configuration and sampling, (iii) trace-checker backend dependence, and (iv) generalization beyond our benchmark.}
    
    \emph{Internal Validity}. We compared \NAME’s diagnoses against a reference diagnosis produced by an expert familiar with the benchmark. The reference is grounded in trace inspection and validated by trace-checking representative mutations within the configured mutation space. 
    {Because the expert is a co-author, this reference may be affected by confirmation bias, we mitigate this in three ways.}
    {First, we enforced role separation, the expert derived the reference diagnosis independently, without access to \NAME's outputs.
    Second, the reference is not an absolute ground truth, obtaining one would require trace-checking all mutated requirements, which is infeasible because mutation domains include real-valued tokens and even large discretizations are prohibitive (e.g., the three-axis grid used for \emph{exp2} contains \num{1030301} candidates).
    Third, other engineers could derive different DTs; while invariant-style cases reduce to a largely non-subjective threshold (from trace extrema/crossings), for the remaining cases the expert may still be wrong. We therefore ground the reference in observable trace witnesses (e.g., extrema/violation windows) and re-trace-check representative mutations, and interpret precision/recall as agreement with this trace-grounded reference rather than a soundness guarantee.}
    A replication with external experts, independent of the authors, would further mitigate this threat, and we leave it to future work.
    
    {For RQ1, the metric used to compare the DTs produced by the tool and the expert could threaten the internal validity of our results.
    While a direct error on the identified threshold would suffice when a single value is mutated, it would not apply to experiments with multiple values; our approach enables us to consider these two cases seamlessly.}

    For RQ1 and RQ2, the values selected for the configuration parameters of our tool (\Cref{tab:configurationParameters}) threaten the internal validity of our results.
    While our experiments provide the results for a specific configuration (defined by selecting configuration values from the literature), in practice, engineers should configure the SBTD tool depending on their domain-specific needs and the desired precision and recall. 
    {Moreover, we report precision and recall to quantify how reliably the learned diagnosis matches the trace-grounded reference under the chosen configuration.}
    
    The dominant cost stems from repeatedly evaluating mutated properties, and is therefore largely inherited from the underlying trace-checker configuration.
    SBTD is modular with respect to this backend: replacing \texttt{ThEodorE}/Z3 with a faster trace-checking engine (for the same property fragment) would reduce end-to-end runtime without changing the diagnosis logic. {In our setting, \texttt{ThEodorE} can be a bottleneck on some instances~\cite{menghi2021trace}: {trace-checking dominates end-to-end runtime, and two chasing-cars configurations reach the time budget (\Cref{sec:efficiency}).}}
    We integrated a second checker (RTAMT) to demonstrate this modularity, and we plan to support further trace-checkers. Other trace-checkers (e.g., dp-Taliro~\cite{dpTaliro}) are more efficient, but support less expressive languages.

    Moreover, this limitation is not unique to \NAME, any approach that ranks candidate explanations via repeated property evaluation is expected to exhibit a similar bottleneck. 
    In addition, the selection of HLS could threaten the internal validity of our results. Considering other languages (e.g., SB-TemPsy-DSL~\cite{boufaied2020trace}, Restricted Signals First Order Logic~\cite{menghi2019generating}) for specifying the requirements could lead to different results.
    
    {The \emph{Generator of Mutations} is stochastic. We therefore ran every RQ1/RQ2 configuration with ten seeds and report seed-level outcomes and robust aggregates.}
    The coverage check in \Cref{sec:discussionsub} shows that every operator is applicable, but our effectiveness and efficiency results rest on experiments in which numeric-threshold mutations dominate. Precision, recall, and runtime may therefore differ on requirements whose diagnosis turns on the operators our benchmark does not exercise. The mutation-space configurations used in our experiments were authored by the research team rather than by independent practitioners, so our results do not establish whether typical CPS engineers would select comparable tokens and ranges; a user study with external engineers would be required to assess this.

    \emph{External Validity}. The set of trace-requirement combinations we considered in our experiments could threaten the external validity of our results, as considering other trace-requirement combinations may lead to different results. However, the requirements of our benchmark refer to different case studies and use different logical operators.
    
    {Overfitting and hyperparameter tuning could threaten external validity, i.e., a configuration that works well on our benchmark might behave differently on other systems. We share the GA, stopping, caching, and time-quantization profile across all RQ1/RQ2 experiments. The requirement-local mutation spaces necessarily differ and are evaluated separately in RQ3. Moreover, the fitness function and the informative-diagnosis cycle guide the search toward parameter changes that consistently explain observed violations across traces, yielding diagnoses that are interpretable and trace-grounded. As with all search-based techniques under practical time bounds (and with SMT-based trace checking in the loop), the method provides high-quality solutions within a budget rather than formal optimality guarantees; in our setting, we quantify this behavior empirically via repeated runs and by reporting effectiveness metrics.}
 
\section{Related Work}
\label{sec:related}
Property violations are typically explained by exploiting some notion of causality (e.g.,~\cite{beer2012,Diehl:RAL22,dou2018})  to extrapolate the causes of the failure (e.g., an event $A$ is said to be a cause of event $B$ if, had $A$ not happened then $B$ would not have happened). 
These causes typically refer to portions of the trace or portions of the property responsible for the violation.

Approaches that extrapolate information coming from the \emph{trace} typically isolate slices of the traces that contain the causes for the property violation (e.g., ~\cite{stratan2024diagnosing,ferrere2015trace,mukherjee2012computing,beer2012,nivckovic2020amt,dou2018}).
Other approaches explain the property violation by checking for traces showing common behaviors that lead to the satisfaction and violation of the property (e.g.,~\cite{luo2014rv,dawes2019}).
Delta-debugging approaches~\cite{zeller2002} instead minimize the failing input itself, iteratively reducing it to a minimal failure-inducing fragment.
Unlike these approaches, \NAME uses small, structured mutations of the property as \emph{diagnostic probes}: by analyzing which minimal mutations {change}
the verdict on the same trace(s), it localizes the falsification to specific sub-formula fragments and signal patterns, yielding an interpretable diagnosis (rather than a recommendation to refine the requirement).

Approaches that extrapolate information coming from the \emph{property} (e.g.,~\cite{boufaied2023,chechik2007framework}) typically exploit its structure to provide viable diagnoses. 
For example, pattern-based diagnostic approaches (e.g.,~\cite{boufaied2023}), enrich trace-checking verdicts (i.e.,~\cite{boufaied2020trace}) by exploiting the syntactical structure of the property (i.e., the patterns used to define the property of interest) to compute viable diagnoses. 
These approaches require engineers to define a predefined set of possible violation causes and corresponding diagnoses upfront or assume a library of violation causes and corresponding diagnoses to be available.
Unlike these approaches, \NAME relies on a novel evolutionary approach that can dynamically generate new diagnoses by applying the mutation and cross-over operators. The mutations in \NAME are not proposed as repaired specifications; they are used to learn an explanatory model (a decision tree) that identifies the likely cause of the observed violation.
{Pointwise diagnoses, such as those produced by TD-SB-TemPsy~\cite{boufaied2023}, report a violating time step and the signal values at that instant, and are effective at directing the engineer to a location in the trace. Change-driven diagnoses instead characterize the behavior the implementation exhibits over the whole trace---the extent of the violation, its margin with respect to the requirement, and the sub-conditions on which it depends. The two are complementary, the former answers \emph{where to look first}, the latter \emph{what the implementation does and by how much it deviates}.}

{A related line of work learns \emph{invariants} or \emph{monitoring oracles} from CPS traces and uses them for monitoring or anomaly detection (e.g.,~\cite{DBLP:journals/tse/AfzalGT22,JiangED17,DBLP:conf/icra/ToledoWED24}), or to support specification/contract repair in verification settings (e.g.,~\cite{AbreuMM23}). 
These approaches are related in the broad sense that they infer interpretable conditions from executions; however, their goal is typically to infer properties that \emph{hold} for (normal) behaviors or to detect anomalies, whereas \NAME starts from a \emph{given violated} HLS requirement and produces a \emph{change-driven diagnosis} that localizes which requirement fragments drive the observed falsification on the given trace(s).
Thus, invariant/oracle approaches are best viewed as complementary (e.g., for monitoring or anomaly detection), while \NAME focuses on diagnosing a known requirement violation.}

{Approaches that can explain property violations are also common within the context of model-checking, where the goal is to explain violations with respect to a model rather than post-hoc diagnosis on traces.}
Most of the existing approaches (e.g.,~\cite{peled2001model,bernasconi2017model,peled2001falsification,mebsout2016proof,basin2018optimal,pnueli2002deductive,balaban2010proving}) are based on deductive reasoning techniques that start from some initial assertions and examine how logical operators support the conclusion that the property is violated. 
Other approaches extract information from the model (e.g., model slices) to explain the model-checking verdict (e.g.,~\cite{menghi2020integrating,schuppan2012towards,hantry:hal-01354475,zheng2021flack,chechik2007framework,bochot2010paths,griggio2018certifying,funke2020farkas,timm2020model,gurfinkel2003proof}).
Explainability was also studied in the context of anomaly detection (see~\cite{li2023survey} for a recent survey). 
Recent work also considered how to explain spurious failures detected by test case generation frameworks~\cite{jodat2024test} and
via feature engineering~\cite{de2024explainability}.
Unlike these approaches, \NAME  produces informative diagnosis in the context of the trace-checking problem domain, a significantly different problem. Additionally, \NAME relies on an evolutionary approach.

Overall, \NAME is complementary to trace-checking: it targets post-hoc explanation of violations by localizing causes in terms of signals and property fragments.
{The individual ingredients \NAME builds on---requirement mutation, black-box trace-checking, and decision-tree learning---are established; its contribution is their integration into a trace-diagnostic framework that generates candidate diagnoses dynamically, without a predefined library, and remains agnostic to the underlying checker, rather than a new trace-checking or diagnostic formalism.}

\section{Conclusion}
\label{sec:conclusion}

In this paper we proposed a search-based trace diagnostic (SBTD) technique to support engineers understanding the cause of violation of CPS requirements. The technique starts from a set
of candidate diagnoses and iteratively applies an evolutionary algorithm that exploits mutation, recombination, and selection to generate new candidate diagnoses. A fitness function determines the qualities of the identified solutions. The technique is implemented in a tool named \NAME, which takes as input signal-based temporal logic requirements expressed using the Hybrid
Logic of Signals (HLS). \NAME is evaluated with respect to {\em effectiveness}, {\em efficiency}, and {\em cost-effectiveness} in producing an informative diagnosis. The results of the evaluation, which considers 17 trace-requirement combinations leading to a property violation, confirm that \NAME can produce informative diagnoses within the time budgets reported in \Cref{sec:evalaution}. 
{Moreover, on the subset where TD-SB-TemPsy is applicable, \NAME provides comparable expert-aligned diagnoses (while TD-SB-TemPsy returns diagnoses for 5/17 requirements).}

In future work, we plan to experiment on different case studies, e.g., in the space domain, to check the generality of \NAME, as well as to better assess its effectiveness and efficiency. We also plan to experiment with languages different from HLS, like  Restricted Signals First-Order Logic, to check to what extent the results in this paper will be confirmed. We also plan to investigate alternative representations of change-driven diagnoses (e.g., ranked rule sets or interactive summaries) that retain the DT’s structure while reducing the cognitive overhead of inspecting deep trees.

\section*{Acknowledgments}
 This work was supported in part by project SERICS (PE00000014) under
the NRRP MUR program funded by the EU - NGEU, by the European Union - Next Generation EU. ``Sustainable Mobility Center (Centro Nazionale per la Mobilità Sostenibile - CNMS)'', M4C2 - Investment 1.4, Project Code CN\_00000023.
In addition, the work is partially supported by (a) the PNRR MUR project VITALITY (ECS00000041), Spoke 2 ASTRA - ``Advanced Space Technologies and Research Alliance'', (b) the European HORIZON-KDT-JU research project MATISSE ``Model-based engineering of Digital Twins for early verification and validation of Industrial Systems'', HORIZON-KDT-JU-2023-2-RIA , Proposal number:  101140216-2, KDT232RIA\_00017, %
and (c) the MUR (Italy) Department of Excellence 2023 - 2027 for GSSI. %
The work of Genaína Rodrigues was financed in part by FAPDF under Call 04/2021 and CNPq process 313215/2021-9. 
This work was enabled in part by support provided by Compute Ontario (\href{https://www.computeontario.ca/}{computeontario.ca}) and the Digital Research Alliance of Canada (\href{https://alliancecan.ca/}{alliancecan.ca}).

\section*{Data Availability}
A complete replication package containing the traces, the {tool- and expert-generated decision trees for all experiments}, and the tool is publicly available \cite{Diagnosistool}.\\
The replication package will be made available on Zenodo upon acceptance.

\bibliography{bibliography}

@InProceedings{STL,
author="Maler, Oded
and Nickovic, Dejan",
title="Monitoring Temporal Properties of Continuous Signals",
booktitle="Formal Techniques, Modelling and Analysis of Timed and Fault-Tolerant Systems",
year="2004",
publisher="Springer Berlin Heidelberg",
address="Berlin, Heidelberg",
pages="152--166",
}

@ARTICLE{Diehl:RAL22,  
author={Diehl, Maximilian and Ramirez-Amaro, Karinne},  
journal={IEEE Robotics and Automation Letters},   
title={{Why Did I Fail? a Causal-Based Method to Find Explanations for Robot Failures}},   
year={2022},  
pages={1-8},  
doi={10.1109/LRA.2022.3188889}}

@article{zeller2002,
  title={Simplifying and isolating failure-inducing input},
  author={Zeller, Andreas and Hildebrandt, Ralf},
  journal={IEEE Transactions on software engineering},
  volume={28},
  number={2},
  pages={183--200},
  year={2002},
  publisher={IEEE}
}

@inproceedings{fse2024,
  title={{ATheNA-S: a Testing Tool for Simulink Models Driven by Software Requirements and Domain Expertise}},
  author={Formica, Federico and Mahboob, Mohammad Mahdi and  Askarpour, Mehrnoosh and Menghi, Claudio},
  booktitle={Companion Proceedings of the 32nd ACM International Conference on the Foundations of Software Engineering (FSE Companion ’24),},
  year={2024},
  publisher={ACM},
  address={New York, NY, USA},
}

@article{DBLP:journals/tse/AfzalGT22,
  author       = {Afsoon Afzal and
                  Claire {Le Goues} and
                  Christopher Steven Timperley},
  title        = {Mithra: Anomaly Detection as an Oracle for Cyberphysical Systems},
  journal      = {{IEEE} Trans. Software Eng.},
  volume       = {48},
  number       = {11},
  pages        = {4535--4552},
  year         = {2022},
  url          = {https://doi.org/10.1109/TSE.2021.3120680},
  doi          = {10.1109/TSE.2021.3120680}
}

@inproceedings{AbreuMM23,
  author    = {Alexandre Abreu and Nuno Macedo and Alexandra Mendes},
  title     = {Exploring Automatic Specification Repair in Dafny Programs},
  booktitle = {2023 38th IEEE/ACM International Conference on Automated Software Engineering Workshops (ASEW)},
  pages     = {105--112},
  year      = {2023},
  publisher = {IEEE},
  doi       = {10.1109/ASEW60602.2023.00019}
}

@article{JiangED17,
  author  = {Hengle Jiang and Sebastian Elbaum and Carrick Detweiler},
  title   = {Inferring and monitoring invariants in robotic systems},
  journal = {Autonomous Robots},
  volume  = {41},
  pages   = {1027--1046},
  year    = {2017},
  doi     = {10.1007/s10514-016-9576-y}
}

@inproceedings{DBLP:conf/icra/ToledoWED24,
  author       = {Felipe Toledo and
                  Trey Woodlief and
                  Sebastian G. Elbaum and
                  Matthew B. Dwyer},
  title        = {Specifying and Monitoring Safe Driving Properties with Scene Graphs},
  booktitle    = {{IEEE} International Conference on Robotics and Automation, {ICRA}
                  2024, Yokohama, Japan, May 13-17, 2024},
  pages        = {15577--15584},
  publisher    = {{IEEE}},
  year         = {2024},
  url          = {https://doi.org/10.1109/ICRA57147.2024.10610973},
  doi          = {10.1109/ICRA57147.2024.10610973}
}

@inproceedings{boufaied2020trace,
  title={Trace-checking signal-based temporal properties: A model-driven approach},
  author={Boufaied, Chaima and Menghi, Claudio and Bianculli, Domenico and Briand, Lionel and Parache, Yago Isasi},
  booktitle={International Conference on Automated Software Engineering},
  publisher={IEEE/ACM},
  pages={1004--1015},
  year={2020}
}

@article{balaban2010proving,
  title={Proving the refuted: Symbolic model checkers as proof generators},
  author={Balaban, Ittai and Pnueli, Amir and Zuck, Lenore D},
  journal={Concurrency, Compositionality, and Correctness: Essays in Honor of Willem-Paul de Roever},
  pages={221--236},
  year={2010},
  publisher={Springer}
}

@inproceedings{pnueli2002deductive,
  title={A deductive proof system for CTL},
  author={Pnueli, Amir and Kesten, Yonit},
  booktitle={International Conference on Concurrency Theory},
  pages={24--40},
  year={2002},
  organization={Springer}
}

@inproceedings{basin2018optimal,
  title={Optimal proofs for linear temporal logic on lasso words},
  author={Basin, David and Bhatt, Bhargav Nagaraja and Traytel, Dmitriy},
  booktitle="Automated Technology for Verification and Analysis",
  pages={37--55},
  year={2018},
  publisher={Springer},
  address="Cham",
}

@inproceedings{mebsout2016proof,
  title={{Proof certificates for SMT-based model checkers for infinite-state systems}},
  author={Mebsout, Alain and Tinelli, Cesare},
  booktitle={2016 Formal Methods in Computer-Aided Design (FMCAD)},
  pages={117--124},
  year={2016},
  organization={IEEE}
}

@inproceedings{peled2001falsification,
  title={From falsification to verification},
  author={Peled, Doron and Pnueli, Amir and Zuck, Lenore},
  booktitle="FST TCS 2001: Foundations of Software Technology and Theoretical Computer Science",
  pages={292--304},
  year={2001},
  publisher="Springer",
  address="Berlin, Heidelberg",
}

@inproceedings{bernasconi2017model,
  title={From model checking to a temporal proof for partial models},
  author={Bernasconi, Anna and Menghi, Claudio and Spoletini, Paola and Zuck, Lenore D and Ghezzi, Carlo},
  booktitle={Software Engineering and Formal Methods},
  pages={54--69},
  year={2017},
  organization={Springer},
  address="Cham",
  series={SEFM 2017}
}

@inproceedings{peled2001model,
  title={From model checking to a temporal proof},
  author={Peled, Doron and Zuck, Lenore},
  booktitle="Model Checking Software",
  year="2001",
  publisher="Springer",
  address="Berlin, Heidelberg",
  pages={1--14},
}

@inproceedings{luo2014rv,
  title={RV-Monitor: Efficient parametric runtime verification with simultaneous properties},
  author={Luo, Qingzhou and Zhang, Yi and Lee, Choonghwan and Jin, Dongyun and Meredith, Patrick O’Neil and {\c{S}}erb{\u{a}}nu{\c{t}}{\u{a}}, Traian Florin and Ro{\c{s}}u, Grigore},
  booktitle={International Conference on Runtime Verification (RV)},
  pages={285--300},
  year={2014},
  organization={Springer}
}

@article{chechik2007framework,
  title={A framework for counterexample generation and exploration},
  author={Chechik, Marsha and Gurfinkel, Arie},
  journal={International Journal on Software Tools for Technology Transfer},
  volume={9},
  pages={429--445},
  year={2007},
  publisher={Springer}
}

@article{witten2002data,
  title={Data mining: practical machine learning tools and techniques with Java implementations},
  author={Witten, Ian H and Frank, Eibe},
  journal={ACM SIGMOD Record},
  volume={31},
  number={1},
  pages={76--77},
  year={2002},
  publisher={ACM New York, NY, USA}
}

@book{o2009riccardo,
    author = "Riccardo Poli and William B. Langdon and Nicholas Freitag McPhee",
    title = "A field guide to genetic programming",
    publisher = "Published via \url{http://lulu.com} and freely available at \url{http://www.gp-field-guide.org.uk}",
    year = "2008",
    note = "(With contributions by J. R. Koza)",
}

@article{nunez2007fitting,
  title={Fitting the control parameters of a genetic algorithm: An application to technical trading systems design},
  author={Nunez-Letamendia, Laura},
  journal={European journal of operational research},
  volume={179},
  number={3},
  pages={847--868},
  year={2007},
  publisher={Elsevier}
}

@article{hall2009weka,
  title={The WEKA data mining software: an update},
  author={Hall, Mark and Frank, Eibe and Holmes, Geoffrey and Pfahringer, Bernhard and Reutemann, Peter and Witten, Ian H},
  journal={ACM SIGKDD explorations newsletter},
  volume={11},
  number={1},
  pages={10--18},
  year={2009},
  publisher={ACM New York, NY, USA}
}

@inproceedings{witten2005practical,
  title={Practical machine learning tools and techniques},
  author={Witten, Ian H and Frank, Eibe and Hall, Mark A and Pal, Christopher J and Data, Mining},
  booktitle={Data mining},
  volume={2},
  number={4},
  pages={403--413},
  year={2005},
  publisher={Elsevier},
    address = {Amsterdam, The Netherlands}
}

@online{Diagnosistool,
  title        = {{Diagnosis}},
  author       = {Araujo, Gabriel and Caldas, Ricardo and Formica, Federico},
  year         = {2026},
  month        = mar,
  url          = {https://github.com/rdinizcal/Diagnosis/releases/tag/v1.0.0},
  urldate      = {2026-03-06},
  note         = {GitHub release v1.0.0}
}

@book{quinlan2014c4,
  title={C4. 5: programs for machine learning},
  author={Quinlan, J Ross},
  year={2014},
  publisher={Elsevier}
}

@inproceedings{menghi2023arch,
  title={ARCH-COMP 2023 Category Report: Falsification},
  author={Menghi, Claudio and Arcaini, Paolo and Baptista, Walstan and Ernst, Gidon and Fainekos, Georgios and Formica, Federico and Gon, Sauvik and Khandait, Tanmay and Kundu, Atanu and Pedrielli, Giulia and others},
  booktitle={International Workshop on Applied Verification of Continuous and Hybrid Systems (ARCH23)},
  volume={96},
  pages={151--169},
  year={2023}
}

@inproceedings{menghi2021trace,
  title={{Trace-checking CPS properties: Bridging the cyber-physical gap}},
  author={Menghi, Claudio and Vigan{\`o}, Enrico and Bianculli, Domenico and Briand, Lionel C},
  booktitle={International Conference on Software Engineering (ICSE)},
  pages={847--859},
  year={2021},
  organization={IEEE/ACM}
}

@online{NNFal,
    title={{NNFal}},
    howpublished={\url{https://gitlab.com/Atanukundu/NNFal}},
    month         = "April",
    year          = "2023 [Online]",
    lastaccessed = {April 2023},
    key = {NNFal}
}

@article{gaaloul2021combining,
  title={Combining genetic programming and model checking to generate environment assumptions},
  author={Gaaloul, Khouloud and Menghi, Claudio and Nejati, Shiva and Briand, Lionel C and Parache, Yago Isasi},
  journal={IEEE Transactions on Software Engineering},
  volume={48},
  number={9},
  pages={3664--3685},
  year={2021},
  publisher={IEEE}
}

@inproceedings{menghi2021theodore,
  title={Theodore: A trace checker for cps properties},
  author={Menghi, Claudio and Vigan{\`o}, Enrico and Bianculli, Domenico and Briand, Lionel C},
  booktitle={2021 IEEE/ACM 43rd International Conference on Software Engineering: Companion Proceedings (ICSE-Companion)},
  pages={183--184},
  year={2021},
  organization={IEEE}
}

@inproceedings{nivckovic2020rtamt,
  title={RTAMT: Online robustness monitors from STL},
  author={Ni{\v{c}}kovi{\'c}, Dejan and Yamaguchi, Tomoya},
  booktitle={International Symposium on Automated Technology for Verification and Analysis},
  pages={564--571},
  year={2020},
  organization={Springer}
}

@InProceedings
{
    psytalirotool,
    author="Thibeault, Quinn
    and Anderson, Jacob
    and Chandratre, Aniruddh
    and Pedrielli, Giulia
    and Fainekos, Georgios",
    title={{PSY-TaLiRo: A Python Toolbox for Search-Based Test Generation for Cyber-Physical Systems}},
    booktitle="Formal Methods for Industrial Critical Systems",
    year="2021",
    publisher="Springer",
    pages="223--231",
    isbn="978-3-030-85248-1"
}

@inproceedings{annpureddy2011s,
  title={{S-TaLiRo}: A tool for temporal logic falsification for hybrid systems},
  author={Annpureddy, Yashwanth and Liu, Che and Fainekos, Georgios and Sankaranarayanan, Sriram},
  booktitle={International Conference on Tools and Algorithms for the Construction and Analysis of Systems},
  pages={254--257},
  year={2011},
  organization={Springer}
}

@article{peltomaki2023requirement,
  title={Requirement falsification for cyber-physical systems using generative models},
  author={Peltom{\"a}ki, Jarkko and Porres, Ivan},
  journal={arXiv preprint arXiv:2310.20493},
  year={2023}
}

@inproceedings{falsQBRobCAV2021,
author = {Zhang, Zhenya and Lyu, Deyun and Arcaini, Paolo and Ma, Lei and Hasuo, Ichiro and Zhao, Jianjun},
title = {{Effective Hybrid System Falsification Using Monte Carlo Tree Search Guided by {QB}-Robustness}},
booktitle={Computer Aided Verification},
year = {2021},
publisher = {Springer},
pages = {595--618},
doi = {10.1007/978-3-030-81685-8_29}
}

@article{jodat2024test,
  title={Test Generation Strategies for Building Failure Models and Explaining Spurious Failures},
  author={Jodat, Baharin A and Chandar, Abhishek and Nejati, Shiva and Sabetzadeh, Mehrdad},
  journal={ACM Transactions on Software Engineering and Methodology},
  volume={33},
  number={4},
  pages={1--32},
  year={2024},
  publisher={ACM New York, NY}
}

@InProceedings{Waga20,
  author       = {Masaki Waga},
  title        = {Falsification of cyber-physical systems with
                  robustness-guided black-box checking},
  year         = 2020,
  booktitle    = {International Conference on
                  Hybrid Systems: Computation and Control ({HSCC})},
    paper = {11},
  doi          = {10.1145/3365365.3382193},
  publisher = {{ACM}}
  }

@article{li2023survey,
  title={A survey on explainable anomaly detection},
  author={Li, Zhong and Zhu, Yuxuan and Van Leeuwen, Matthijs},
  journal={Transactions on Knowledge Discovery from Data},
  volume={18},
  number={1},
  pages={1--54},
  year={2023},
  publisher={ACM}
}

@article{formica2022search,
    title={Search-based Software Testing Driven by Automatically Generated and Manually Defined Fitness Functions},
    author={Formica, Federico and Tony, Fan and Menghi, Claudio},
    journal = {ACM Transactions on Software Engineering and Methodology},
    doi = {10.1145/3624745},
    address = {New York, NY, USA},
    year={2023},
    volume = {33},
    number = {2},
    paper = {40},
    articleno = {40},
    publisher = {Association for Computing Machinery}
}

@article{de2024explainability,
  title={Explainability for Property Violations in Cyber-Physical Systems: An Immune-Inspired Approach},
  author={de Araujo, Jo{\~a}o Paulo Costa and Rodrigues, Gena{\'\i}na Nunes and Carwehl, Marc and Vogel, Thomas and Grunske, Lars and Caldas, Ricardo and Pelliccione, Patrizio},
  journal={IEEE Software},
  year={2024},
  publisher={IEEE}
}

@article{stratan2024diagnosing,
  title={Diagnosing Violations of State-based Specifications in iCFTL},
  author={Stratan, Cristina and Mandrioli, Claudio and Bianculli, Domenico},
  journal={IEEE Transactions on Software Engineering},
  year={2026},
  publisher={IEEE}
}

@inproceedings{menghi2019generating,
  title={{Generating automated and online test oracles for Simulink models with continuous and uncertain behaviors}},
  author={Menghi, Claudio and Nejati, Shiva and Gaaloul, Khouloud and Briand, Lionel C},
  booktitle={ACM Joint Meeting on European Software Engineering Conference and Symposium on the Foundations of Software Engineering},
  pages={27--38},
  year={2019}
}

@InProceedings{ARIsTEO,
  author    = {Menghi, Claudio and Nejati, Shiva  and Briand, Lionel and Isasi Parache, Yago},
  title     = {Approximation-Refinement Testing of Compute-Intensive Cyber-Physical Models: An Approach Based on System Identification},
  booktitle = {International Conference on Software Engineering (ICSE)},
  year      = {2020},
    pages = {372–384},
  publisher = {{IEEE} / {ACM}},
}

@article{Zhao2017,
author = {Zhao, Xiao-Wen and Guan, Zhi-Hong and Li, Juan and Zhang, Xian-He and Chen, Chao-Yang},
title = {Flocking of multi-agent nonholonomic systems with unknown leader dynamics and relative measurements},
journal = {International Journal of Robust and Nonlinear Control},
volume = {27},
number = {17},
pages = {3685-3702},
doi = {https://doi.org/10.1002/rnc.3762},
year = {2017}
}

@inproceedings{Reynolds87,
author = {Reynolds, Craig W.},
title = {Flocks, herds and schools: A distributed behavioral model},
year = {1987},
publisher = {Association for Computing Machinery},
address = {New York, NY, USA},
doi = {10.1145/37401.37406},
booktitle = {Proceedings of the 14th Annual Conference on Computer Graphics and Interactive Techniques},
pages = {25–34},
numpages = {10},
series = {SIGGRAPH '87}
}

@article{SMITH1981195,
title = {Identification of common molecular subsequences},
journal = {Journal of Molecular Biology},
volume = {147},
number = {1},
pages = {195-197},
year = {1981},
issn = {0022-2836},
doi = {https://doi.org/10.1016/0022-2836(81)90087-5},
author = {T.F. Smith and M.S. Waterman}
}

@article{boufaied2023,
  title={Trace Diagnostics for Signal-based Temporal Properties},
  author={Boufaied, Chaima and Menghi, Claudio and Bianculli, Domenico and Briand, Lionel C},
  journal={IEEE Transactions on Software Engineering},
  year={2023},
  publisher={IEEE},
  volume={49},
  number={5},
  pages={3131-3154},
  doi={10.1109/TSE.2023.3242588}
}

@inproceedings{dou2018,
author = {Dou, Wei and Bianculli, Domenico and Briand, Lionel},
title = {Model-Driven Trace Diagnostics for Pattern-Based Temporal Specifications},
year = {2018},
isbn = {9781450349499},
publisher = {ACM/IEEE},
doi = {10.1145/3239372.3239396},
booktitle = {International Conference on Model Driven Engineering Languages and Systems},
pages = {278–288},
numpages = {11},
series = {MODELS}
}

@inproceedings{gurfinkel2003proof,
  title={Proof-like counter-examples},
  author={Gurfinkel, Arie and Chechik, Marsha},
  booktitle={International Conference on Tools and Algorithms for the Construction and Analysis of Systems (TACAS)},
  pages={160--175},
  year={2003},
  organization={Springer}
}

@article{timm2020model,
  title={Model checking safety and liveness via k-induction and witness refinement with constraint generation},
  author={Timm, Nils and Gruner, Stefan and Nxumalo, Madoda and Botha, Josua},
  journal={Science of computer programming},
  volume={200},
  pages={102532},
  year={2020},
  publisher={Elsevier}
}

@inproceedings{funke2020farkas,
  title={Farkas certificates and minimal witnesses for probabilistic reachability constraints},
  author={Funke, Florian and Jantsch, Simon and Baier, Christel},
  booktitle={International Conference on Tools and Algorithms for the Construction and Analysis of Systems},
  pages={324--345},
  year={2020},
  organization={Springer}
}

@inproceedings{griggio2018certifying,
  title={{Certifying proofs for LTL model checking}},
  author={Griggio, Alberto and Roveri, Marco and Tonetta, Stefano},
  booktitle={2018 Formal Methods in Computer Aided Design (FMCAD)},
  pages={1--9},
  year={2018},
  organization={IEEE}
}

@inproceedings{bochot2010paths,
  title={Paths to property violation: A structural approach for analyzing counter-examples},
  author={Bochot, Thomas and Virelizier, Pierre and Waeselynck, H{\'e}l{\`e}ne and Wiels, Virginie},
  booktitle={2010 IEEE 12th International Symposium on High Assurance Systems Engineering},
  pages={74--83},
  year={2010},
  organization={IEEE}
}

@inproceedings{zheng2021flack,
  title={Flack: Counterexample-guided fault localization for alloy models},
  author={Zheng, Guolong and Nguyen, ThanhVu and Brida, Sim{\'o}n Guti{\'e}rrez and Regis, Germ{\'a}n and Frias, Marcelo F and Aguirre, Nazareno and Bagheri, Hamid},
  booktitle={2021 IEEE/ACM 43rd International Conference on Software Engineering (ICSE)},
  pages={637--648},
  year={2021},
  organization={IEEE}
}

@inproceedings{hantry:hal-01354475,
  TITLE = {{Handling Conflicts in Depth-First-Search for LTL Tableau to Debug Compliance Based Languages}},
  AUTHOR = {Hantry, Fran{\c c}ois and Hacid, Mohand-Said},
  BOOKTITLE = {{Fifth Workshop on Formal Languages and Analysis of Contract-Oriented Software (FLACOS)}},
  publisher={Open Publishing Association},
  address={M{\'{a}}laga, Spain},
  PAGES = {39-53},
  YEAR = {2011},
  DOI = {10.4204/EPTCS.68.5},
}

@article{schuppan2012towards,
  title={Towards a notion of unsatisfiable and unrealizable cores for LTL},
  author={Schuppan, Viktor},
  journal={Science of Computer Programming},
  volume={77},
  number={7-8},
  pages={908--939},
  year={2012},
  publisher={Elsevier}
}

@inproceedings{menghi2020integrating,
  title={Integrating Topological Proofs with Model Checking to Instrument Iterative Design},
  author={Menghi, Claudio and Rizzi, Alessandro Maria and Bernasconi, Anna},
  booktitle = {Fundamental Approaches to Software Engineering},
  pages={53--74},
  year={2020}
}

@article{nivckovic2020amt,
  title={AMT 2.0: qualitative and quantitative trace analysis with extended signal temporal logic},
  author={Ni{\v{c}}kovi{\'c}, Dejan and Lebeltel, Olivier and Maler, Oded and Ferr{\`e}re, Thomas and Ulus, Dogan},
  journal={International Journal on Software Tools for Technology Transfer},
  volume={22},
  pages={741--758},
  year={2020},
  publisher={Springer}
}

@article{mukherjee2012computing,
  title={Computing minimal debugging windows in failure traces of AMS assertions},
  author={Mukherjee, Subhankar and Dasgupta, Pallab},
  journal={IEEE Transactions on Computer-Aided Design of Integrated Circuits and Systems},
  volume={31},
  number={11},
  pages={1776--1781},
  year={2012},
  publisher={IEEE}
}

@inproceedings{dawes2019,
  title={Explaining violations of properties in control-flow temporal logic},
  author={Dawes, Joshua Heneage and Reger, Giles},
  booktitle={International Conference on Runtime Verification (RV)},
  pages={202--220},
  year={2019},
  organization={Springer}
}

@inproceedings{ferrere2015trace,
  title={Trace diagnostics using temporal implicants},
  author={Ferr{\`e}re, Thomas and Maler, Oded and Ni{\v{c}}kovi{\'c}, Dejan},
  booktitle={International Symposium on Automated Technology for Verification and Analysis},
  pages={241--258},
  year={2015},
  organization={Springer}
}

@article{beer2012,
  title={Explaining counterexamples using causality},
  author={Beer, Ilan and Ben-David, Shoham and Chockler, Hana and Orni, Avigail and Trefler, Richard},
  journal={Formal Methods in System Design},
  volume={40},
  pages={20--40},
  year={2012},
  publisher={Springer}
}

@article{pearl2003statistics,
  title={Statistics and causal inference: A review},
  author={Pearl, Judea},
  journal={Test},
  volume={12},
  pages={281--345},
  year={2003},
  publisher={Springer}
}

@article{mitchell1999machine,
  title={Machine learning and data mining},
  author={Mitchell, Tom M},
  journal={Communications of the ACM},
  volume={42},
  number={11},
  pages={30--36},
  year={1999},
  publisher={ACM New York, NY, USA}
}

@article{goldberg1994genetic,
  title={Genetic and evolutionary algorithms come of age},
  author={Goldberg, David E},
  journal={Communications of the ACM},
  volume={37},
  number={3},
  pages={113--120},
  year={1994},
  publisher={Association for Computing Machinery, Inc.}
}

@INPROCEEDINGS{dpTaliro,
    author={Fainekos, Georgios E. and Sankaranarayanan, Sriram and Ueda, Koichi and Yazarel, Hakan},
    booktitle={2012 American Control Conference (ACC)}, 
    title={{Verification of automotive control applications using S-TaLiRo}}, 
    year={2012},
    volume={},
    number={},
    pages={3567-3572},
    doi={10.1109/ACC.2012.6315384}
}

@online{ZenodoAppendix,
    title={{Appendix: Tool vs Prediction}},
    howpublished={\url{https://doi.org/10.5281/zenodo.12520834}},
    month         = "June",
    year          = "2024 [Online]",
    lastaccessed = {June 2024},
    key = {Appendix}
}
\bibliographystyle{IEEEtran}

\setcounter{figure}{0}
\setcounter{table}{0}

\end{document}